\documentclass[%
reprint,
nofootinbib,
 amsmath,amssymb,+--
 prl
]{revtex4-2}

\usepackage{graphicx}
\usepackage{dcolumn}
\usepackage{bm}
\usepackage{graphicx}
\usepackage{epstopdf, epsfig}
\usepackage{float}
\usepackage{subfig}
\usepackage{amsmath,bm}
\usepackage{bbold}
\usepackage{dsfont}
\usepackage{comment}
\usepackage[dvipsnames]{xcolor}
\usepackage{mathtools}
\usepackage{float}
\usepackage{subfig}
\usepackage{amsmath,amssymb}
\usepackage{comment}
\usepackage[dvipsnames]{xcolor}
\usepackage{adjustbox}
\usepackage{multirow}
\usepackage{cases}
\usepackage{scalerel,stackengine}
\usepackage{appendix}
\usepackage{mathrsfs}
\usepackage{enumerate}

\stackMath

\newcommand\reallywidehat[1]{%
\savestack{\tmpbox}{\stretchto{%
  \scaleto{%
    \scalerel*[\widthof{\ensuremath{#1}}]{\kern.1pt\mathchar"0362\kern.1pt}%
    {\rule{0ex}{\textheight}}
  }{\textheight}%
}{2.4ex}}%
\stackon[-6.9pt]{#1}{\tmpbox}%
}

\newcommand{\sgn}{{\rm sgn}}
\newcommand{\be}{\mathbf{e}}
\newcommand{\bu}{\boldsymbol{u}}

\newcommand{\bU}{\boldsymbol{U}}
\newcommand{\bforc}{\boldsymbol{f}}

\newcommand{\bx}{\boldsymbol{x}}

\newcommand{\bomega}{\boldsymbol{\omega}}
\newcommand{\bk}{\boldsymbol{k}}
\newcommand{\bq}{\boldsymbol{q}}
\newcommand{\bp}{\boldsymbol{p}}

\newcommand{\bh}{\boldsymbol{h}}

\newcommand{\uv}{\langle uv \rangle}

\newcommand{\Roe}{Ro_{\epsilon}}

\newcommand{\Vp}{V_{\bp\bk\bq}}

\newcommand{\pyf}{p_y^f}

\newcommand{\Up}{{U'}}
\newcommand{\Vol}{\mathcal{V}}

\newcommand{\Ttwo}{T_{\rm 2D}}
\newcommand{\Tthree}{T_{\rm 3D-2D}}
\newcommand{\Ttwotwo}{T_{\rm 2D-2D}}

\newcommand\SG[1]{{\color{black}#1}}

\newcommand\blue[1]{{\color{blue}#1}}

\newcommand\resub[1]{{\color{black}#1}}

\begin{document}
\preprint{APS/123-QED}

\author{S\'ebastien Gom\'e
}
\author{Anna Frishman}

\affiliation{
Department of Physics, Technion Israel Institute of Technology, 32000 Haifa, Israel
}

\date{\today}
\title{Waves maintain large-scale 2D flows in rotating turbulence and cause their demise.}

\begin{abstract}
Turbulence follows a few well-known organizational principles, rooted in conservation laws.
One such principle states that a system conserving two sign-definite invariants self-organizes into large-scale structures.
Ordinary three-dimensional turbulence does not fall within this paradigm, but is profoundly altered when subject to rotation. In rotating turbulence,
%
3D inertial waves coexist alongside emergent two-dimensional structures, which tend to take the form of domain-scale flows called condensates.
This interplay raises a fundamental question: why and when are 2D flows sustained if only 3D waves are excited?
We develop a quasi-linear wave–kinetic theory to answer this question. We show that near-resonant 
interactions between 3D waves and a large-scale 2D flow impose an additional conservation law: 
waves must conserve their helicity separately for each helicity sign.
This emergent sign-definite invariant constrains the waves to transfer their energy to large-scale 2D motions\resub{, which maintains the latter in statistical steady state}.
We derive analytical expressions for the 3D–2D energy transfer as a function of rotation, Reynolds number and domain geometry \resub{in a rotation dominated regime}, and compare them with extensive numerical simulations of the rotating 3D Navier–Stokes equations.
As rotation increases,
the energy transfer from 
the waves to the 2D flow
progressively vanishes \resub{as the two decouple}, leading to a 
transition between distinct 
classes of turbulence: from 2D-dominated to 
3D-dominated wave
turbulence. Our theory shows that this gradual transition is caused by
a depletion of modes satisfying the resonance conditions, and exhibits good agreement with numerical simulations when the number of near-resonant modes is not too small. We discuss such limitations of our theory,
as well as the validity range of its underlying assumptions.
Together, these results suggest a mechanism underlying two-dimensionalization in rotating turbulence, and, more broadly, illustrate how
non-linear systems sustaining waves can self-organize into anisotropic, zero-frequency structures.
\end{abstract}

\maketitle

Rotating turbulent flows abound in nature, from industrial applications to atmospheric \citep{callies2014transition}, oceanic \citep{balwada2022direct} and even astrophysical
settings \citep{guervilly2019turbulent}.
The distinct properties of rotating turbulence feature prominently in climate and weather modeling
\cite{ghil2020physics}: namely, the spontaneous formation of two-dimensional flows and their effective decoupling from three-dimensional motions. The latter take the form of 
waves, while two-dimensional structures do not behave as waves, instead corresponding to zero-frequency
modes. 

The concentration of energy into zero-frequency modes, implying
the self-organization of turbulence into anisotropic structures, is not unique to rotating turbulence. Rather, such self-organization is 
shared by a broader class of nonlinear hydrodynamical systems that support waves. Examples range from
zonal flows in magnetized plasmas \citep{diamond2005zonal}, to jets in the atmospheres of planets \citep{vallis1993generation, connaughton2015rossby}, layering in stratified flows \citep{caulfield2021layering,smith2002generation,labarre2026distinguished} and structure formation in chiral fluids \citep{de2024pattern}.  
Yet, despite its ubiquity, this phenomenon lacks a first-principles explanation.

The classical 
understanding of how various degrees of freedom interact in turbulence relies on inviscid conservation laws, which lead to the formation of out-of-equilibrium fluxes.
A remarkable manifestation is provided by
systems which conserve two sign-definite, scale-related quantities, 
causing turbulence to self-organize into
large-scale structures.
The archetype system following this principle is the two-dimensional Navier-Stokes equations (2DNSE), in which both energy and enstrophy (the vorticity squared) are conserved and flow in opposite directions:
 energy cascades upscale and enstrophy downscale \citep{fjortoft1953changes, kraichnan1967inertial}.
In a finite domain and with low dissipation, energy is ultimately concentrated on the lowest modes and forms a condensate: a strong large-scale coherent flow fueled by small-scale eddies.

Three-dimensional fluids, in contrast, conserve energy and sign-indefinite helicity (the scalar product between velocity and vorticity \citep{moffatt1969degree})
and are organized very differently:
both energy and helicity are transferred to small scales \citep{chen2003joint, DitlevsenCascades}.
However, when a 3D fluid is rotated, its phenomenology fundamentally changes, while its invariants remain the same.
Rotation produces 3D inertial waves \citep{yarom2013experimental}, each with an energy and a helicity of a given sign, while also tending to homogenize the flow along
the rotation
axis, generating two-dimensional motions \citep{godeferd2015structure}.
Once energized, 2D modes seem to exhibit an inverse energy cascade, similarly to 2DNSE, 
which progressively generates
larger-scale flows \citep{smith1996crossover, smith1999transfer, smith2005near,chen2005resonant, 
mininni2009scale,
mininni2010rotating,
yarom2013experimental, deusebio2014dimensional,
campagne2015disentangling, biferale2016coherent,
buzzicotti2018inverse, alexakis2018cascades, 
van2020critical,
lam2023supply,
shaltiel2024direct}.
With low dissipation, this leads to the generation of large-scale condensates in these rotating 3D flows \citep{ kolvin2009energy,rubio2014upscale,guervilly2014large,
le2017inertial,  clark2020phase,guzman2020competition,kolokolov2020structure,
de2022discontinuous,shaltiel2024direct}.

How and why two-dimensionalization occurs in rotating turbulence
remains unsettled.
Generally, for fast-enough rotation, 
 nonlinear interactions should be restricted to resonances, 
 but such 3D-2D interactions have a strictly vanishing coupling \citep{waleffe1992nature,  cambon1997energy, bordes2012experimental}.
This causes 3D and 2D modes to decouple in the limit of infinite rotation, as established mathematically in
\citep{babin1997regularity, babin1999global}, which stabilizes 2D motions when those are directly 
excited \citep{gallet2015exact, billant2021taylor}.

However, most experiments and numerical simulations do not reach this asymptotic limit, and energy is observed to be transferred from 3D to 2D modes \citep{smith1996crossover, smith1999transfer, smith2005near,chen2005resonant, bourouiba2007intermediate,
mininni2010rotating,
yarom2013experimental, deusebio2014dimensional,
campagne2015disentangling, biferale2016coherent,
buzzicotti2018inverse,  le2017inertial, alexakis2018cascades, 
van2020critical,
lam2023supply,
shaltiel2024direct}. This possibly occurs via near-resonant triads \citep{waleffe1992nature, 
smith1999transfer, 
smith2005near, alexakis2015rotating, di2016quantifying, le2020near} or four-wave interactions \citep{brunet2020shortcut},
but it is unclear why energy is transferred directionally
\emph{from} 3D \emph{to} 2D.
To date, no mechanism for such a transfer based on the flow's global inviscid invariants has been established.

Here we focus on two-dimensional condensates in rotating 3D turbulence, and identify such a mechanism in this context.
Large-scale steady condensates
alter the nature of turbulence.
They favor non-locality and a different hierarchy of correlations.
Contrary to homogeneous and isotropic turbulence, the dynamics is \resub{expected to be} dominated by mean-flow-turbulence interactions.
Quasi-Linear (QL) theory (inherited from Rapid Distortion Theory \citep{batchelor1954effect})
\resub{can then} provide a self-consistent perturbative statistical framework for this kind of inhomogeneous turbulence \citep{marston2023recent}. 
\resub{This approach has proved successful when applied to the 2DNSE for} 
analytical predictions of both mean-flow profiles and turbulent correlations \citep{laurie2014universal,kolokolov2016structure,frishman2017culmination,frishman2018turbulence,doludenko2021coherent}, and is able to capture many aspects of 
2D geophysical and plasma systems \citep{diamond2005zonal, srinivasan2012zonostrophic,parker2013zonal,marston2016generalized,woillez2017theoretical,svirsky2023two}. This is the approach we will follow here, focusing on solutions of the QL equations in statistical steady state.

We investigate the sustainment of condensates in rotating 3D turbulence using direct numerical simulations (\S \ref{sec:numerics}) and a quasi-linear wave–kinetic theory, describing non-linear interactions between a large-scale 2D mean flow and 3D inertial waves
(\S \ref{sec:homoA} \& \ref{sec:near_reso}).
We show that in this set-up near-resonant interactions enforce a conservation of helicity by sign, causing inertial waves to energize the large-scale 2D flow, and present a qualitative argument for this mechanism for scale-separated 3D-2D interactions more generally (\S \ref{sec:conservation}). 
As rotation increases, we find that the 3D-2D energy transfer becomes vanishingly-small and the flow transitions from
a condensate-dominated turbulence to an asymptotic turbulence of inertial waves.
We argue that this transition is due to resonances becoming increasingly stringent, and derive scaling predictions for the 3D–2D energy transfer and condensate amplitude, in good agreement with numerical simulations for our highest Reynolds numbers in this regime. However, our theory and its predictions become invalidated when near-resonant wave-condensate interactions are too sparse. We offer a qualitative description of this regime and discuss the limitations of our theory more broadly (\S \ref{sec:decoupling}).
Taken together, these results provide an important first step for the understanding of the maintenance and evolution of 
two-dimensional condensates.

\section{Numerical set-up and observations}
\label{sec:numerics}

We consider an incompressible rotating 3D fluid governed by the Navier-Stokes equations (3DNSE)
\begin{align}
	\partial_t \bu + \bu \cdot \nabla \bu &=- 2\Omega \be_z \times \bu  - \nabla p + \nu \nabla^2 \bu + \nu_h \nabla^{2h} \bu +  \bforc,
	\label{eq:3DNSE}
\end{align} 
in a periodic domain of dimensions $(L_x,L_y,L_z)$. We assume solid body rotation in the $\be_z$ direction, with rotation rate $\Omega$, entering the equations through the Coriolis force.
The forcing $\bforc$ is a random white noise centered on a Fourier shell of radius $k_f$ and width $dk=1$. 
The force has zero mean and a correlation function given by  
\begin{equation}
  \langle f_{\bp,i} (t) f_{\bq,j} (t') \rangle = \left( \delta_{i,j} - \frac{p_i p_j}{p^2} \right)  \epsilon \chi_{\bp} \delta_{\bp+\bq} \delta(t-t'),
\end{equation}
where, here and in the following, $\delta_{g(x)}$ is the discrete delta function equal to one when $g(x)=0$ and to zero otherwise. We set 
$\chi_{\bp} = \delta_{p - k_f \pm 1} /(4 \pi k_f^2 dk \Vol)  $ with volume $\Vol = L_x L_y L_z /(2\pi)^3$, such that energy is injected isotropically at a total rate $\epsilon$.
Note that, in this set-up, energy is injected in both 3D and 2D modes (i.e., $z$-invariant modes, with $p_z=0$) at rates $\epsilon_{\rm 3D}$ and $\epsilon_{\rm 2D}$, respectively, such that
\begin{align}
    \epsilon_{\rm 2D} + \epsilon_{\rm 3D} = \epsilon, ~~~ \epsilon_{\rm 2D}\ll \epsilon_{\rm 3D}.
    \label{eps_divided}
\end{align}
%
%
%
Nondimensional Reynolds and Rossby numbers defined at the forcing scale $l_f\equiv 2\pi/k_f$ are given by
$Re= \epsilon^{1/3} k_f^{-4/3} /\nu$
and
$Ro = \epsilon^{1/3} k_f^{2/3} /(2\Omega)$.
We simulate Eq.~\eqref{eq:3DNSE} with the pseudospectral code GHOST \citep{mininni2011hybrid}, with resolution $(N_x,N_y,N_z)=128 (L_x,L_y,L_z) /\pi$.
The inertial range of the forward energy cascade is controlled by an eighth-order hyperviscosity $\nu_8= 6 \times 10^{-29}$.

\begin{figure}[t]
\subfloat{\includegraphics[width=\columnwidth]{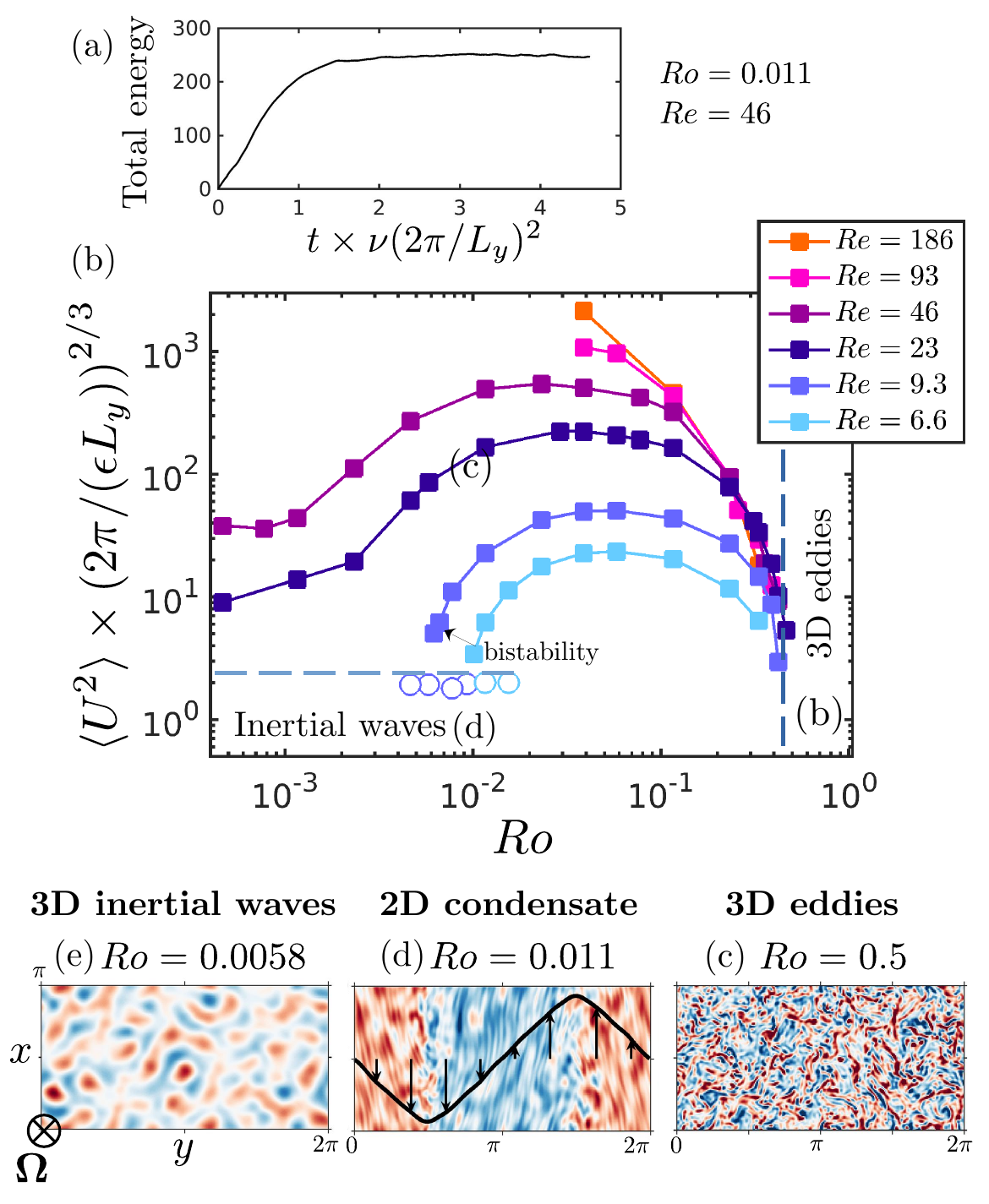}}
\caption{
Formation of large-scale 2D condensates in rotating 3DNSE \eqref{eq:3DNSE}
(a) Temporal evolution of the total energy for $Ro=0.011$ and $Re= 46$, plotted in viscous time scale.
(b) Condensate energy (normalized by $(\epsilon L_y/2\pi)^{2/3}$) as a function of $Ro$ and $Re$.
(b-d) Flow visualizations of vertical vorticity at various values of $Ro$ and fixed $Re=9.3$.
At low rotation $Ro \geq 0.5$ (c), the flow is not rotationally-constrained and only exhibits a turbulence of 3D eddies.
With decreased $Ro$, the flow becomes $z$-invariant and takes the form of box-filling jets (d).
At very low $Ro$ for this $Re$ (e), the two-dimensionalization stops ($\langle U^2 \rangle=0$), and the flow consists of 3D inertial waves (circles in (a), which show the total energy).
}
   \label{fig1}
\end{figure}
Consistently with previous observations \citep{godeferd2015structure, clark2020phase}, a large-scale 2D flow spontaneously emerges in our simulations for a range of parameters, see Fig.~\ref{fig1}. Here we focus on the statistical steady state, and do not address the build-up stage during which energy is initially transferred from 3D modes to 2D modes, which could be due to triadic near-resonances or quartetic resonances \citep{brunet2020shortcut, le2020near}. The saturation of the condensate amplitude, as illustrated in Fig.~\ref{fig1}(a), is here controlled by regular viscosity, since we do not include Ekman drag (see also \cite{guervilly2014large,guervilly2017jets,seshasayanan2018condensates,clark2020phase}).
We use a rectangular box with $L_y=2L_x = 2\pi$, 
chosen such that the condensate takes the form of jets~\cite{bouchet2009random,frishman2017jets,guervilly2017jets},
which allows for a simpler theoretical treatment.
Note that the theory can be easily adapted to the vortex case $L_x=L_y$ and to the inclusion of friction $\alpha$ by replacing $\nu k_y^2$ by $\alpha$.
Note finally that at very low rotation, the condensate takes the form of localized vortices instead of jets in this rectangular domain, which we further discuss in \citep{gome2025helicity}.

We first explore the variation of the energy of the condensate with the external control parameters $Ro$ and $Re$, using the energy in 2D modes, $U^2 = \langle |\int\bu \text{d}z|^2 \rangle$, as a proxy (Fig.~\ref{fig1}(b)). 
The condensate emerges when rotation is strong enough, for $Ro \lesssim 0.5$, below which the forcing excites rotation-dominated modes with a period faster than the eddy-turnover time ($\tau_\Omega\equiv 1/\Omega < \tau_{nl} (k_f)=\epsilon^{-1/3}k_f^{-2/3}$). 
In contrast, for $Ro \gtrsim 0.5$, the forcing shell is not rotationally-constrained and the flow only exhibits the usual 3D forward energy cascade, see Fig.~\ref{fig1}(c).

When rotation is increased, the condensate grows in amplitude. 
%
However, notably, the energy of the condensate starts decaying with rotation below a $Re$-dependent  value of $Ro$ (e.g $Ro = 0.02$ for $Re=23$). At very large rotation and for low enough $Re$, the condensate is replaced by a state of 3D inertial waves, shown in Fig.~\ref{fig1}(e).
The hollow circles in Fig.~\ref{fig1}(b) show the total energy for this flow. 
For higher $Re$, the amplitude of the condensate seems to instead asymptotically saturate at low $Ro$ (around $Ro = 0.001$ for $Re=23$). 
Note that the system at low $Ro$ and low $Re$ exhibits bistability and hysteresis: 
if initialized with $\bu=0$, the inertial-wave state is obtained, as the flow does not spontaneously two-dimensionalize. On the other hand, the condensate state is obtained when traced from
high to low $Ro$ by an adiabatic decrease
(the squares obtained at the same $Ro$ as hollow circles in Fig.~\ref{fig1}(b) for $Re=6.6$ and $9.3$ are obtained in this way).

The decrease in condensate amplitude with $Ro$ in our DNS, and its complete disappearance for low $Re$, 
is a signature of \emph{decoupling} between 3D and 2D modes, expected in rotating 3DNSE when $Ro\to 0$ ($\Omega\to \infty$).
In this limit, 
the horizontal components of
$z-$averaged solutions of rotating 3DNSE are expected to converge to solutions of 2DNSE, 
as established mathematically in \citep{babin1997regularity}.
Our numerical results provide compelling evidence of this phenomenon, thereby complementing previous numerical work \citep{chen2005resonant}, 
as well as stability analyses of a single 3D wave \citep{le2020near} or of a 2D flow \citep{gallet2015exact, seshasayanan2020onset, billant2021taylor, lohani2024effect}.
In the following sections, we will provide a quantitative theoretical description of the gradual nature of this 2D-3D decoupling, and the non-dimensional parameters controlling it. 

\begin{figure}[t]
\subfloat{\includegraphics[width=0.97\columnwidth]{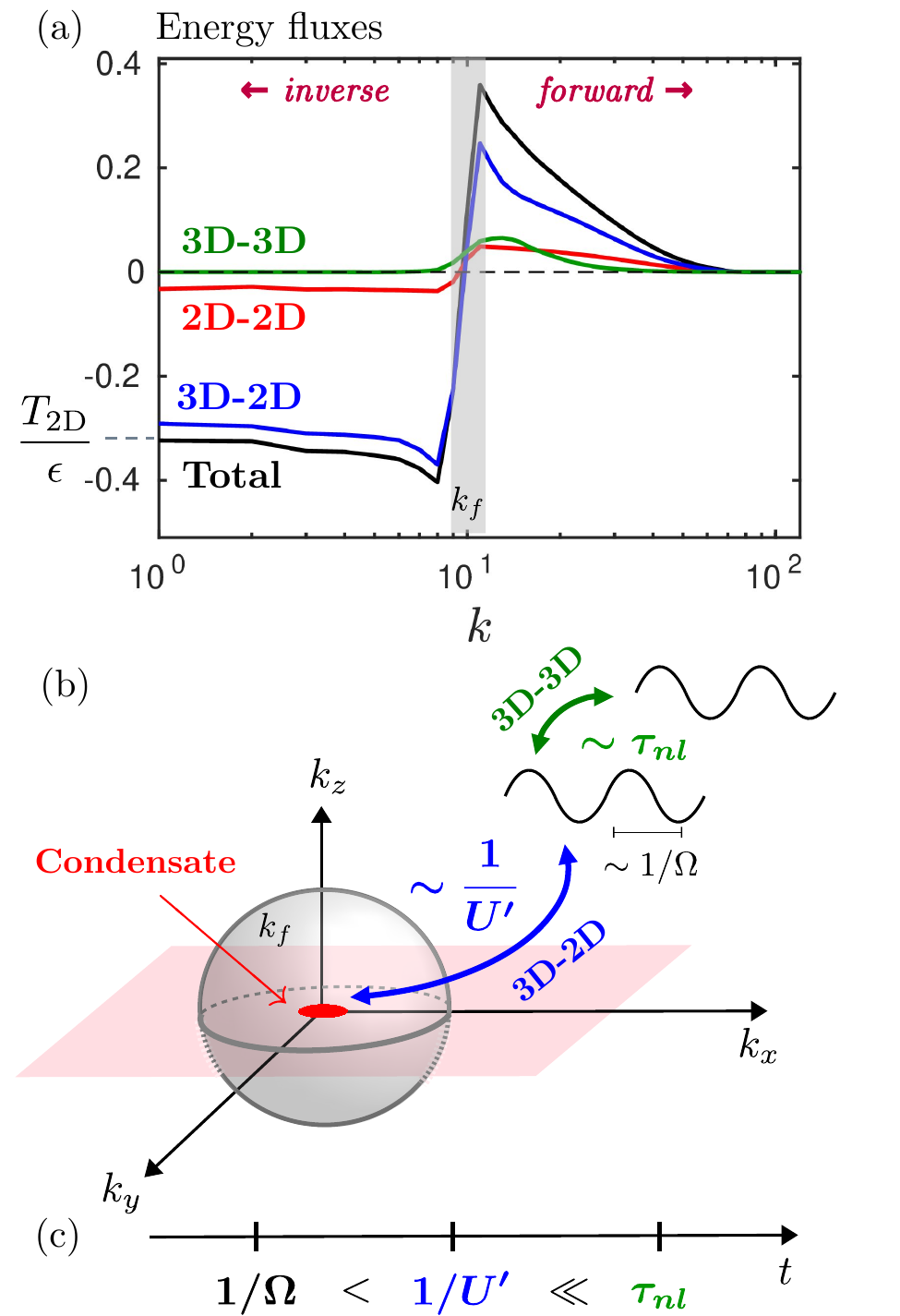}}
\caption{
(a) Energy fluxes across scales, measured from the DNS at $Ro= 0.011$, $Re=23$. 
(b) Interaction types and (c) time-scale hierarchy in rotating 3D NSE 
sustaining a condensate of shear-rate $\Up$.
The 3D-forced flow is dominated by 3D-2D interactions occurring over a time scale $\sim 1/\Up \ll \tau_{nl}$, the characteristic time scale of 3D-3D interactions.
Under fast rotation $ \Up \ll 2\Omega$, 3D modes consist in inertial waves and all interactions are restricted by wave resonances.
}
   \label{fig:flux_cond}
\end{figure}

Our modeling approach will rely on a Quasi-Linear (QL) type of approximation, justifiable in the presence of a strong condensate 
with which 3D fluctuations primarily interact --- instead of with any intermediate 2D mode. We discuss the range of validity of this approach in \S\ref{sec:homoA}.
In our DNS, the relevance of this picture is confirmed by analyzing 
the spectral energy fluxes, decomposed into different interaction types, 3D-2D, 2D-2D and 3D-3D (see \citep{buzzicotti2018energy}),
as showcased in 
Fig.~\ref{fig:flux_cond}(a) for $Ro = 0.011$ and $Re= 23$.
The large scales, dominated by the condensate, are primarily energized by 3D modes excited around $k_f$, while 2D-2D interactions are subdominant.
Meanwhile, only a small portion of the injected energy reaches small scales for this $Ro$. Such a directional transfer of energy \textit{from} 3D modes \textit{to} 2D modes appears to be ubiquitous in rotating flows \citep{smith1999transfer, kolvin2009energy,rubio2014upscale,guervilly2014large,clark2020phase,guzman2020competition,kolokolov2020structure,de2022discontinuous,shaltiel2024direct}.
One goal of the present work is to suggest a first-principles explanation of this phenomenon.

\section{Quasi-linear approximation and effect of rotation}
\label{sec:homoA}

We consider a stationary mean flow 
$\bU=\langle \bu\rangle= U(y) \be_x$ consisting of two $x$-invariant jets, corresponding to the condensate in the 
rectangular
domain of Fig.~\ref{fig1}. 
The average $\langle \cdot \rangle$ denotes ensemble (over forcing realizations), temporal, and vertical averaging.
In the following, we denote by 
\begin{align}
    \Up \equiv \sqrt{\frac{1}{L_y}\int (\partial_y U)^2 dy }
\end{align}
the root-mean-square  shear rate, averaged over $y$.
This defines a typical time scale $1/\Up$
over which turbulent fluctuations $\bu'$ interact with the condensate.

Interactions with the condensate are expected to dominate over eddy-eddy or wave-wave interactions if the mean-flow shear is fast enough, 
$1/\Up \ll \tau_{nl}$, implying that
$\bu' \cdot \nabla \bu' \ll \bu' \cdot \nabla \bU $, i.e., a
Quasi-Linear (QL) dynamics for the fluctuations \citep{frishman2017culmination,marston2023recent}.
At large enough rotation, the relevant nonlinear time-scale is the wave-wave interaction time, $\tau_{nl}^{\rm wave} \sim (2 \Omega)^{1/2} \epsilon^{-1/2} k_z^{1/2} k_\perp^{-3/2} $  \citep{galtier2003weak}, which is scale dependent. Consequently, this Quasi-Linear (QL) approximation \citep{frishman2017culmination,marston2023recent} should hold for large enough scales $k<k_U$, with the cutoff wavenumber $k_U$ given by 
\begin{align}  
\frac{k_U}{k_f}  &= 
              \frac{1}{Ro}
              \left(\frac{U'}{\Omega}\right)^{2/3} \left(\frac{k_z}{k_f}\right)^{1/3}.
              \label{eq:kU}
\end{align}
We will assume in the following that $k_U/k_f\gg 1$ (which will be verified self-consistently in \S \ref{sec:validity}),
so that 3D waves dominantly interact with the condensate.
The momentum equations for the mean flow and fluctuations $\bu'$ then read
\begin{align}
\partial_t U &= - \partial_y \uv + \nu \partial_y^2 U =0 \label{eq:mf}
\\
\partial_t \bu' + U(y) \partial_x \bu' &= - \partial_y U v \be_x - 2 \Omega \be_z \times \boldsymbol{u'} -  \nabla p' + \bforc,
\label{eq:QLNSE}
\end{align}
where $u$ and $v$ denote the $x$ and $y$ components of the fluctuating velocity $\bu'$.
In writing the fluctuations equation \eqref{eq:QLNSE} we have omitted
both regular and hyper-viscosity: 
our theory will focus on the range of scales where dissipative effects are negligible for the fluctuations, corresponding to an inertial range where wave-condensate interactions dominate.
From Eq.~\eqref{eq:mf}, the mean flow obeys the global steady-state energy balance
\begin{align}
    \nu (U')^2= \frac{1}{L_y}\int  \uv \partial_y U \text{d}y \equiv \Ttwo,
    \label{eq:mf_bal}
\end{align}
where $\nu (U')^2 = \nu \int (\partial_y U)^2 dy/L_y $ is the energy dissipation by the 2D condensate and $T_{2D}$ is the energy transfer between the fluctuations and the condensate, both spatially averaged.

We set out to answer the most basic question for this system: how does the condensate amplitude (i.e.~energy in 2D modes) depend on the control parameters $Ro,~Re$ and the domain geometry, encoded in the ratios $l_f/L_y,l_f/L_z,l_f/L_x$?
To do so, we must determine the energy transfer between 3D modes and the 2D flow, $\Ttwo$, and, crucially, its direction.
This requires obtaining the Reynolds stress $\uv$ appearing in Eq.~\eqref{eq:mf}, 
by solving the dynamical equation \eqref{eq:QLNSE} for a given mean-flow profile $U(y)$.
Here we will not be interested in the detailed mean-flow profile, but rather look for a global observable encoding the condensate energy content.
For this purpose, the mean flow can be approximated as the lowest single trigonometric mode $U(y) = \Up \sqrt{2}/k_y ~  \sin( k_y y)$  with $k_y= 2\pi /L_y$, where $U'$ is the corresponding rms shear rate. 
Indeed, we find that more than $80\%$ of the mean-flow energy is contained within this mode, see Fig.~\ref{fig:mf_spec} in Appendix \ref{app:simplifying}. This is to be expected on phenomenological grounds, as the emergence of the condensate is due to energy accumulating in the gravest mode\footnote{We do not consider the build-up of the condensate here, but this process seems to follow the 2D inverse cascade phenomenology, energy flowing to the largest available mode, at least at low $Ro$.}. 
Within this single-mode approximation,
the condensate amplitude 
is proportional to the rms shear rate $U' $, which therefore encodes all the relevant information: both the time-scale associated with the mean flow and its energy content. 
It will hence serve as the observable in the following, sufficient to capture 
the evolution of the mean flow energy with $Ro$ and $Re$ 
(Fig.~\ref{fig1}(b)), and to be
determined from the coupled system \eqref{eq:mf}-\eqref{eq:QLNSE} self-consistently.

Having described the mean flow, we now turn to the 3D fluctuations. In the following, we exploit their wave-like character, assuming a time-scale separation so that $U'/2\Omega \ll 1$, similarly to what is done in wave-turbulence theory. 
In the absence of a condensate ($\Up = 0$), the Coriolis force generates inertial waves, $\bh_{\bp}^s  e^{i\omega_{\bp}^s t + i\bp \cdot \bm{x}}$, with dispersion relation
\begin{align}
\omega_{\bp}^{s}= s 2\Omega  \frac{p_z}{p}, \qquad s=\pm1.
    \label{eq:dispersion}
\end{align}
This dispersion relation implies that slow motions are close to being $z$-invariant,
a correspondence embodying the celebrated Taylor-Proudman theorem \citep{proudman1916motion, taylor1917motion, 
 greenspan1969theory}. Here, the vectors
$\bh_{\bp}^s$ form the helical basis
\begin{equation}
\bh_{\bp}^s = \frac{\bp \times [\bp \times \be_z]   - is p \bp \times \be_z}{\sqrt{2} p p_\perp} 
= \frac{1}{\sqrt{2} p p_\perp}\begin{bmatrix}
      	p_x p_z - i s p p_y	 \\
		p_z p_y  + i  s p p_x  \\
	   - p_x^2  -  p_y^2    	
         \end{bmatrix}.          
 \label{eq:helical_basis}
\end{equation}
with $\bp=(p_x,p_y,p_z)$, $p=|\bp|$ and $p_\perp^2 = p_x^2+p_y^2$
\citep{lesieur1987turbulence}, showing that the waves are circularly polarized. 
In addition, this basis satisfies $i \bp \times \bh_{\bp}^s = s p \bh_{\bp}^s $, i.e.~diagonalizes the curl operator with an associated eigenvalue equal to $s p $. This, in turn, implies that inertial waves carry a sign-definite helicity, $ \bu\cdot(\nabla \times \bu)$, 
i.e~that a mode $\bp$ with energy $E_{\bp}$ carries helicity equal to $spE_{\bp}$.

We decompose the fluctuation field as 
\begin{equation}
    \bu'(\bx,t) =\sum_{\bp,s=\pm 1} \bh_{\bp}^s ~a_{\bp}^s (t) ~  e^{i\omega_{\bp}^s t + i\bp \cdot \bm{x}},
\end{equation} 
where $a_{\bp}^s(t)$ is the time-varying amplitude of a helical mode $(\bp,s)$ and $\omega_{\bp}^s$ is the frequency \eqref{eq:dispersion} of the corresponding inertial  wave.
Each Fourier mode $\bp$ thus supports two inertial waves $a_{\bp}^+$ and $a_{\bp}^-$, 
each associated with a \emph{chirality} $s=\pm 1$, corresponding to the handedness of its circular polarization around the wavevector $\bp$. 
This chirality simultaneously determines the sign of the helicity carried by the wave, $s p |a_{\bp}^s|^2$, and its direction of propagation in the $z$ direction, given by $ s ~{\rm{sgn}}(\mp p_z)\be_z$. Waves of opposite chiralities coexist, which causes the total helicity of the flow
$H = \langle \bu\cdot (\nabla \times \bu) \rangle 
 =\sum_{\bp} (p| a_{\bp}^+|^2 - p|a_{\bp}^-|^2) \equiv H^+ + H^-$  to be sign-indefinite.

Projecting Eqs.~\eqref{eq:mf}-\eqref{eq:QLNSE} on the helical basis \eqref{eq:helical_basis}
leads to the triadic system
\begin{align}
\nu k_y^2 U_{\bk} &= - i k_y  \sum_{\bp,\bq} \Big[ C_{\bk\bp\bq}^{ss} 
 \langle a_{\bp}^{s*} a_{\bq}^{s*} ~ e^{-i s \omega_{\bp\bq}^{ss} t} \rangle \nonumber \\
 &~~~~~~~+  C_{\bk\bp\bq}^{s,-s} \langle a_{\bp}^{s*} a_{\bq}^{-s*} ~ e^{-i \omega_{\bp\bq}^{s,-s} t} \rangle  \Big]  \delta_{\bk\bp\bq} \label{eq:triadMF}  \\
 \partial_t a_{\bp}^s &=  \sum_{\bq, \bk}\Big[ \Vp^{ss}  U_{\bk}^* ~ a_{\bq}^{s*}  e^{-i \omega_{\bp\bq}^{ss} t} \nonumber \\
& ~~~ +\Vp^{s,-s}  U_{\bk}^* ~ a_{\bq}^{-s*}  e^{-i  \omega_{\bp\bq}^{s,-s} t}  \Big] \delta_{\bk\bp\bq} + \bforc_{\bp}^s e^{-i\omega_{\bp}^s t}
\label{eq:triad3D}
\end{align}
with $\delta_{\bk\bp\bq} = \delta_{\bk+\bp+\bq}$ and beating parameters
\begin{align}
 \omega_{\bp\bq}^{ss} &\equiv \omega_{\bp}^s + \omega_{\bq}^s= s2 \Omega p_z \Big(\frac{1}{p} - \frac{1}{q} \Big), \\
 \omega_{\bp\bq}^{s,-s} &\equiv \omega_{\bp}^s + \omega_{\bq}^{-s}= s2 \Omega p_z \Big(\frac{1}{p} + \frac{1}{q}\Big).
\end{align}
The coupling coefficients arising from the projection are given by
$ C_{\bk\bp\bq}^{s_1,s_2} =\frac{1}{2} (h_{\bp,x}^{-s_1} ~ h_{\bq,y}^{-s_2} + h_{\bq,x}^{-s_1} ~ h_{\bp,y}^{-s_2})$ and
$\Vp^{s_1,s_2} =  - i p_x ~( \bh_{\bp}^{-s_1}  \cdot \bh_{\bq}^{-s_2})  + i k_y ~ (h_{\bp,x}^{-s_1} ~ h_{\bq,y}^{-s_2})$, for a $y-$dependent shear flow in the $x$ direction.
In this paper, we always refer to $\bk=(0,\pm k_y,0)$ as the single 2D mean-flow mode, and $\bp,\bq$ as 3D modes such that $q_z=-p_z, q_x =-p_x$, $q_y = -p_y - k_y$. 

A 3D mode $a_{\bp}^s$ interacts with the condensate over a characteristic time $1/\Up$ 
via Eq.~\ref{eq:triad3D}.
This interaction involves either a mode $a_{\bq}^s$ of the same chirality $s$ (which we call a \emph{homochiral-wave interaction}), or a mode $a_{\bq}^{-s}$ of opposite chirality (\emph{heterochiral-wave interaction})\footnote{Note that
we do not decompose the 2D mode into different chiralities.
Our classification of homochiral-wave and heterochiral-wave interactions therefore differs from those established in \citep{waleffe1992nature, biferale2012inverse, alexakis2018cascades, 
 buzzicotti2018energy} based on the chirality of the three triadic modes.}.
Waves exchange energy with the condensate through both types of interactions, as seen in Eq.~\ref{eq:triadMF}.

However, in the presence of rotation, the oscillating factors in Eqs.~\eqref{eq:triadMF}-\eqref{eq:triad3D}, $ e^{-i \omega_{\bp\bq}^{ss} t}$, will suppress interactions between waves that satisfy $\omega_{\bp\bq}^{s \tilde{s}}\gg\Up$. This is a type of resonance condition for the condensate-wave system, similar to that in wave turbulence, where wave interactions are restricted to triadic resonances in the limit of fast rotation. As a result, interactions between two 3D waves with wavenumbers $\bp, \bq$ are possible only if 
\begin{align}
    2 p_z \Big(\frac{1}{p} - \frac{1}{q} \Big) &\lesssim \frac{\Up}{\Omega};  \label{eq:general_res}\\
    2 p_z \Big(\frac{1}{p} + \frac{1}{q} \Big) &\underset{k\ll p}{\simeq} \frac{2 \omega_{\bp}^s}{\Omega} \lesssim \frac{\Up}{\Omega}; 
     \label{eq:general_res2}
\end{align}
We will establish these conditions in more detail and discuss their consequences in \S \ref{sec:near_reso}.

Rotation is therefore seen to select only the interactions obeying \eqref{eq:general_res}-\eqref{eq:general_res2} out of the full set of wave-condensate interactions in the 3DNSE, depending only on the value of the rescaled condensate amplitude $\frac{\Up}{\Omega}$. This implies that, in the limit of fast rotation, the energy transfer $\Ttwo$ is a function of $\frac{\Up}{\Omega}$ only. Hence Eq.~\eqref{eq:mf_bal} can be recast in the form
\begin{align}
     \frac{{\Up}^2}{\Omega^2} = 4 \Roe^2 ~ \frac{\Ttwo}{\epsilon} \Big[ \frac{\Up}{\Omega} \Big],
 \label{eq:global_bal}
\end{align}
where
\begin{equation}
  \Roe \equiv \frac{1}{2\Omega} \sqrt{\frac{\epsilon}{\nu}} = Ro \times Re^{1/2}  
\end{equation}
 is a large-scale Rossby number emerging when rescaling Eq.~\eqref{eq:mf_bal} by $\Omega$, see also \cite{kolokolov2020structure,parfenyev2021velocity}.

\begin{figure}[t]
\includegraphics[width=\columnwidth]{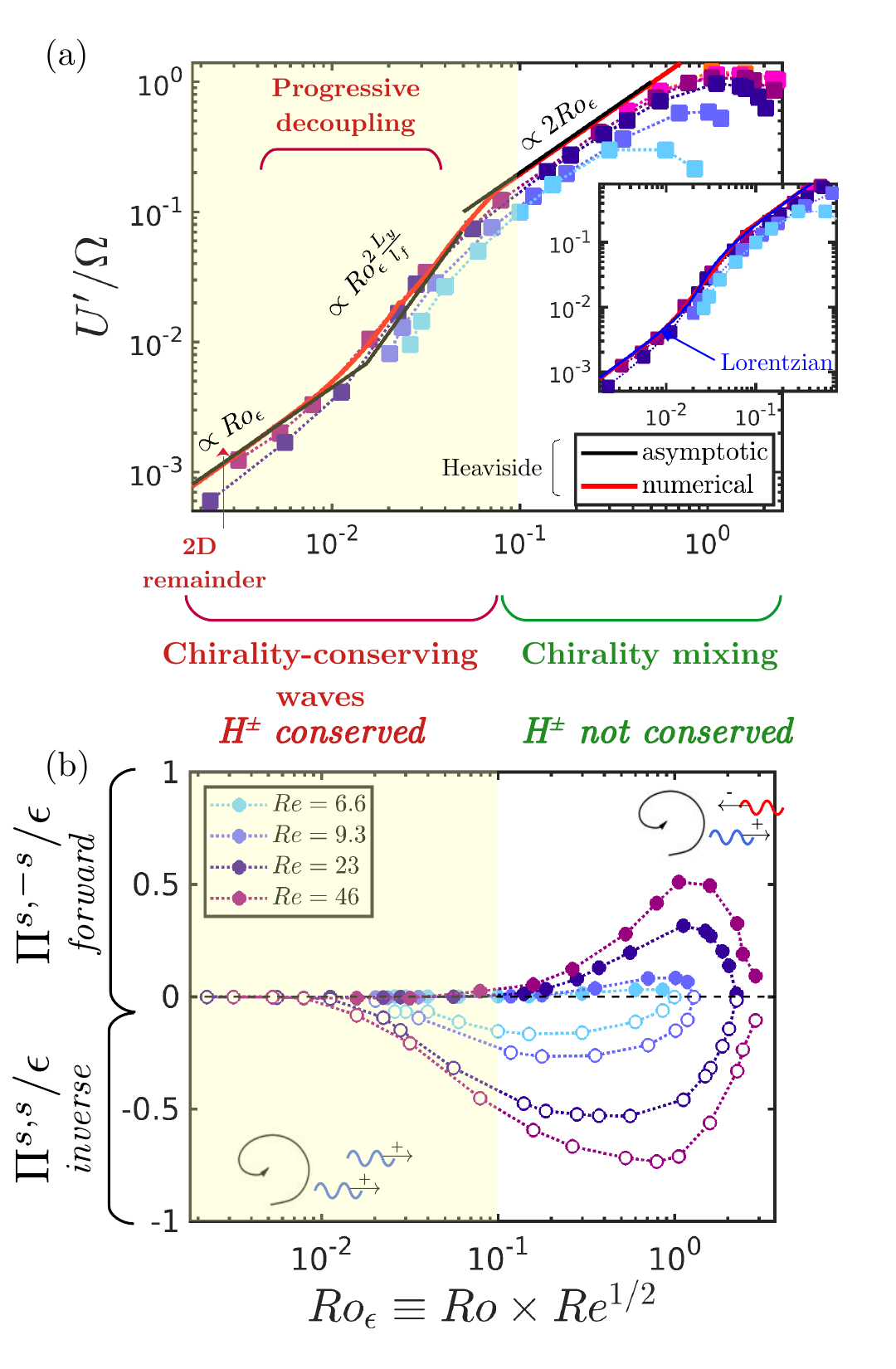} 
\caption{
(a) Rescaled condensate mean shear rate $\Up/\Omega$ from Navier-Stokes simulations (squares), illustrating the collapse with $\Roe=\sqrt{\epsilon/\nu}/(2\Omega) = Ro Re^{1/2}$ (same colors as Fig.~\ref{fig1}).
(b) Energy fluxes to the condensate due to 3D waves of various chiralities.
At high rotation, the condensate-wave interactions do not mix wave chiralities, leading to the conservation of sign-definite helicity $H^{\pm}$ for the waves. As a consequence, waves energize the condensate ($\Pi^{s,s}<0$). The 2D flow eventually decouples from the waves when $\Roe\to0$ following scaling \eqref{eq:U_O_w2d} (black lines in panel (a)).
The red line in (a) shows the numerical closure of \eqref{eq:global_bal} with $\Tthree$ in \eqref{eq:dec_scalings2}.
Inset: comparison with the numerical QL solution obtained with a Lorentzian approximation of the oscillating factor (blue line).
}
\label{fig:cond_rescaled}
\end{figure}

Equation~\eqref{eq:global_bal} describes a self-tuning of the rescaled condensate amplitude, $\Up/\Omega$, such that the condensate exactly dissipates the energy input from the waves. 
It also implies that $\Up/\Omega$ is a function of the single parameter $\Roe$. 
This is confirmed in our DNS data, for which the rescaled condensate amplitude $(\int\bU^2 dy/L_y)^{1/2}/(L_y \Omega)\approx U' /\Omega $ collapses as a function of the control parameter $\Roe$ for all  $\Roe \lesssim 1$ and high $Re$, see Fig.~\ref{fig:cond_rescaled}(a).

To obtain the condensate amplitude as a function of $\Roe$, i.e.\ solve equation~\eqref{eq:global_bal} for $U'/\Omega$, we need to consider the interactions selected by a given value of $U'/\Omega$ in Eqs.~\eqref{eq:general_res}-\eqref{eq:general_res2}. In particular, for heterochiral-wave interactions, condition \eqref{eq:general_res2} reads
\begin{align}
      \omega_{\bp\bq}^{s,-s} \underset{k\ll p}{\simeq} 2 \omega_{\bp}^s = \frac{4 \Omega p_z}{p} &\lesssim \Up.
      \label{eq:hetero_condition}
\end{align}
At sufficiently-large rotation,
when $\Up/\Omega \ll l_f /L_z$, no waves $\bp$
satisfy condition \eqref{eq:hetero_condition} and heterochiral interactions therefore decorrelate.
Thus, for fast enough rotation, we expect that only homochiral-wave interactions would be present.

To verify this in our DNS, we decompose the energy flux to the condensate into contributions from homochiral and heterochiral interactions, by measuring $\Pi^{s_1,s_2} = -\sum_{|\bk| < 5} \sum_{\bk\in 2D, \bp,\bq \in 3D} \langle \bu_{\bk} \cdot (\bu_{\bp}^{s_1} \times {\bomega_{\bq}^{{s_2}})} \rangle \delta_{\bk\bp\bq}$. Here $\Pi^{s,s}$ 
is the transfer between 3D waves of same chirality and 2D modes (the latter not decomposed into the two helical modes). 
Similarly, $\Pi^{s,-s}$ is the energy transfer involving waves of opposite chiralities.

We find that $\Pi^{s,-s} = 0$ below $\Roe \simeq 0.1$, see Fig.~\ref{fig:cond_rescaled}(b). 
In the following, we will focus on the regime $\Roe\lesssim 0.1$, 
where condensate-wave interactions do not couple waves of opposite helicity signs.
The low-rotation range $\Roe\gtrsim 0.1 $ 
will be analyzed in a forthcoming publication~\citep{gome2025helicity}.

\section{Kinetic theory for the mean-wave system}
\label{sec:near_reso}
\subsection{Kinetic equation and near-resonances}

The remaining homochiral interactions require additional treatment: 3D waves and 2D modes are expected to decouple in the infinite rotation limit, making a kinetic theory for the mean-wave system subtle. Here we develop such a theory, showing how the constraint \eqref{eq:general_res} arises.
Assuming $\Up \ll \Omega$, Eq.~\eqref{eq:triad3D} 
can be treated following the procedural steps of wave-kinetic theory \citep{zakharov2012kolmogorov,
 galtier2003weak, newell2011wave, buckmaster2021onset, newell2025augmenting, nazarenko2011wave}, here applied to an inhomogeneous system.
We focus on correlators at an arbitrary time $t=0$, 
$\langle a_{\bp}^s a_{\bq}^s  \rangle (t=0) = 2 \delta_{\bp+\bq} ~ e_{\bp}^s(0) + \langle a_{\bp}^s a_{-\bp+\bk}^s \rangle (0) \delta_{\bk+\bp+\bq}$, 
with $ e_{\bp}^s \equiv \frac{1}{2}\langle |a_{\bp}^s|^2 \rangle$ the energy of a 3D mode $\bp$ and $\langle a_{\bp}^s a_{-\bp + \bk}^s \rangle$ an inhomogeneous correlator associated to the single condensate wavenumber $\bk$.
We consider the wave dynamics on a time scale $t = o (1/\Up)$ which is slow compared to the characteristic wave period $1/\Omega$, and 
obtain a kinetic equation
describing the energy evolution from $t=0$ (see Appendix \ref{app:KE}):
\begin{align}
&\frac{e_{\bp}^s(t) - e_{\bp}^s(0)}{t} 
 = \epsilon \chi_{\bp}^s 
 + \frac{1}{2} \frac{1 - e^{- i \omega_{\bp\bq}^{ss}t}}{i\omega_{\bp\bq}^{ss} t}  U_{\bk}^* \Vp^{ss} \langle a_{\bp}^s a_{\bq}^s \rangle^*  \delta_{\bk\bp\bq}
\nonumber\\
&~~~~~~~~~~~~~~~~~~~~~~~~~~~~~~~~~~
+ (\bk \to -\bk) + c.c.,
\label{eq:KE}
\end{align}
written at first order in the time-scale ratio $U'/\Omega$.

Because the oscillating factor 
$\Delta(t) \equiv \frac{1- e^{-i \omega_{\bp\bq}^{ss}t }}{i\omega_{\bp\bq}^{ss}t}$
in Eq.~\eqref{eq:KE} decays with $\omega_{\bp\bq}^{ss}t$, 
mean-wave interactions are restricted to triads such that $\omega_{\bp\bq}^{ss} t \ll 1$ for $t= o(1/\Up)$, with two possibilities:
\begin{align}
    \begin{cases}
    \text{Exact resonances: }
    \omega_{\bp\bq}^{ss} = 0 \nonumber\\
    \text{Near-resonances: }
    \omega_{\bp\bq}^{ss} \neq 0 \text{ and }
    |\omega_{\bp\bq}^{ss} | t \ll 1.
    \end{cases}
\end{align}

Exact resonances are the only contribution to Eq.~\eqref{eq:KE} in the asymptotic limit $\Omega t \to \infty$ (meaning the limit $Ro\to0$ taken before any other).
However, system \eqref{eq:mf}-\eqref{eq:QLNSE} 
obeys the Greenspan property \citep{greenspan1969non, waleffe1993inertial}:
the coupling to the 2D mode in Eq.~\eqref{eq:triadMF}, proportional to 
\begin{align}
i k_y C_{\bk\bp\bq}^{ss} &= \frac{i k_y}{4 p p_\perp q q_\perp} 
\Big[ (p_y- q_y) p_x (   p_z^2 - pq  ) \nonumber \\
&- i s (p - q) p_z ( p_x^2  + p_y q_y) )  \Big],
\label{eq:Cpq+Cqp}
\end{align}
is null when $\omega_{\bp\bq}^{ss} = 0$, i.e when either $(k_y=0, q_y = - p_y)$ or $(k_y=-2p_y, q_y= p_y)$.
This is a consequence of the triad being constrained by the resonant condition and the conservation of energy and helicity
\citep{greenspan1969non, waleffe1993inertial, shavit2024sign}. 
Therefore, exactly-resonant triads do not transfer energy to the 2D flow. 

However, 3D modes that satisfy the near-resonance condition rather than the exact one, $\omega_{\bp\bq}^{ss} t\ll1$ and in particular $\omega_{\bp\bq}^{ss} /U'\ll1$, \textit{can} exchange energy with 2D modes. 
Secular growth in Eq.~\eqref{eq:KE} indeed occurs on the slow time scale $ t = o( 1/\Up)$ for such triads.
As the condensate occupies the lowest, box-scale, mode $k_y=\pm 2\pi/L_y$, it couples with waves such that $q_y \simeq  - p_y$, hence the corresponding homochiral interaction is naturally near-resonant (i.e.\ close to the exactly-resonant triads $k_y=0, \bp=-\bq$). Indeed, the beating parameter
\begin{align}
    \omega_{\bp\bq}^{ss} = \frac{2\pi}{L_y p}  \frac{2p_y p_z \Omega}{p^2 }+ O \Big( \frac{1}{(L_y p)^2} \Big)
    \label{eq:beating}
\end{align}
can be much smaller than $\Omega$ (in absolute value), suppressed by a factor of $1/(pL_y)$ for such triads. Note that as these interactions are never exactly resonant, the transfer coefficient Eq.~\eqref{eq:Cpq+Cqp} is always finite, proportional to $2\pi/L_y$.

In contrast, modes such that $1/|\omega_{\bp\bq}^{ss}| < t \ll 1/U'$ do not generate a secular growth: their contribution instead approximately averages out on the slow time scales $t=o(1/\Up)$, consistently with the multiple-scale expansion in \citep{bretherton1964resonant}.
The oscillating factor can therefore be modeled as
\begin{align}
    \Delta(t) = \frac{1 - e^{-i \omega_{\bp\bq}^{ss} t } }{\omega_{\bp\bq}^{ss} t} 
   & \simeq  \Theta \left(  1
    - \frac{ |\omega_{\bp\bq}^{ss}|}{\Up}
    \right) \nonumber\\
   & = \begin{cases}
    1 ~~~ \text{if} ~~
    |\omega_{\bp\bq}^{ss}| \leq  \Up  \\
    0 ~~~ \text{if} ~~
    |\omega_{\bp\bq}^{ss}| > \Up, 
    \end{cases}
    \label{eq:filter}
\end{align}
using a 
Heaviside filter equal to one
when $|\omega_{\bp\bq}^{ss}| / \Up < 1$ (near-resonance) and zero otherwise (off-resonance).
In doing so, we entirely neglect
contributions from off-resonant modes,
the latter being
sub-leading compared to near-resonances. The distinction between the two is implemented in the Heaviside filter as a sharp cutoff at $|\omega_{\bp\bq}^{ss}|/U'=1$ (see Appendix \S\ref{app:energy}).

The geometry of a given triad enters condition \eqref{eq:filter} through the geometric factor $|\omega_{\bp\bq}^{ss}|/\Omega$, which determines how close to resonance the triad is. For $\Omega /\Up$ large but finite, triads 
with a sufficiently small geometric factor are near-resonant ($|\omega_{\bp\bq}^{ss} | <\Up$) and
give a secular contribution to \eqref{eq:KE} (see \S \ref{app:continuum}). 
However, any such triad becomes off-resonant if $\Omega/\Up$ is too large ($|\omega_{\bp\bq}^{ss} | \geq \Up$), since $|\omega_{\bp\bq}^{ss}|/\Omega>0$. Such triads will therefore not contribute to \eqref{eq:KE} within our approximation, and will decouple from the 2D flow at leading order.

The sharp distinction between near-resonant and off-resonant modes can seem like a crude approximation, and can be smoothed out using,
e.g., a Lorentzian form given by 
\begin{align}
    \Delta(t) =\frac{1 - e^{-i \omega_{\bp\bq}^{ss} t } }{\omega_{\bp\bq}^{ss} t} \simeq  \frac{1}{1+ \left(\frac{\omega_{\bp\bq}^{ss}}{U'}\right)^2},
    \label{eq:Lorentzian_main}
\end{align}
see \citep{lvov1997statistical, lvov2012resonant}.
Within such an approximation,
off-resonances
$\omega_{\bp\bq}^{ss} \gg U'$ effectively contribute to the secular growth at time $t=o(1/U')$ in Eq.~\eqref{eq:KE}, with an arbitrary weight in $O((U'/\omega_{\bp\bq}^{ss})^2)$.
Our results should be mostly independent of this modeling choice as long as near-resonant terms provide the main contribution and 
non-secular terms can be neglected. The latter are either put to zero within the Heaviside approximation, or to an order $O((\Up/\omega_{\bp\bq}^{ss})^2)$ (for modes $\Up/\omega_{\bp\bq}^{ss}\ll1$) within the Lorentzian approximation.
Either way, our modeling captures reliably only near-resonant contributions, and should be treated with caution when such contributions are scarce. In particular, results obtained using either the Lorentzian or the Heaviside filters need not agree in a regime dominated by contributions from off-resonant modes $\omega_{\bp\bq}^{ss} > U'$.

In most of this paper we will use the sharp Heaviside filter \eqref{eq:filter}, as it enables a fully analytical treatment, and discuss the effect of the Lorentzian filter in \S \ref{app:energy}.

With our modeling of the oscillating factor in \eqref{eq:filter}, we finally obtain a kinetic equation at $O(\Up/\Omega)$ (see the discussion of the order in Appendix \ref{app:continuum}),
\begin{align}
    \partial_t e_{\bp}^s 
    =&
    \epsilon \chi_{\bp}^s
+  \frac{1}{2}  U_{\bk}^* \Vp^{s s} \langle a_{\bp}^{s} a_{\bq}^{s} \rangle^* 
\delta_{\bk\bp\bq} 
~\mathbb{1}_{\omega_{\bp\bq}^{ss} < \Up}
\nonumber\\& + (\bk \to -\bk)
+ c.c.,
\label{eq:KE2}
\end{align}
describing the evolution of the wave energy of mode $\bp$ due to a near-resonant coupling with the condensate mode $\bk$, where most of the 2D energy is concentrated.

Similarly to classical wave-turbulence theory, here the order of limits $l_f/L\to 0$ and $\Omega t \to \infty$ matters. Taking the infinite box limit first, followed by $\Omega t \to \infty$, leads to $|\omega_{\bp\bq}^{ss}|/\Omega\to 0$ in \eqref{eq:beating}, so that the 2D condensate and 3D waves remain coupled. On the contrary, in the opposite order of limits, near-resonances disappear and the 2D mode decouples from the 3D waves. The former limit is the classical weak wave-turbulence limit \citep{nazarenko2011wave, galtier2003weak}, and can also be considered in the absence of a condensate, for a homogeneous 2D flow. As interactions are dominated by near-resonances in this limit, we expect 2D and 3D modes to remain coupled, 
as they do here with a condensate.
However, within the standard weak wave-turbulence derivation \citep{nazarenko2011wave, galtier2003weak}, the transfer coefficient between 3D and 2D modes entering the kinetic equation is evaluated on the resonant manifold, $\omega_{\bp\bq}^{ss} = 0$, where it vanishes, resulting in \resub{triadic resonances} decoupling instead. This implies that wave-turbulence theory may need to be refined to properly capture the coupling between waves and 2D modes in this limit, as we discuss in \ref{app:decoupling_WT}. 
\resub{Resonant 2D-3D quartets interactions could also be important in that case, and are expected to enter at the same order, see \citep{newell1969rossby,smith1999transfer,brunet2020shortcut} .}

In our derivation we have only included contributions from near-resonant triads, excluding quartets.
Our neglect of such interactions is justified for the special 2D mode we are considering, the box scale condensate. The interaction of the condensate with \textit{any} 3D mode that forms a triad with it is near-resonant in the classical wave turbulence limit described above. 
Related to this, the coupling coefficient between the 2D condensate and 3D waves \eqref{eq:Cpq+Cqp} is small in the same sense for all such triads due to scale separation, and is proportional to $2\pi/L_y$.
Thus, quartet interactions involving a condensate-wave triad (with the \emph{same} order of the coupling coefficient \eqref{eq:Cpq+Cqp}) and
mediated by an additional wave-wave triad always contribute an order higher in the non-linearity compared with near-resonances, similar to classical wave turbulence \citep{nazarenko2011wave}.
This is in contrast with a situation where a given 2D mode can be involved in both near-resonant interactions and off-resonant interactions through triads of different types, hence with different magnitudes of the 2D-3D coupling coefficient. The former type of interactions have a small 2D-3D coupling coefficient, while the latter type have a finite 2D-3D coupling coefficient but occur at the next order in the expansion (corresponding to a resonant quartet). Thus, near-resonant triads and resonant quartets could contribute at the same order in such a case, in contrast with our condensate-wave system.

\subsection{Kinetic equation in the limit of scale separation $l_f/L_y \ll 1$}

In the limit $l_f/L_y \ll 1$, where there is scale separation between the 2D flow and the 3D waves, the kinetic equation Eq.~\eqref{eq:KE2} can be greatly simplified.  
In this limit, nearly-resonant interactions such that $\omega_{\bp\bq}^{ss} < \Up$ (evaluated at the forcing scale $k_f$) obey the energy balance
\begin{align}
\partial_t e_{\bp}^s =
\frac{\Up}{\sqrt{2}} \Bigg[  p_x \partial_{p_y}  +  \frac{p_x p_y}{p^2} \Bigg] \Phi_{\bp\bk}^s 
+  \epsilon \chi_{\bp}^s,\label{eq:Econs} \\
\text{with } \Phi_{\bp\bk}^s \equiv   \frac{1}{2}\Big(\langle a_{\bp}^s a_{-\bp+\bk}^s \rangle  + \langle a_{\bp}^s a_{-\bp-\bk}^s \rangle \Big) + c.c, \nonumber
\end{align}
obtained from Eq.~\eqref{eq:KE2} by expanding the coupling coefficient $\Vp^{ss}$ 
at the lowest order in $l_f/L_y$ (see Appendix \ref{app:simplifying} and \citep{kolokolov2020structure} for a similar treatment).
The correlator 
$\Phi_{\bp\bk}^s$ in Eq.~\eqref{eq:Econs}
is responsible for
the energy transfer between the 3D waves and the 2D condensate, 
as it determines the Reynolds stress correlator between modes $\bp$ and $\bp \pm \bk$ with $\bk=2\pi/L_y ~\be_y$:
\begin{align}
    \uv_{\bp\bk}^{ss} &\equiv \frac{\langle u_{\bp}^s v_{-\bp+\bk}^s\rangle + \langle u_{\bp}^s v_{- \bp - \bk}^s \rangle}{2}  + c.c\nonumber\\
    &=-   \frac{p_x p_y}{p^2} \Phi_{\bp\bk}^s  + O \Big(\frac{1}{L k_f} \Big).
\end{align} 
The correlator $\uv_{\bp\bk}^{ss}$ contributes to the total Reynolds stress 
$\uv =\left(\sum_{\bp,s} \langle uv\rangle_{\bp\bk}^{ss} \right)\cos(k_y y) $ and enters the mean-flow energy balance:
\begin{align}
 &\nu \Up^2 = \Tthree  + \Ttwotwo,  ~~\text{with}~~ \\
 &\Tthree
 = \Vol \int {\rm d}\bp \sum_{s=\pm1} \frac{\Up}{\sqrt{2}}
 \uv_{\bp\bk}^{ss} ~\mathbb{1}_{\omega_{\bp\bq}^{ss} < \Up}\nonumber \\
&= -   \frac{ \Up \Vol}{\sqrt{2}}  \int_{p_z \neq 0 } 
 {\rm d}\bp \sum_{s=\pm1}  \frac
{p_x p_y}{  p^2} ~\Phi_{\bp\bk}^s ~\mathbb{1}_{\omega_{\bp\bq}^{ss} < \Up} + O \Big(\frac{1}{L k_f} \Big), 
\label{eq:uv_homo}
 \end{align}
the total energy transfer between the 3D waves and the 2D condensate, and 
$\Ttwotwo$ the energy input from 2D modes $p_z=0$, which must be treated separately (Appendix \ref{app:2D-2D}).

In steady state, Eq.~\eqref{eq:Econs} takes the form
\begin{align}
    &\partial_{p_y} \Pi_{\rm adv}^s(\bp) = -   \frac{\Up}{\sqrt{2}} \uv_{\bp\bk}^{ss}
    +   \epsilon \chi_{\bp}^s,     \label{eq:ODE_homo}
 \\
   &\Pi_{\rm adv}^s(\bp) \equiv - \frac{\Up}{\sqrt{2}} p_x \Phi_{\bp \bk}^s \nonumber,
\end{align}
where $\Pi_{\rm adv}^s(\bp)$
is the spectral energy flux in the $p_y$ direction, due to shearing of the waves by the mean flow, and 
$- \frac{\Up}{\sqrt{2}} \uv_{\bp\bk}^{ss}$
is the energy transfer between the condensate and wave modes $\bp$,
which enters the mean-flow energy balance  Eq.~\eqref{eq:uv_homo}.

Equation \eqref{eq:ODE_homo} reflects the exact relationship between the spectral wave energy flux and the energy transfer to the condensate in statistical steady state: 
outside of the forcing scale, the energization of the condensate is determined by the dependence of the energy flux $\Pi_{\rm adv}^s$ on $p_y$.
A wave $\bp$ can lose (gain) energy to (from) the condensate, in which case $\Up \uv_{\bp\bk}^{ss} > 0$ ($\Up \uv_{\bp\bk}^{ss} < 0$), and $\Pi_{\rm adv}^s(\bp)$ decreases (increases) with $p_y$. A vanishing transfer term naturally implies a constant energy flux $\Pi_{\rm adv}^s (\bp)$.

Note that Eq.~\eqref{eq:Econs} 
does not apply to modes excited at $k_f$ that do not 
resonate with the condensate (i.e.\ such that $\omega_{\bp\bq}^{ss} \delta_{p- k_f} > \Up$). 
The contribution of these modes to the 2D condensate is neglected
within our Heaviside approximation of $\Delta(t)$ in \eqref{eq:filter}.
Their interaction with the condensate impeded, these modes are rather assumed to interact with other 3D modes via wave-wave interactions,
which act as a sink by which most of the wave energy is ultimately dissipated  --- a process outside the QL framework, and which we do not explicitly model here.
By writing our QL energy balance \eqref{eq:Econs} as a closed system, 
we have also ignored the scenario by which some modes $p>k_f$, energized from off-resonant modes at $k_f$ by 3D-3D interactions,
become nearly-resonant with the condensate.
Instead, we consider that for each  $(p_x,p_z)$, if the mode $p_y$ at the forcing scale ($p_y^2= k_f^2 - p_x^2 - p_z^2$) is off-resonant,
larger wavenumbers $p_y^2 > k_f^2 - p_x^2 - p_z^2$ also remain so,
and 
do not exchange energy with the condensate.\footnote{
This assumption is difficult to assess.
It could be that waves with very large $p$ always reenergize the 2D manifold (and then eventually the condensate) via near resonances.
However, wave-wave interactions or viscous effects might dominate over such a 2D-3D interaction if $p$ is very large, preventing this scenario.
We have not found evidence for this scenario in our DNS, which however is subject to a finite viscous cutoff.}

\section{Emergent conservation law and 
energy transfer to 2D
}
\label{sec:conservation}

Before analyzing the kinetic equation \eqref{eq:Econs} in detail, 
we first discuss 
why energy is transferred directionally \emph{from} the 3D waves \emph{to} the large-scale 2D flow.
We show that this is due to $(i)$ the emergent conservation of the waves' single-sign helicity $H^s \equiv \sum_{\bp} s p e_{\bp}^s$, which implies that $(ii)$ energy is transferred from the 3D waves to the large-scale 2D flow.
Note that the argument does not require the presence of a stationary 2D mean flow and applies for any scale-separated 2D-wave interaction in statistical steady state, given the existence of a strong-enough 2D field $U(\bx,t)$ at large scales.

Wave interactions mediated by a 
2D flow are restricted to be between same-helicity-sign waves at large rotation, as established in \S \ref{sec:homoA}. 
Given such a restriction,
energy (resp. helicity) injected into a sector $s$ is only transferred to 2D or to the $s$-waves, and 
the balance of wave energy (resp. helicity) in the inertial range
can be written separately for each sign $s$:
\begin{alignat}{3}
   \epsilon^s          &{}={}& \mathcal{E}^s
   &{}+{}& \sum_{\bp,\, p\neq k_f} T_{\rm 3D}^{s,\bp} \\[6pt]
   s k_f \epsilon^s    &{}={}& \underbrace{\mathcal{H}^s}_{\mathclap{3+3\leftrightarrow 2}}
   &{}+{}& \underbrace{\sum_{\bp,\, p\neq k_f} s p T_{\rm 3D}^{s,\bp}}_{\mathclap{3+2\leftrightarrow 3}}
   \label{eq:hel_cons}
\end{alignat}
with $\epsilon^s$ the energy injection rate into modes of helicity sign $s$, 
$T_{3D}^{s,\bp}$ the energy transfer to a 3D mode $(\bp,s)$ mediated by the 2D flow, and
$ \mathcal{E}^s$ the energy transfer between the 2D flow and the $s-$waves. 
$\mathcal{H}^s $ is the helicity transfer rate between the $s$-waves and the 2D flow. 
(Note that $T_{\rm 3D}^{s,\bp}$ and  $\mathcal{E}^s$ generalize, respectively, the terms $\partial_{p_y} \Pi_{\rm adv}^s$ in Eq.~\eqref{eq:ODE_homo} and $\int d\bp \Up \uv_{\bp\bk}^{ss} /\sqrt{2}$ in Eq.~\eqref{eq:uv_homo}, derived in the presence of a condensate.)
The 2D flow is assumed to be of characteristic wavenumber $k$, and $\mathcal{H}^s $ can be further decomposed into the contribution from each chirality of the 2D flow:
$\mathcal{H}^s = k (\mathcal{E}^{s,+} - \mathcal{E}^{s,-} )$.

The only way for waves of opposite helicities to exchange helicity is via $\mathcal{H}^s$. 
However, when the 2D flow occupies a scale much larger than the energy injection scale, $k/k_f \ll1$, 
this 2D-3D helicity exchange is negligible: $\mathcal{H}^s \lesssim k\epsilon^s\ll k_f \epsilon^s$, 
assuming that 
the energy transfer to each 2D chirality is 
bounded by
$\epsilon^s$ ($|\mathcal{E}^{s,\pm}|/\epsilon^s \leq O(1)$).
Therefore, from \eqref{eq:hel_cons},
\begin{align}
    k_f \epsilon^s \simeq \sum_{\bp, p\neq k_f} s p   T_{\rm 3D}^{s,\bp},
    \label{eq:Hcons_quali}
\end{align}
and all the helicity injected in sector $s$ is transferred to the $s$-waves.
Waves of chirality $s$ that interact with the large-scale 2D flow therefore conserve their single-sign helicity $H^s = \sum_{\bp} s p e_{\bp}^s$ 
at the leading order in 
$k/k_f$. This establishes $(i)$.

Due to this additional conservation of a sign-definite quantity $H^s$,
we expect the energy of the waves to be blocked from reaching arbitrarily small scales (large $p$), similarly to the argument by
Fj{\o}rtoft \citep{fjortoft1953changes}.
Indeed, assuming that all modes at arbitrarily large $p$ have the same sign of transfer $T_{\rm 3D}^{s,\bp}$,
and that $|T_{\rm 3D}^{s,\bp}|$ remains bounded,
we have that $T_{\rm 3D}^{s,\bp}\to 0$ as $p\to\infty$, otherwise the sum in equation \eqref{eq:Hcons_quali} does not converge and the single-sign helicity balance cannot be satisfied. 

The wave energy can then either be transferred to large-scale 3D modes $p\ll k_f$, or to the large-scale 2D mode $k\ll k_f $. 
However, the former scenario is not expected due to the advection of the waves by the large-scale 2D flow, which tends to shear them and generally
transfer the wave energy to small scales. This generates a positive forward energy flux, which however is vanishingly small at small scales due to the constraint of conservation of $H^s$.
Thus, energy is dominantly transferred from the waves to the large-scale 2D flow, $\mathcal{E}^s \sim \epsilon^s$, showing $(ii)$. 
This explains how large-scale 2D flows can be irreversibly energized by 3D inertial waves at large rotation.

Note that the conservation of single-sign helicity for scale-separated triads requires the presence of a strong-enough large-scale 2D flow with which some sector of waves interacts (wavenumbers below $k_U$). 
%
If the 2D flow is initially weak, or zero, 2D modes can be excited via instabilities due to near-resonant triads or resonant quartets \citep{brunet2020shortcut, le2020near}. Then, the argument explaining how waves would maintain this 2D flow could be generalized if the 2D flow can interact quickly-enough with some scale-separated waves.
Wavenumber $k_f$ in our argument would then represent a typical wave scale for which
such interactions occur, and  $\epsilon^s$ would correspond to the energy flux received by these waves, e.g. from other waves.
Note also that the argument does not depend on the exact treatment of triadic near-resonances but would be invalidated if the 2D mode in question has tetrad interactions at the same order in the expansion (e.g. is not sufficiently scale-separated from 3D modes), breaking the conservation of helicity by sign.

We now derive these results for our condensate-wave system, starting from Eq.~\eqref{eq:Econs}.
When multiplying Eq.~\eqref{eq:Econs} by $sp$, we obtain a conservation law for the wave helicity $s p e_{\bp}^s$, for each pair $(p_x,p_z)$ that obeys the near-resonant condition $\omega_{\bp\bq}^{ss} < \Up$
at the forcing scale $p=k_f$:
\begin{align}
 \partial_t (s p ~ e_{\bp}^s) +  \partial_{p_y} (s p ~\Pi_{\rm adv}^s  )  =   sp ~ \epsilon   \chi_{\bp}^s,
 \label{eq:Hcons}
\end{align}
with $ s p \Pi_{\rm adv}^s $ a helicity flux in the $p_y$ direction due to the advection by the mean flow.

The advection of the waves by the mean shear flow $\bU = U(y) \be_x$ causes the motion of energy along characteristics in the variable $p_y$, and hence generates an energy flux $\Pi_{\rm adv}^s $ to small scales $|p_y|\to \infty$. 
However, the downscale energy flux $\Pi_{\rm adv}^s$ 
decays with $p$ as a consequence of the conservation of single-sign helicity.
Indeed, integrating Eq.~\eqref{eq:Hcons} in steady state with consistent boundary conditions at $|p_y|\to \infty$ (See Appendix \ref{app:stationary}), 
we obtain that all the injected helicity is transferred to small scales via a finite helicity flux, so that
\begin{align}
    |\Pi_{\rm adv}^s(\bp)| &=
     \left\{
    \begin{aligned}
       & \chi^s_{p_x,p_z} \frac{k_f}{p },  & \text{ for } p> k_f \text{ and } \Up p_x p_y < 0 \\
   &  0,  &\text{ for } p> k_f \text{ and } \Up p_x p_y > 0
    \end{aligned}
    \right.
  \label{eq:sol_homo_kf}
\end{align}
with 
$\chi^s_{p_x,p_z} \equiv  \int_{-\infty}^{\infty} \chi^s_{\bq} ~dq_y$ 
the energy injection rate in each $(p_x,p_z)$ line, assuming that the near-resonant condition 
$\omega_{\bp\bq}^{ss} < \Up $ 
is satisfied at the forcing shell.
Because $\Pi_{\rm adv}^s(\bp)$ decreases with $p_y$ at small scales, the energy balance
\eqref{eq:ODE_homo} implies that
waves lose energy to the condensate, i.e.\ that the energy 
transfer $\Up \uv_{\bp\bk} = - \Up \frac{p_x p_y}{2p^2} \Phi_{\bp}^s > 0$ when $p>k_f$.
Now, we can integrate Eq.~\eqref{eq:ODE_homo} to obtain the energy transfer to the condensate from each resonant $(p_x,p_z \neq 0)$ line:
\begin{align}
       \int\frac{\Up}{\sqrt{2}} \uv_{\bp\bk}^{ss} dp_y &=  \int_{-\infty}^{\infty} dp_y~ \epsilon \chi_{\bp}^s  
    - 
     \Pi_{{\rm adv}}^s(\bp)\Big|_{p_y=-\infty}^{p_y=\infty} 
     \nonumber\\
     &=\epsilon \chi^s_{p_x,p_z},
     \label{eq:eps2D_pxpz}
\end{align}
using Eq.~\eqref{eq:sol_homo_kf} when $|p_y|\to \infty$.
Equation \eqref{eq:eps2D_pxpz} shows that all the energy from such a near-resonant line is transferred to the condensate. 
In contrast, if the line is off-resonant, we expect condensate-wave interactions to be subdominant, and most of the remaining energy to be transferred to smaller scales via wave-wave interactions and be dissipated there, following a wave-turbulent forward cascade \citep{galtier2003weak}.

The conservation of the wave helicity $H^s$ 
in the background of a large-scale mean flow can also be understood as the conservation of wave action $\sum_{\bp}e_{\bp}^s/\omega_{\bp}^s$ \citep{andrews1978wave}, applied to each wave species $s$ separately.
Indeed, for each $p_z\neq0$, wave action is proportional to helicity:
$\sum_{p_y}e_{\bp}^s /\omega_{\bp}^s =  ( \sum_{p_y}s p e_{\bp}^s) /p_z$. The directional transfer of energy to the mean flow can then be understood as being due to the motion of energy along characteristics in $p_y$-space, towards large $p$. 
This decreases the wave frequency $\omega_{\bp}^s$, and hence requires a decrease in wave energy $e_{\bp}^s$. A similar phenomenon occurs in plasma
or Rossby-wave systems \citep{diamond2005zonal} and in streaming of acoustic waves \citep{lighthill2001waves}.
See also \citep{ivchenko2025absorption} for a similar behavior with a wave packet.

\section{Solution for the condensate as $\Omega\to \infty$ and 2D-3D decoupling}
\label{sec:decoupling}

We now explicitly compute how much energy is transferred from the 3D waves to the large-scale 2D flow as a function of control parameters $\Roe$ and $l_f/L_i$ ($i=x,y,z$), and thus determine the condensate amplitude $\Up/\Omega$, using our QL theory.
The strategy goes as follows: we first determine the total energy transfer from the 3D waves to the condensate for a given mean-shear rate $\Up$, $\Tthree[\Up/\Omega]$.
Next, we use the mean-flow energy balance \eqref{eq:global_bal} to close the system and determine the condensate amplitude $\Up/\Omega$, and thereby the energy transfer to the 2D flow, $\Ttwo$, as a function of $\Roe =\frac{1}{2\Omega} \sqrt{\frac{\epsilon}{\nu}} $ and $l_f/L_i$ ($i=x,y,z$).

\subsection{Energy transfer as a function of $\Up/\Omega$}
\label{sec:T_UO}

For a given mean-shear rate $\Up$, the total transfer to the condensate is obtained by summing Eq.~\eqref{eq:eps2D_pxpz} over all near-resonant lines $(p_x,p_z \neq 0)$, which gives
\begin{align}
 \Tthree  \left[ \frac{\Up}{\Omega}\right]=  \epsilon \Vol \int_{ \omega_{\bp\bq}^{ss} < \Up, p_z\neq 0} {\rm d}\bp ~ \chi_{\bp},
 \label{eq:eps2D_closed}
\end{align}
where integration is restricted to
near-resonant 
waves excited in the forcing shell.
The explicit dependence of $\Tthree$ on $\Up/\Omega$ is found by
computing the number of contributing near-resonant modes,
i.e.\ such that
\begin{align}
    \begin{cases}
    \frac{p_y p_z}{k_f^2}  < \frac{\Up}{\Omega} \frac{k_f}{2 k_y},  \qquad p_y \geq k_y   \\
    \frac{p_z}{k_f} < \frac{\Up}{\Omega} \frac{k_f^2}{k_y^2}, \qquad p_y = 0,
    \end{cases}
    \label{eq:beating_ineq}
\end{align}
at the forcing scale, corresponding to condition \eqref{eq:filter} using the approximation $ p_y \gg k_y$ up to $p_y =k_y$.

\begin{figure}[t]
    \centering \includegraphics[width=\linewidth]{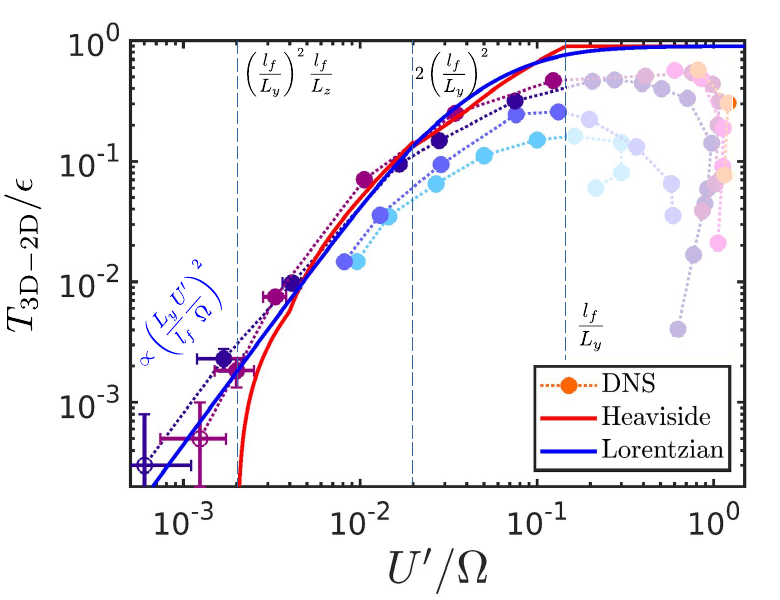}
    \caption{Energy transfer from the 3D waves to the 2D flow, $\Tthree$, as a function of the mean shear rate $U'/\Omega$, as measured from the DNS (solid points) and predicted from the QL theory. In the theory, near-resonances are treated either with a Heaviside (red line) or Lorentzian function of $U'/\Omega$ (blue line). White points correspond to values of $\Tthree$ below statistical error.
    }
    \label{fig:T32_UO}
\end{figure}

At low rotation, provided that all interactions are with homochiral waves, 
all the excited waves transfer their energy to the condensate: 
\begin{align}
    \Tthree = \epsilon_{\rm 3D},  \qquad  
    \frac{l_f}{L_y} < \frac{\Up}{\Omega}, 
\end{align}
where $\epsilon_{\rm 3D}$ is the energy injection rate into the 3D modes.
However, below $\Up/\Omega= l_f/L_y$, waves start to progressively decouple from the 2D flow as rotation is increased (i.e.\ do not obey condition \eqref{eq:beating_ineq}), 
with fewer and fewer modes contributing, being restricted to low $p_z$ and $p_y$.
Consequently, the energy transfer $\Tthree [\Up/\Omega]$
deviates from 
$\epsilon_{\rm 3D}$ and decreases with rotation as
\begin{equation} \frac{\Tthree}{\tilde{\epsilon}}=
  \begin{cases}  
       \frac{2}{\pi} \beta \ln\left( \frac{2}{\beta} \right) + \beta - \frac{l_f}{L_z}, \quad \frac{l_f}{L_z}  < \beta <\frac{1}{2}  \\\\
      \frac{2}{\pi} \beta \ln\left(\frac{ 2L_z}{l_f} \right), \quad \quad \frac{l_f}{L_y} < \beta < \frac{l_f}{L_z}, \\ \\ 
      \frac{2}{\pi}\left(\beta -\frac{1}{2}\frac{l_f}{L_z}\frac{l_f}{L_y}\right)+
      \frac{2}\pi \beta \ln \left(\beta \frac{L_y}{l_f}\frac{ L_z}{l_f}\right), \\
   \quad\quad\quad \qquad\qquad\qquad\qquad \frac{l_f}{L_y}\frac{l_f}{L_z} < \beta < \frac{l_f}{L_y} \\\\
   \frac{2}{\pi}\left(\beta -\frac{1}{2}\frac{l_f}{L_z}\frac{l_f}{L_y}\right), \quad\frac{1}{2}\frac{l_f}{L_y}\frac{l_f}{L_z} < \beta <\frac{l_f}{L_y}\frac{l_f}{L_z} \\\\
      0, \qquad \qquad   \beta <\frac{1}{2}\frac{l_f}{L_y}\frac{l_f}{L_z},
  \end{cases}  
  \label{eq:dec_scalings_main}
\end{equation}
with $\beta \equiv \frac{U'}{2 \Omega} \frac{L_y}{l_f}$,  $\tilde{\epsilon} = \epsilon \frac{1 - l_f/(2 L_z )}{1 - l_f/(2L_y) - l_f/L_z}$, $\frac{l_f}{L_z}, \frac{l_f}{L_y} \ll 1$, and for $L_y > L_z$
--- see Appendix \ref{app:QR}. For $U'/\Omega\ll l_f/L_y$, the energy transfer first decays 
as a $O(U'/\Omega \times L_y/l_f)$, but eventually goes to zero faster than that and vanishes at the finite value $\frac{\Up}{\Omega}= \frac{l_f}{L_z} (\frac{l_f}{L_y})^2 $, as shown in Fig.~\ref{fig:T32_UO} (red line).
Below this threshold, even waves with the lowest values of $p_z$ and $p_y$, 
$p_z \sim 2\pi/L_z$ and $p_y = 0 $, are off-resonant ($\omega_{\bp\bq}^{ss} > U'$).
Hence, within our Heaviside treatment \eqref{eq:filter},
they decouple from the 2D flow and $\Tthree=0$. 
Thus, off-resonant contributions, not considered here, are bound to become important in this range.
In particular, for $U'/\Omega$ below this value, results using either the Heaviside \eqref{eq:filter} or the Lorentzian \eqref{eq:Lorentzian_main} for the condensate-wave coupling will necessarily disagree, and our theory, which relies on the dominance of near-resonant 2D-3D triads, should not hold.

Using the smooth Lorentzian function \eqref{eq:Lorentzian_main} to compute the transfer from the 3D waves to the condensate can be done numerically on a discrete grid, see Appendix 
\ref{app:lorentzian}. For most of the range, $(l_f/L_y)^2 l_f/L_z < U'/\Omega \ll l_f/L_y$, there is a good agreement between the predictions from the Heaviside, the Lorentzian, and the measurements from DNS at high $Re$ (respectively, red line, blue line, and circles in Fig.~\ref{fig:T32_UO}). This indicates that near-resonances dominate in this range, implying that our theory applies.
For $U'/\Omega < (l_f/L_y)^2 l_f/L_z$, near-resonances $\omega_{\bp\bq}^{ss} < U'$ do not exist, and our theory cannot apply.
Off-resonant terms start being important for $U'/\Omega $ slightly above $ (l_f/L_y)^2 l_f/L_z$ and we find empirically that their order of magnitude can be captured using the Lorentzian: DNS measurements of $\Tthree$ as a function of $U'/\Omega$ roughly agree with results obtained using the Lorentzian also in this regime. When using the Lorentzian, the contribution to $\Tthree$ in this region comes from its tails, of order $O( (U'/\omega_{\bp\bq}^{ss})^2 )\sim O ((U'/\Omega \times L_y/l_f)^2)$.
This suggests that the order of magnitude of off-resonant contributions is $O ((U'/\Omega \times L_y/l_f)^2)$.
Note that in the absence of near-resonances in this regime, 2D-3D quartets involving the condensate, neglected in our QL framework and which do not conserve the wave helicity by sign, may now enter at leading order and exchange energy with the condensate.

As the energy transfer due to near-resonances decays with rotation, and the contribution of off-resonant terms eventually comes to dominate, it is important to determine the value of $U'/\Omega$ below which the latter are no longer negligible. In particular, it is not a-priori clear that off-resonant terms remain sub-leading when the transfer is of order $ O \left(\frac{\Up}{\Omega}\frac{L_y}{l_f} \right) $, as the results above seem to indicate.
In this case, while the contribution of individual off-resonant modes may be small, 
their number is of $O(1)$ since the number of near-resonances is vanishingly small, of the order $ O \left(\frac{\Up}{\Omega}\frac{L_y}{l_f} \right)$. 
Hence, if the individual contributions of off-resonances are suppressed by a factor $ O \left(\frac{\Up}{\Omega}\frac{L_y}{l_f} \right)$, 
near-resonances and off-resonances 
would have a net contribution to $\Tthree$ at the same order, $O\left( \epsilon \frac{\Up}{\Omega}\frac{L_y}{l_f} \right) $. Note that this is a larger contribution than the one captured by the Lorentzian approximation, and would lead to deviations from our theory already when $\Tthree/\epsilon\sim  O \left(\frac{U'}{\Omega } \frac{L_y}{l_f} \right)$.

Naively, if off-resonant terms were to transfer a fraction of their energy of the order of $O\left(  \frac{\Up}{\Omega}\frac{L_y}{l_f} \right)$ in the typical time $1/U'$, this would correspond to a slow interaction time $\frac{1}{\Up} \frac{\omega_{\bp\bq}^{ss}}{\Up} \sim \frac{\Omega}{\Up^2}\frac{l_f}{L_y}$.
Such contributions would be further depleted
if wave-wave interactions around $k_f$ were to occur faster than this time scale, taking away most of the energy available for off-resonant 2D-3D interactions, i.e.\ if
\begin{align}
    \tau_{nl}^{\rm wave}<  \frac{\Omega}{\Up^2}\frac{l_f}{L_y}  \Longleftrightarrow
    \frac{U'}\Omega< \sqrt{\frac{l_f}{L_y}} Ro^{3/4}.
    \label{eq:off-reso_condition}
\end{align}
This provides a sufficient condition for our predictions to give the 
correct
leading order even when $\Tthree/\epsilon \sim O(U'/\Omega \times L_y/l_f)$. 
Note that the QL condition requires in addition that $\frac{U'}\Omega\gg \left(\frac{l_f}{L_z}\right)^2 Ro^{3/2}$.
The DNS data points \resub{for our two highest Reynolds numbers in this regime} mostly obey both these conditions when $\Tthree/\epsilon\sim O(U'/\Omega \times L_y/l_f)$ , see \S \ref{sec:validity} and Appendix \ref{app:off-reso}. This could explain the 
relatively close agreement
between the theory and the DNS in the range $4\times 10^{-3}< U'/\Omega < 4\times 10^{-2}$, where $\Tthree/\epsilon$ is at most $ O(U'/\Omega \times L_y/l_f)$, shown in Fig.~\ref{fig:T32_UO},.

Finally, the remaining missing piece to determine the total energy transfer is $\Ttwotwo$, the energy input from 2D modes $p_z=0$ due to 2D-2D interactions.
The above framework does not apply to such modes, which are not affected by rotation.
In particular, 2D modes do not conserve single-sign helicity but instead horizontal enstrophy \citep{biferale2017two}.
It can be shown that half the energy injected into the 2D modes is transferred to the condensate, because of the conservation of horizontal enstrophy, 
\begin{align}
    \Ttwotwo = \frac{1}{2} \epsilon_{\rm 2D}  = \frac{l_f}{4L_z }.
\end{align}
The other half is transferred to small 3D scales due to the vertical velocity component
behaving as a passive scalar, hence transferring its energy to small scales (see Appendix \ref{app:2D-2D}).

\subsection{Closed solution with isotropic forcing}
\label{sec:w2D}

We now obtain a closed solution $\frac{\Up}{\Omega} = f(\Roe)$ by solving the mean-flow energy balance \eqref{eq:global_bal} with $\Ttwo(U'/\Omega) = \Tthree (U'/\Omega) + \Ttwotwo$.
For isotropic forcing, like in our DNS, 
the closed solution at low rotation reads
\begin{align}
    \frac{\Up}{\Omega}&= 2 \left( \frac{\epsilon_{\rm 3D}}{\epsilon} + \frac{\epsilon_{\rm 2D}}{2 \epsilon} \right)  \Roe \Longleftrightarrow \nu \Up^2 = \epsilon_{\rm 3D} + \frac{1}{2}\epsilon_{\rm 2D}, \nonumber \\ &\qquad  \qquad \text{ for }   \Roe > \frac{l_f}{2L_y}.
    \label{eq:2D-like}
\end{align}
This solution is similar to condensates in 2DNSE, where all the injected energy reaches the domain scale \citep{laurie2014universal, frishman2017culmination}.
In contrast, at large rotation, the solution deviates from the 2D-like scaling \eqref{eq:2D-like} and an approximate asymptotic solution when $\Roe \lesssim \frac{l_f}{L_y} \left(\frac{l_f}{L_z} \right)^{\frac{1}2}$ can be obtained
from the number of near-resonances computed in Eq.~\eqref{eq:dec_scalings_main} (i.e.\ within our Heaviside treatment \eqref{eq:filter}):
\begin{subequations} \label{eq:U_O_w2d}
\begin{align}
 \frac{\Up}{\Omega} &= \frac{4}{\pi} \Roe^2 \frac{L_y}{l_f} \ln \left( \frac{2L_z}{l_f}\right), \qquad \Roe > \Roe^*  \label{eq:U_Ow2da}\\
     \frac{\Up}{\Omega} &= 2\Roe \sqrt{ \frac{\epsilon_{\rm 2D}}{2 \epsilon}},  \qquad \qquad  \Roe < \Roe^*, \label{eq:U_Ow2db}
\end{align}
\end{subequations}
where $\Roe^* = \frac{\pi}{2} \frac{l_f}{L_y}  \sqrt{\frac{\epsilon_{\rm 2D}}{2\epsilon}}   \frac{1}{\ln (2L_z/l_f)}$, with $\epsilon_{\rm 2D} = \frac{l_f}{2L_z} \ll \epsilon_{\rm 3D}$ for our isotropic forcing,
is a crossover $\Roe$ below which 
$\Ttwotwo = \frac{\epsilon_{\rm 2D}}{2}$ dominates over $\Tthree$.
From solution \eqref{eq:U_O_w2d},
the amplitude of the 2D flow saturates to a finite value $\Up \sim (\frac{\epsilon_{\rm 2D}}{2\nu})^{1/2}$ when $\Roe \ll \Roe^*$.
The corresponding transfer $\Tthree$ is found as:
\begin{widetext}
\begin{align}
    \frac{\Tthree}{\epsilon} \left(\Roe \right)=
  \begin{cases}
      \frac{\epsilon_{\rm 3D}}{\epsilon} +  \frac{\epsilon_{\rm 2D}}{2\epsilon} \qquad\qquad 
        \frac{l_f}{2L_y} 
        < \Roe <  \frac{l_f}{2L_z}  \\
       \frac{4}{\pi^2} \left(\frac{L_y}{l_f}\right)^2 \Roe^2 (\ln \frac{ 2 L_z}{l_f} )^2, \qquad\qquad 
    \sqrt{\frac{\pi}{2 \ln ( 2 L_z/l_f)}}(\frac{l_f}{L_y})^{3/2} < \Roe \lesssim  \frac{l_f}{L_y} \sqrt{\frac{l_f}{L_z}}\\ \\
    \frac{4}{\pi^2} \Roe^2 \left(\frac{L_y}{l_f}\right)^2 \ln \left(\frac{2 L_z}{l_f}\right)
    \ln \left[ \frac{2}{\pi} \Roe^2 \left(\frac{L_y}{l_f}\right)^3 \frac{2L_z}{l_f} \ln \left( \frac{2 L_z}{l_f}\right)\right], \qquad\qquad  \Roe^* < \Roe  < 
     \sqrt{\frac{\pi}{2 \ln (\blue{2} L_z/l_f)}}(\frac{l_f}{L_y})^{3/2}
    \\ \\
      \frac{1}\pi \frac{L_y}{l_f} \Roe \left( \frac{l_f}{L_z}\right)^{\frac{1}{2}}  \ln \left( \frac{\Roe}{2 \Roe^c}\right)  + 
       \frac{2}{\pi} \frac{L_y}{l_f} \left(\frac{l_f}{4 L_z} \right)^{\frac{1}{2}} \left(\Roe - \Roe^c \right), 
   \quad\quad\quad\qquad 
   2 \Roe^c
     < \Roe < \Roe^*\\\\
    \frac{2}{\pi} \frac{L_y}{l_f} \left(\frac{l_f}{4 L_z} \right)^{\frac{1}{2}} \left(\Roe - \Roe^c \right),  \qquad\qquad  \Roe^c < \Roe  <
    2 \Roe^c \\\\
      0, \qquad \qquad   \Roe < \Roe^c, ~~~~~~~~~~~~~~~~~~~~~~~~~~~~~~~~~~~~~~~~~~~~~~~
       \text{ with }
    \Roe^c=  \left(\frac{l_f}{L_y} \right)^2  \left(\frac{l_f}{L_z} \right)^{\frac{1}2}. 
   \label{eq:eps32_leading_main} 
  \end{cases} 
  \end{align}    
\end{widetext}
Below $\Roe^c$, no near-resonances $\omega_{\bp\bq}^{ss} < U'$ remain.

\begin{figure}
    \centering
    \includegraphics[width=\columnwidth]{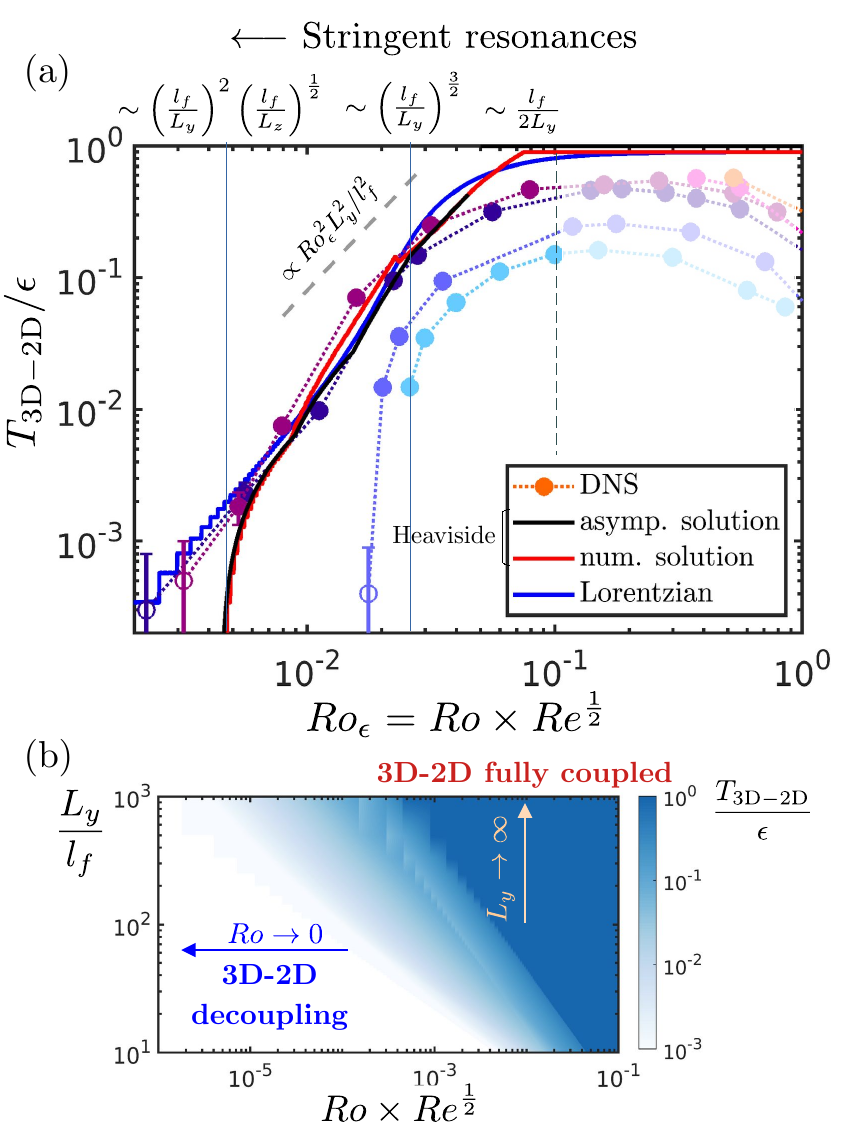}
    \caption{(a) Energy transfer from the 3D waves to the 2D flow, $\Tthree$, measured in the DNS of rotating 3DNSE as a function of $\Roe = Ro \times Re^{\frac{1}{2}}$.
    As waves progressively decouple from the 2D flow, $\Tthree \underset{\Roe \to 0}{\to}0$. 
    White points correspond to values of $\Tthree$ below statistical error.
    Solid lines: predictions from the QL theory. 
    Red: numerical solution with the exact function $\Tthree(\Up/\Omega)$ in \eqref{eq:dec_scalings_main}, predicted with a Heaviside treatment of near-resonances; black: 
    approximate expressions derived in \eqref{eq:eps32_leading_main}.
    Below $\Roe \sim \left(\frac{l_f}{L_y}\right)^2 \left(\frac{l_f} {L_z}\right)^{\frac{1}2}$, the modes closest to $p_z=0$ are off-resonant and $T_{\rm 3D-2D} = 0$.
    Blue: numerical solution with a Lorentzian treatment of near-resonances. 
    (b) Predictions for the 3D-2D energy transfer with the Heaviside approximation, shown in the ($\Roe, L_y/l_f$) plane.} 
    \label{fig:eps32}
\end{figure}

We compare the condensate amplitude, Eqs.~\eqref{eq:2D-like}-\eqref{eq:U_O_w2d}, to our DNS results in Fig.~\ref{fig:cond_rescaled}(a), 
and the corresponding energy transfer $\Tthree (\Roe) $, Eq.~\eqref{eq:eps32_leading_main}, in Fig.~\ref{fig:eps32}(a) (black solid lines).
The energy balance \eqref{eq:global_bal}, closed with $\Tthree(\Up/\Omega)$ in \eqref{eq:dec_scalings_main}, can also be solved directly via a numerical root-finding procedure, leading to the red lines in Figs.~\ref{fig:cond_rescaled}(a) and \ref{fig:eps32}(a), which our asymptotic solutions approximate.
With the Lorentzian approximation of the oscillating factor, Eq.~\eqref{eq:global_bal} can be solved numerically in a similar way, resulting in the blue lines in Figs.~\ref{fig:cond_rescaled}(a) (inset) and \ref{fig:eps32}(a). Contrary to the Heaviside treatment, the Lorentzian approximation produces a transfer $\Tthree = O(\Roe^2)$ below $\Roe^c$, which does not lead to a complete decoupling at any finite $\Roe$.

The QL solution that we derive is compatible with the DNS data points for $\Roe < l_f/(2L_y) \simeq 0.05$ at high $Re$, without any fitting parameter:
below $\Roe = l_f/(2L_y) \simeq 0.05$,
the condensate progressively disappears with increasing rotation rate, following solution \eqref{eq:U_O_w2d} (black line in Fig.~\ref{fig:cond_rescaled}(a)). Similarly, $\Tthree$ decays with rotation, in a manner close to but steeper than $\Roe^2$, following Eq.~\eqref{eq:eps32_leading_main} --- see Fig.~\ref{fig:eps32}(a). 
For sufficiently small $\Roe$, $\Up/\Omega$ approaches a linear scaling $\frac{\Up}{\Omega}\propto \Roe$, at which point $\Tthree \ll \Ttwotwo$, eventually approaching statistical error (white points in Fig.~\ref{fig:eps32}(a)). The condensate amplitude in this regime is mainly set 
by
2D-2D interactions, while most of the 3D wave energy is dissipated to small scales via wave-wave interactions.
Note, however, that observing the linear scaling \eqref{eq:U_Ow2db} for $U'/\Omega$ when $\Roe \to 0 $ requires very large $Re$ and small $Ro$, and it is difficult to completely confirm it from the DNS. 
In this range (corresponding to $\Roe \lesssim \Roe^c$), the contribution of off-resonant terms $\omega_{\bp\bq}^{ss} > U'$ becomes important and our theory no longer applies, as seen through the discrepancy with the Heaviside prediction in Fig.~\ref{fig:eps32}(a) (see also Fig.~\ref{fig:T32_UO}). Instead, we find empirically that the contribution of these off-resonances is consistent with the scaling $\Tthree = O(\Roe^2)$, 
which is the order of such terms within the 
Lorentzian approximation \eqref{eq:Lorentzian_main}, see the discussion in \S \ref{sec:T_UO}.
This discrepancy does not alter the leading-order scaling of the condensate amplitude, $U'/\Omega \sim 2 \Roe \sqrt{\frac{\epsilon_{\rm 2D}}{2\epsilon}}$, set by 2D-2D interactions.

Note that since $\epsilon_{\rm 2D} \ll \epsilon$,
the value of $\Ttwotwo$ in the DNS can be reduced and even vanish due to viscous effects at the forcing scale. This is why the condensate disappears in the lowest-$Re$ cases in the DNS, replaced by a state of inertial waves.
Note also that our predictions 
overshoot the DNS data points in the regime $\Roe \sim l_f/(2L_y)$.

\subsection{Other forcing scenarios}

When forcing only the 3D waves ($\epsilon_{\rm 2D} = 0$),
the notable difference with the isotropic case in \S \ref{sec:w2D} is that there are no remaining 2D-2D interactions in the limit $\Roe \to 0$.
Then, for $\Roe \ll \frac{l_f}{2L_y}$ the solution with the Heaviside approximation essentially takes the same form as in \eqref{eq:U_Ow2da} 
(see \eqref{eq:closure_wo2D} for the full result), 
but vanishes below a critical $\Roe^c$. 
This is because the energy transfer in Eq.~\eqref{eq:dec_scalings_main} vanishes at a finite $U'/\Omega$, so the solution of Eq.~\eqref{eq:global_bal} for $U'/\Omega$ with $\epsilon_{\rm 2D}= 0 $ switches at this point from a finite value to a zero condensate value.
The condensate amplitude and the 3D-2D energy transfer then exhibit a jump near the transition:  
\begin{align}
    \frac{\Up}{\Omega} &\simeq  
      \frac{2}{\pi}\Roe^2 \frac{L_y}{l_f}
      \left(1+\sqrt{ 1 - \frac{\Roe^{c2}}{\Roe^{2}}} \right),
      \label{eq:UO_3df}
    \\
    \frac{\Tthree}{\epsilon} &\simeq \frac{1}{\pi^2} \left( \frac{L_y}{l_f} \right)^2 \Roe^2
    \left(1+\sqrt{ 1 - \frac{\Roe^{c2}}{\Roe^{2}}} \right)^2,   \\
     &  \qquad \qquad \text{ for } \Roe \geq \Roe^c = \sqrt{\pi} \left(\frac{l_f}{L_y}\right)^{3/2} \left(\frac{l_f}{L_z} \right)^{1/2},
     \nonumber
\end{align}
while both $U'/\Omega$ and $\Tthree$ vanish for $\Roe < \Roe^c$ (Appendix \ref{app:closure_wo2D}).

Therefore, in the absence of forcing in the 2D manifold, we predict a complete disappearance of the 2D flow below a finite $\Roe^c$
with our Heaviside treatment of near-resonances,
and an associated first-order transition occurring at $\Roe^c$.
Note that this discontinuous transition 
is a generic feature if the energy transfer can become as small as $\Tthree=O( (\Up/\Omega \times L_y/l_f)^2)$, as, e.g., it does within the Lorentzian approximation.
In this case, a solution to the mean-flow energy balance \eqref{eq:mf_bal} ceases to exist below the value of $\Roe$ at which $\Tthree=O( (\Up/\Omega \times L_y/l_f)^2)$.
This discontinuous transition is expected to be 
independent of the particular 3D forcing and large-scale dissipation mechanism.

In contrast, when forcing only the 2D manifold ($\epsilon_{\rm 3D} = 0, \epsilon=\epsilon_{\rm 2D}$), $\Ttwo= \epsilon/2 $: Eq.~\eqref{eq:global_bal} therefore leads to 
\begin{align}
   \frac{\Up}{\Omega} = \Roe \Longleftrightarrow \nu \Up^2 = \frac{1}{2} \epsilon, \text{ for } 
    \Roe < \frac{l_f}{2 L_z},
\end{align}
a rotation-independent scaling similar to that of condensates in 2DNSE.
This regime was observed in numerical simulations of the 2D-forced rotating 3DNSE when $Ro\to 0$ \citep{seshasayanan2018condensates}.
We expect this regime to be realized below $Ro \sim (l_f/2L_z) Re^{-1/2}$ .

\subsection{Remarks on lower rotations}

At large enough rotation ($\Roe \lesssim 0.1)$,
we have shown that
only condensate-wave interactions 
involving homochiral waves survive, and that these waves transfer all their energy to the condensate: rotation has turned sign-definite an otherwise sign-indefinite invariant, thereby fundamentally altering the energy distribution through scales \citep{biferale2012inverse, shavit2024sign}.
However, at lower rotation, ($\Roe \gtrsim 0.1$) interactions are not limited to be homochiral, and waves of opposite helicity interact through the condensate (Fig.~\ref{fig:cond_rescaled}a).
In this case, waves can also extract energy from the 2D flow.
We dedicate a forthcoming publication to this higher-$Ro$ regime \citep{gome2025helicity}, \resub{where we show that this helicity-sign mixing generates a finite flux of energy to small scales, which we evaluate analytically.}

As the sustainment of 2D condensates in rotating 3D flows requires some homochiral wave-condensate interactions, it relies on the existence of inertial waves (or, more loosely, rotation-dominated modes) at the forcing scale.
This is why the condensate is lost when $Ro \gtrsim 0.5$, above which the forcing scale is not constrained by rotation (see Fig.~\ref{fig1}(b)).

\subsection{Order of limits}

We have established that 3D-2D decoupling occurs progressively, starting from a fully-coupled regime at intermediate rotation ($\frac{l_f}{2L_y } < \Roe $), 
to a regime of vanishingly-small coupling when $\Roe \ll \left(\frac{l_f}{L_y} \right)^2  \left(\frac{l_f}{L_z}\right)^{\frac{1}2} $, corresponding to a 3D-2D transfer of at most
$o(\Roe)$.
Crucially, the limits $L_y, L_z \to \infty$ and $\Roe \to 0$ do not commute.

Taking the infinite box limit in the horizontal direction followed by $\Roe\to 0$ (the weak wave turbulence limit in this direction), 3D and 2D modes remain fully coupled, see Fig.~\ref{fig:eps32}(b). For our $y$-dependent condensate, this corresponds to $L_y /l_f \to \infty$ first, followed by $\Roe \to 0$, resulting in a vanishing 2D wavenumber $\bk\to 0$ which makes the corresponding 2D-3D triads always near-resonant. As a result,
the scaling $\nu \Up^2 \simeq \epsilon$ \eqref{eq:2D-like}, similar to 2DNSE, is realized in this limit. 
This is consistent with the previous derivation in \citep{kolokolov2020structure}, corresponding to this order of limits.

In the vertical direction, when taking $L_z /l_f \to \infty$ before $\Roe\to 0$,
3D and 2D modes can never completely decouple, as
3D modes $p_z\sim 2\pi/L_z$ are always nearly-resonant with the 2D flow.
This behavior is consistent with the numerical stability analysis in \citep{billant2021taylor} and with the convergence rate between rotating 3DNSE and 2DNSE derived in \citep{babin1997regularity}.
In the other direction, when $L_z /l_f$ is decreased (for $\Roe<l_f/2L_y$), 
there are fewer resonant modes $p_z \sim 2\pi/L_z$, so near-resonant contributions will vanish at larger $\Roe\sim (l_f/L_z)^{1/2}$.

 \subsection{Validity range of the Quasi-Linear predictions}
 \label{sec:validity}

Finally, an important self-consistent verification of our theory is whether the QL approximation $ \frac{1}{\Up} \ll  \tau_{nl}^{\rm wave} $ is still expected to hold even if the condensate is relatively weak when $\Roe \to 0$.
This can be verified a posteriori from the QL solution $U'$.
With isotropic forcing, in the worst case $p_z= 2\pi/L_z$ and when $U'/\Omega \sim 2 \Roe \sqrt{\epsilon_{\rm 2D}/(2\epsilon)}$ (for $\Roe < \Roe^*$), we need that
\begin{align}
    \frac{k_U}{k_f} = \frac{1}{Ro} \left(\frac{l_f}{L_z} \right)^{1/3} \left(\frac{\epsilon_{\rm 2D}}{2\epsilon}\right)^{1/3}
    (2 \Roe)^{2/3} \gg 1
\end{align}
This condition is equivalent to 
$Re \gg  Ro \frac{l_f}{L_z} \left(\frac{2 \epsilon}{\epsilon_{\rm 2D}}\right)^{2}$,
which is relatively well satisfied by our DNS data points for our two highest Reynolds numbers in the relevant regime ($k_U/k_f = O(10)$), and approximately so for the lower Reynolds numbers, for which the agreement with our theory indeed deteriorates. See more details in Appendix \ref{app:off-reso}.

The second self-consistent condition concerns the treatment of the remaining contribution of off-resonant 2D-3D interactions, assumed to be depleted by wave-wave interactions via condition \eqref{eq:off-reso_condition}. 
If this condition were not respected, off-resonances could contribute an $O \left(\frac{\Up}{\Omega} \frac{L_y}{l_f} \right)$ to $\Tthree$, and our predictions for the regime $\Roe \ll l_f/L_y$ would have to be modified.
With isotropic forcing, a sufficient condition for our predictions to apply is therefore that $\Roe <  Ro^{3/8} \left(\frac{l_f}{L_y}\right)^{3/4}$ in the regime $ \Roe^* < \Roe < l_f/(2L_y)$,
and $\Roe < Ro^{3/4}\sqrt{\epsilon/\Ttwotwo}\sqrt{l_f/L_y}$ in the regime dominated by 2D-2D interactions $\Roe < \Roe^*$.
This range covers most DNS data points, see Appendix \ref{app:off-reso}.
Outside this condition, the governing equation \eqref{eq:KE} should be modified to include off-resonant contributions, and, when 2D-2D interactions dominate,
we expect the transfer $\Tthree$ to be bounded
by $O(\Roe)$.

\section{Conclusion and discussion}

This work identifies a nonlinear mechanism by which 3D inertial waves can transfer energy to large-scale two-dimensional structures in rotating turbulence.
Near-resonant, scale-separated, 2D–3D interactions conserve the helicity of sufficiently-fast waves separately by sign.
This 
imposes an upscale flux of energy \resub{which maintains} large 2D scales. This is in sharp contrast with 3D–3D interactions, whether dominated by inertial waves \citep{galtier2003weak} or by eddies \citep{chen2003joint, alexakis2017helically}, which mix helicity signs. There, both energy and sign-indefinite helicity cascade to small scales.
Our work suggests that this additional conservation law is responsible for 
sustaining 
large-scale 2D condensates in numerical simulations of rotating 3D turbulence \resub{in statistical steady state}.

As rotation is increased,
near-resonant conditions become more restrictive, which progressively suppresses the
coupling between 3D waves and 2D motions,
until it becomes vanishingly small in the limit $Ro\ll l_f/L\ll 1$, i.e taking the limit of fast rotation before the infinite-box limit.
When 3D modes are excited, this leads to a regime
dominated by inertial-wave turbulence for the 3D modes.
3D-2D decoupling is a fundamental property of the rotating 3D Navier-Stokes equations \citep{babin1999global, chen2005resonant} 
for which we provide compelling numerical evidence.

We have proposed a mean-wave quasi-linear kinetic theory to capture the coupling between waves and the 2D condensate in a rotation-dominated regime. We have derived its steady-state solutions,  determining how much energy is transferred to the condensate as a function of $Ro$ and $Re$. Our theory relies on three crucial hypotheses:  
$(i)$ the quasi-linear approximation, expected to be valid only if the 2D flow is strong enough, which is satisfied reasonably well for our DNS at large enough $Re$ (see Appendix \ref{app:off-reso});  
$(ii)$ a finite width of the near-resonant broadening, of the order of the mean shear rate $U'$;
$(iii)$ that only near-resonant triadic wave-condensate interactions contribute, in particular excluding resonant quartets which should be subleading in a condensate-wave system. 
We provide an estimate for when assumptions $(ii)$ and $(iii)$ are bound to break: $\Roe \lesssim (l_f/L_y)^2 (l_f/L_z) ^{1/2} $. Below this order of magnitude, 2D-3D near-resonances are absent or too few for quartets to be fully neglected,  and the contribution of 2D-3D triadic off-resonances is no longer negligible, invalidating our theory and predictions. 
In this regime, we find that
the order of the contributions from neglected terms can be captured by empirically adjusting the tails of the broadening function beyond $U'$, see Eq.~\eqref{eq:Lorentzian_main}.
For $\Roe >(l_f/L_y)^2 (l_f/L_z) ^{1/2}$, we find good agreement between our theory and numerical simulations as long as assumption $(i)$ holds. There, the theory quantitatively captures the gradual nature of the 3D-2D decoupling for our two choices of the resonant broadening in $(ii)$.

Developing a unified framework to address the limitations of our theory is an important challenge for future work. 
In this vein, dynamical numerical simulations of a viscous QL system are vital to assess the exact validity bounds of our theory, 
in particular to justify the neglect of
wave-wave interactions and our modeling of the resonant broadening. 
They would provide a definitive proof that the conservation of the waves single-sign helicity is sufficient to maintain the mean flow in a statistical steady state, and validation that the steady state we predict is obtained within QL dynamics \cite{nivarti2025non}.
Finally, while our work suggests that a QL framework is applicable, 
it has not been rigorously established that the full system converges to the QL equations in some distinguished limit -- see \citep{chini2022exploiting} for recent work in this direction.

How our results extend to rotating turbulence without a condensate -- i.e.\ during the initial growth of the 2D flow, or in 
statistical steady state with
strong large-scale dissipation  -- remains to be clarified.
In that case, statistically homogeneous and isotropic 2D flows generically emerge, even in systems forced in 3D~\citep{smith1999transfer, 
yarom2013experimental,deusebio2014dimensional,
godeferd2015structure, brunet2020shortcut}.
Once the energy reaches 2D modes \resub{(e.g., by triadic or quartetic instabilities \citep{le2020near, brunet2020shortcut})}, 
it is transferred to large 2D scales via a classical inverse cascade within 2D modes,
see e.g.~\citep{mininni2010rotating,yarom2013experimental, campagne2014direct}:
 2D–2D interactions are indeed expected to dominate over 3D-2D interactions, as the latter operate on slower near-resonant time scales or even decouple \citep{alexakis2018cascades}.

 Thus, a key remaining question is how and why energy is transferred from 3D to 2D, powering this inverse cascade.
 Our work suggests a possible scenario, 
 through near-resonant triads irreversibly transferring energy to 2D modes when the scales of 2D and 3D modes are well separated.
The extent to which this scenario is realized in statistically homogeneous flows requires further investigation, 
but there is some numerical evidence of non-local 3D-to-2D transfers that
seem to support this picture \citep{bourouiba2012non, mininni2009scale}.
It also remains to be seen whether a 3D-2D decoupling similar to the one we report can be observed in laboratory flows.
So far, a turbulence of purely-inertial waves was experimentally realized only when the 2D modes were strongly damped \citep{monsalve2020quantitative, brunet2020shortcut}.

Our helicity-based mechanism is directly applicable to 3D chiral fluids, and should explain the sustainment of strong 2D flows there \cite{de2024pattern}.
It may also 
play a role in the persistence of such flows in other types of rotating systems, like rotating convection \citep{guervilly2014large, favier2014inverse, rubio2014upscale, guervilly2017jets, VanKan_bridging}, or rotating–stratified flows \citep{bartello1995geostrophic, smith2002generation, alexakis2024large}.
The helicity-by-sign constraint uncovered here places rotating turbulence in a wider class of wave-dominated systems with emergent adiabatic invariants, where resonant interactions give rise to effective conservation laws, that then govern large-scale self-organization \citep{nazarenko2011wave, connaughton2015rossby}. 

Rotating turbulence exhibits two general, and seemingly contradictory, phenomena: the spontaneous generation and dominance of zero-frequency modes in wave systems, and the asymptotic decoupling between these modes and the wave modes \citep{vallis1993generation,diamond2005zonal,caulfield2021layering,smith2002generation,labarre2026distinguished,de2024pattern}. Our work is an important first step in understanding the two. It implies that the dominance of zero-frequency modes may be due to a hidden conservation law, realized only when interactions are restricted to be resonant. In this vein, our results could be applicable to 2D stratified flows \citep{shavit2024sign, labarre2026distinguished}, where internal gravity waves interacting with shear modes should conserve their pseudo-momentum for each sign separately.

At the same time, our work demonstrates that the asymptotic decoupling between waves and zero-frequency modes occurs through a gradual
transition from the coupled to the decoupled state, with each limiting configuration, 2D-dominated or 3D-dominated,
belonging to a distinct turbulence universality class.
In the context of geophysical flows, 
our work suggests distinguished limits under which asymptotic models of rotating turbulence (2D or quasi-geostrophic \citep{charney1971geostrophic}, widely used in climate modeling \citep{ghil2020physics}),
can be approximately realized, and demonstrates the existence of an intermediate
regime, where only part of the waves decouple from the 2D flow.

\acknowledgements
We are grateful to Michal Shavit for sharing her invaluable insights with us.
We also wish to thank Alexandros Alexakis, Eran Sharon, Omri Shaltiel, Paul Billant and Basile Gallet for fruitful discussions. 
We would also like to acknowledge the anonymous Referee 1 for raising several important points that helped us arrive at a more consistent formulation of our theoretical results and clarify their domain of applicability.
%
The authors would like to thank the Isaac
Newton Institute for Mathematical Sciences, Cambridge,
for support and hospitality during the programme ``Antidiffusive dynamics: from sub-cellular to astrophysical scales'', which was critical to the development of this work. This work was supported by BSF grant No. 2022107 and ISF grant No. 486/23. 

The data used to generate the figures of this article are openly available \citep{data_availability}.
The source code GHOST used to generate the simulations is publicly available \citep{ghost}.

\newpage 
\appendix

\section{Derivation of the kinetic equation}
\label{app:KE}

In this section, we derive the dynamical
equation for the cumulant $e_{\bp}^s(t) = \langle a_{\bp}^s a_{\bp}^{s*} \rangle/2$ 
in an expansion in $\Up / \Omega$. Here, $\langle\cdot \rangle$ denotes an ensemble average over forcing realizations. 
Due to the presence of the mean flow, inhomogeneous correlators will be non-zero, and we will consider
\begin{align}
\langle a_{\bp}^{s}  a_{\bq}^{s} \rangle  &= 2\delta_{\bp+\bq} e_{\bp}^s + \Phi^s_{\bp\bk}  \delta_{\bk+\bp+\bq} 
\label{eq:ic}
\end{align}
where $\delta_{f(k)}$ is the discrete delta function:
\begin{equation}
 \delta_{f(k)}=\begin{cases}
     1 \qquad f(k)=0 \\
     0 \qquad \text{otherwise}
 \end{cases}   
\end{equation}
and $\Phi^s_{\bp\bk}$ is the (inhomogeneous) correlator between modes with wavenumbers $\bp$ and $ \bq=-\bp-\bk$.

We consider an isotropic white in time forcing, limited to a shell around a wave number $k_f$:
\begin{align}
\langle f_{\bp}^s (t) f_{\bq}^{s'} (t') \rangle = 2\epsilon \delta(t-t') \delta_{\bp+\bq}  \delta_{s s'} \chi_{\bp}^s.
\end{align}
where $\sum_p \chi^s_{\bp}=1/2$ and $\chi_{\bp}\propto \mathbb{1}_{k_f^2-\Delta \leq p^2\leq k_f^2+\Delta}$, $\Delta k =\left(\frac{2\pi}{L_x}\right)^2+\left(\frac{2\pi}{L_y}\right)^2+\left(\frac{2\pi}{L_z}\right)^2$
, i.e.\ a shell around $k_f^2$ is assumed to be forced such that the total injected energy is equal to $\epsilon$. Note that we assume zero total helicity injection.
In the continuum limit, $\sum_p \chi^s_{\bp}\to \Vol \int dp \chi^s_p $ where $\Vol = (L_xL_yL_z)/(2\pi)^3$ so that the noise correlator becomes $\chi_{\bp}^s  = \chi_{\bp}/2 = \delta(p-k_f) /(8\pi k_f^2 \Vol)$, giving that
$\Vol\int d\bp \chi_{\bp} = 1$, and the forcing injects total energy at a rate $\epsilon$.

\subsection{Wave energy equation}
\label{app:energy}

We begin with the equations for the mean flow and fluctuations (waves) assuming the quasi-linear approximation, a time-independent two-dimensional mean flow $U_{\bk}$, and considering scales where the dissipation is negligible for the waves. We shall work with a finite space (and discrete wavenumbers for the waves) here, and will comment about the continuum limit at the end. Recall that we are working in the interaction picture (working with the amplitudes of inertial waves, $a_p^s$) so that the equations are non-autonomous.
We take the arbitrary initial time to be $t=0$.  The equations then read:
\begin{align}
\nu k_y^2 U_{\bk} &=  i k_y  \sum_{\bp,\bq, s,\tilde{s}} \Big[ C_{\bk\bp\bq}^{s\tilde{s}} 
 \langle a_{\bp}^{s*} a_{\bq}^{\tilde{s}*} ~ e^{-i \omega_{\bp\bq}^{s\tilde{s}} t} \rangle  \Big]  \delta_{\bk\bp\bq} \label{eq:triadMF_app}  \\
 \partial_t a_{\bp}^s &=  \sum_{\bq, \bk,\tilde{s}}\Big[ \Vp^{s\tilde{s}}  U_{\bk}^* ~ a_{\bq}^{\tilde{s}*}  e^{-i \omega_{\bp\bq}^{s\tilde{s}} t} 
 \Big] \delta_{\bk\bp\bq} + \bforc_{\bp}^s e^{-i\omega^s_{\bp} t}
\end{align}
with 
\begin{equation}
   \omega_{\bp\bq}^{s\tilde{s}}=\omega^s_{\bp}+\omega^{\tilde{s}}_{\bq}=2\Omega p_z\left(\frac{ s}{p}-\frac{\tilde{s}}{q} \right)
\end{equation}
where we have used that $U_k$ is two dimensional so that $k_z=0$, implying that $p_z=-q_z$. 

Next we write the energy balance (equivalent of the kinetic equation) for the waves $e_{\bp}^s = \frac{1}{2}\langle a_{\bp}^s a_{\bp}^{s*} \rangle$:
\begin{align}
    & \partial_t e^s_{\bp}=   \frac{1}2\left(\langle a^{s*}_{\bp}\partial_t a^s_{\bp}\rangle + \langle \partial_ta^{s*}_{\bp} a^s_{\bp}\rangle \right)\nonumber  \\
    & =    \frac{1}2\sum_{\bq, \bk, \tilde{s}}\Big[ \Vp^{s\tilde{s}}  U_{\bk}^* ~ \langle a^{s*}_{\bp} a_{\bq}^{\tilde{s}*} \rangle  e^{-i \omega_{\bp\bq}^{s\tilde{s}} t} 
 \Big] \delta_{\bk\bp\bq} \nonumber \\
 &+\frac{1}2\langle a^{s*}_{\bp} \bforc_{\bp}^s\rangle e^{-i\omega^s_{\bp} t} +c.c \nonumber \\
 & =  \frac{1}2\sum_{\bq, \bk, \tilde{s}}\Big[ \Vp^{s\tilde{s}}  U_{\bk}^* ~ \langle a^{s*}_{\bp} a_{\bq}^{\tilde{s}*} \rangle  (t) e^{-i \omega_{\bp\bq}^{s\tilde{s}} t} 
 \Big] \delta_{\bk\bp\bq} +c.c+\epsilon \chi^s_{\bp}
 \label{eq:e_1}
\end{align}
where $c.c$ denotes complex conjugate and we have used that $\langle a^{s*}_{\bp} \bforc_{\bp}^s\rangle =\epsilon \chi^s_{\bp}e^{i\omega^s_{\bp} t} $, since the forcing is white in time (we assume the Stratonovich convention).

We assume that wave-mean flow interactions occur on a typical time scale of the order of $1/\Up$, so that over a much shorter time scale, of order $t\ll 1/U'$, the wave amplitude correlators can be assumed to be constant.
Thus, performing a partial time average up to time $t\ll 1/\Up$ (the same procedure should be applied to the equation for the mean flow, Eq.~(\ref{eq:triadMF_app}), consistently giving contributions from the same triads), 
\begin{align}
    &\frac{e^s_{\bp}(t)-e^s_{\bp}(0)}{t} \nonumber \\
    &\approx\frac{1}2\sum_{\bq, \bk, \tilde{s}}\Big[ \Vp^{s\tilde{s}}  U_{\bk}^* ~ \langle a^{s*}_{\bp} a_{\bq}^{\tilde{s}*} \rangle \frac{1}{t}\int_0^t e^{-i \omega_{\bp\bq}^{s\tilde{s}} t'}dt' 
 \Big] \delta_{\bk\bp\bq}  +c.c+\epsilon \chi^s_{\bp} \nonumber  \\
 &= 
 \frac{1}2\sum_{\bq, \bk, \tilde{s}}\frac{1 - e^{- i \omega_{\bp\bq}^{s\tilde{s}}t}}{i \omega_{\bp\bq}^{s\tilde{s}}t}\Big[ \Vp^{s\tilde{s}}  U_{\bk}^* ~ \langle a^{s*}_{\bp} a_{\bq}^{\tilde{s}*} \rangle 
 \Big] \delta_{\bk\bp\bq}  +c.c+\epsilon \chi^s_{\bp} 
 \label{eq:KE_app1}
\end{align}
up to terms in $O(U'^2 t)$, corresponding to only keeping first-order terms in $O(U'/\Omega)$ in the time-scale expansion.

\begin{figure}[t]
\subfloat{\includegraphics[width=0.9\columnwidth]{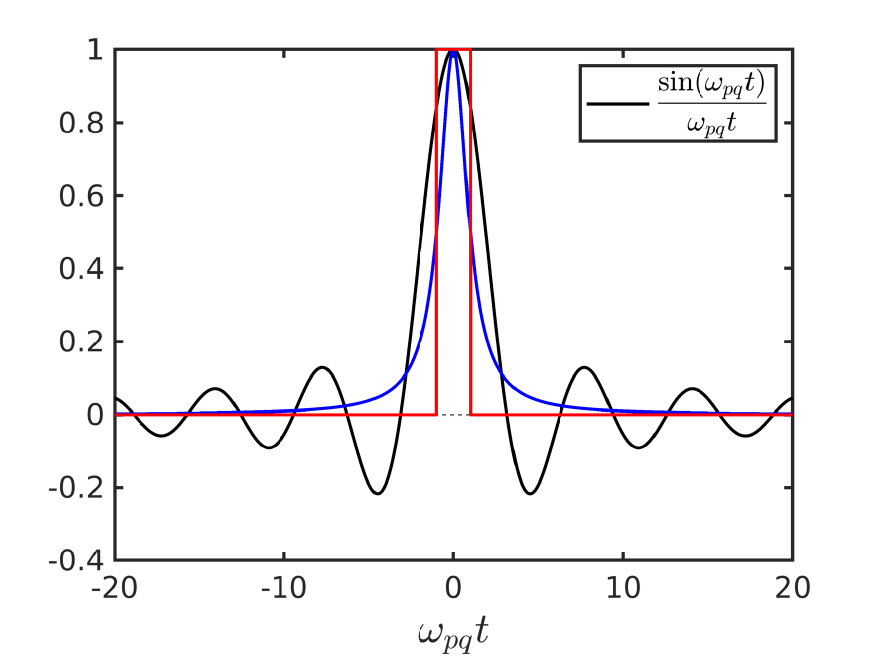}}
\caption{
Oscillating factor $\Re(\Delta(t)) = \frac{\sin(\omega_{\bp\bq}t)}{ \omega_{\bp\bq} t} $ appearing in \eqref{eq:KE} (black line), and its approximation by a Heaviside function \eqref{eq:OF_approx_t} (red line), and by a Lorentzian function \eqref{eq:Lorentzian_t} (blue line).
}
   \label{fig:OF}
\end{figure}

The influence of rotation on the dynamics comes in through the oscillating factor, $\frac{1 - e^{- i \omega_{\bp\bq}^{s\tilde{s}}t}}{i \omega_{\bp\bq}^{s\tilde{s}}}$, whose real part is shown in Fig.~\ref{fig:OF} as a function of $\omega_{\bp\bq}^{s\tilde{s}}t$.
In particular
\begin{align}
\Delta(t) \equiv \Re\left(\frac{1 - e^{- i \omega_{\bp\bq}^{s\tilde{s}}t}}{i \omega_{\bp\bq}^{s\tilde{s}}t} \right)\approx 
      \begin{cases}
    1 \qquad \qquad \omega_{\bp\bq}^{s\tilde{s}} t \ll 1 \\
    0 \qquad \qquad \omega_{\bp\bq}^{s\tilde{s}} t \gg 1
    \end{cases}.
    \label{eq:OF_asymptot}
\end{align}
Indeed, if $\omega_{\bp\bq}^{s\tilde{s}}$ is too large compared to the inverse of the evolution time scale $1/t$, the wave-mean-flow interaction term does not contribute secularly, as the oscillating factor evolves as $\simeq \frac{1}{\omega_{\bp\bq}^{ss}t}$. Triads such that $ \omega_{\bp\bq}^{s\tilde{s}} t \gg 1$ are therefore off-resonant and do not couple with the 2D flow at leading order.
In contrast, for resonant or near-resonant triads, such that $\omega_{\bp\bq}^{s\tilde{s}}t$ is sufficiently close to zero, the oscillating factor approaches its Taylor expansion and the interaction contributes secularly to the waves energy dynamics over the slow time scale $t$.
Note that $\Im\left(\frac{1- e^{-i \omega_{\bp\bq}^{s\tilde{s}t}}}{i \omega_{\bp\bq}^{s\tilde{s}}t} \right)=\frac{\cos(\omega_{\bp\bq}^{s\tilde{s}}t)-1}{\omega_{\bp\bq}^{s\tilde{s}}t}$ 
contributes non-secularly,
hence is neglected here.

There is also a lower bound for the time $t$ we need to consider: the oscillations of the waves will be felt by the system (meaning that some interactions will average out to zero) only for times much larger than $1/\max(\omega_{\bp\bq}^{s\tilde{s}})=1/4\Omega$, which occurs in interactions between opposite-helicity waves, $s=-\tilde{s}$. Hence, such an expansion makes sense for $1/4\Omega\ll t\ll 1/U'$, requiring a sufficient time-scale separation. Thus, the kinetic equation that we derive is an expansion in $U'/4\Omega$ (the ratio between the wave and non-linear time scale), where we keep the leading order term $O(U'/4\Omega)$. 

We now approximate the oscillating factor as
\begin{align}
\Delta (t) \approx 
      \begin{cases}
    1 \qquad \qquad |\omega_{\bp\bq}^{s\tilde{s}} | t < 1 \\
    0 \qquad \qquad |\omega_{\bp\bq}^{s\tilde{s}} | t >1
    \end{cases},
    \label{eq:OF_approx_t}
\end{align}
as visualized in Fig.~\ref{fig:OF} (red line).
By doing so, we extend the secular contribution of near-resonances up to $\omega_{\bp\bq}^{s \tilde{s}} t = 1$, 
and deem off-resonant all values of $\omega_{\bp\bq}^{s \tilde{s}} t$ larger than 1.
Next, given that $0<t<1/\Up$, we further simplify the oscillating factor by estimating it at the upper bound of $t$:
\begin{align}
    \Delta(t)  \approx \Theta \Big( 1 - \frac{|\omega_{\bp\bq}^{s\tilde{s}}|}{\Up} \Big),
    \label{eq:Heaviside}
\end{align}
where $\Theta$ denotes the Heaviside function.
See \citep{bretherton1964resonant} for a similar approach.
Making this approximation amounts to the following:
\begin{itemize}
    \item[-]  When $|\omega_{\bp\bq}^{s \tilde{s}}|/\Up \ll1$, all times $t< \frac{1}{\Up} \ll  \frac{1}{|\omega_{\bp\bq}^{s \tilde{s}}|} $  produce a secular growth in \eqref{eq:KE_app1} --- i.e the RHS in \eqref{eq:KE} is constant for long times, which is consistent with the approximate behavior of $\Delta(t)$ for such times. 
    \item[-] For modes for which $|\omega_{\bp\bq}^{s \tilde{s}}|/\Up \lesssim 1$, and if the time-scale separation between $1/\Omega$ and $1/U'$ is not very large, the time for partial-time averaging cannot be chosen such that $t\ll 1/U'$. This is a borderline case which does not strictly correspond to secular growth, though we set the contribution from these modes to one.
    \item[-] For modes for which $|\omega_{\bp\bq}^{s \tilde{s}}|/\Up \gg1$ the interaction term is set to zero. 
    This is consistent since there is sufficient time scale separation to take the partial-time average over times $\frac{1}{|\omega_{\bp\bq}^{s \tilde{s}}|} \ll t \ll \frac{1}{\Up} $, for which such terms are non-secular.
    \item[-] Modes such that  $|\omega_{\bp\bq}^{s \tilde{s}}|/\Up \gtrsim 1$, are again borderline, since there is not enough time-scale separation between $1/U'$ and $\omega_{\bp\bq}^{s \tilde{s}}$ to make such terms strictly non-secular. They are set to zero within our approximation.
\end{itemize}

Another option, which treats borderline cases more smoothly 
than the Heaviside approximation in \eqref{eq:Heaviside},
is to use a Lorentzian function to model the oscillating factor:
\begin{align}
     \Re\left( \frac{1 - e^{- i \omega_{\bp\bq}^{s\tilde{s}}t}}{i \omega_{\bp\bq}^{s\tilde{s}}t}\right)
    \simeq \frac{1}{ 1+  \left( \omega_{\bp\bq}^{s \tilde{s}} t \right)^2  }.
    \label{eq:Lorentzian_t} 
\end{align}
 See the blue line in Fig.~\ref{fig:OF}.
Like the original oscillating factor, this function 
is $1$ when $\omega_{\bp\bq}^{s \tilde{s}} t = 0$, but is less than one for $|\omega_{\bp\bq}^{s \tilde{s}}  t| > 0$, decaying  as $1/(\omega_{\bp\bq}^{s \tilde{s}} t)^2$.
This weight on the tails makes the function integrable in time, like the original oscillating factor $\Delta(t)$.
%
This type of function is often used to broaden the resonance condition in within the classical wave-turbulence approach. There, the integration in the kinetic equation is over many near-resonances (rather than time), and the amplitude of the Lorentzian is fixed so as to give the same value as the oscillating factor upon this integration, see \citep{lvov1997statistical,lvov2012resonant}.

Next, we evaluate \eqref{eq:Lorentzian_t} at $t = 1/\Up$:
\begin{align}
      \frac{1 - e^{- i \omega_{\bp\bq}^{s\tilde{s}}t}}{i \omega_{\bp\bq}^{s\tilde{s}}t}
    \simeq \frac{1}{ 1+  \left( \frac{ \omega_{\bp\bq}^{s \tilde{s}}}{\Up} \right)^2 }.
    \label{eq:Lorentzian}
\end{align}
By doing so, for modes for which $\frac{1}{\omega_{\bp\bq}^{s \tilde{s}}} <  \frac{1}{\Up} $ we set the oscillating factor to a value between 0 and 1, instead of setting it at $0$ as in \eqref{eq:Heaviside}. 
Freezing the Lorentzian at $t=1/\Up$ puts the contributions from non-secular terms, assumed sub-leading, to a small but non-zero value. It effectively treats those terms as secular but with a small amplitude.
Therefore, unlike the Heaviside filter, the Lorentzian filter produces 
a finite 
contribution of off-resonant modes as $\Omega\to\infty$, including when near-resonant modes disappear.
Note also that for modes for which $ t < 1/\omega_{\bp\bq}^{s\tilde{s}}$ the oscillating factor is taken to be less than 1, unless there is significant time scale separation $\omega_{\bp\bq}^{s \tilde{s}}/U'\ll1$.

Altogether, due to the contribution of near-resonant triads modeled by our choice of filter, the energy equation reduces at $O(\Up/\Omega)$ in the time-scale expansion to
\begin{align}
    \partial_t e_{\bp}^s 
    =
    \epsilon \chi_{\bp}^s
+  \frac{1}2\sum_{\bq, \bk, \tilde{s}}  U_{\bk}^* \Vp^{s\tilde{s}} \langle a_{\bp}^{s*} a_{\bq}^{\tilde{s}*} \rangle F_{\bp\bq}
+ c.c,
\label{eq:E_full}
\end{align}
where $F_{\bp\bq}$ is either the Heaviside filter \eqref{eq:Heaviside} or the Lorentzian filter \eqref{eq:Lorentzian}.
In most of this paper, we use the Heaviside filter \eqref{eq:Heaviside}, where Eq.~\eqref{eq:E_full} is the quasi-linear version of the energy balance for the Euler equations with only part of the interactions included, based on the quasi resonant condition $\frac{|\omega_{\bp\bq}^{s\tilde{s}}|}{\Up}<1$.
We discuss the use of the Lorentzian filter \eqref{eq:Lorentzian} in \S \ref{app:lorentzian}.

Both filters \eqref{eq:Heaviside}-\eqref{eq:Lorentzian} 
select exact resonances $\omega_{\bp\bq}^{s \tilde{s}} = 0$ when time-scale separation between the non-linear time $1/U'$ and the wave period is infinite $\frac{1}{\Omega}  \ll\frac{1}{\Up}$ so that the resonant condition $\omega_{\bp\bq}^{s \tilde{s}} /U'<1$ becomes extremely stringent.
They therefore reproduce the behavior of the oscillating factor in this limit:
$\frac{1 - e^{-i \omega_{\bp\bq}^{s \tilde{s}} t}}{i\omega_{\bp\bq}^{s \tilde{s}} t } = \delta_{\omega_{\bp\bq}^{s \tilde{s}}}$ when $\omega_{\bp\bq}^{s \tilde{s}} t \to \infty$.

Unlike quasi-resonances, exact resonances $\omega_{\bp\bq}^{s \tilde{s}} = 0$ exhibit the
anomalous property of zero 2D-3D coupling, $i k_y C_{\bk\bp\bq}^{ss}= 0$ for our $x$-invariant 2D flow $U_{\bk}$. For such a 2D flow, there are two exactly-resonant poles: 
$k_y = 0, q_y = -p_y$ (however, $U_0 = 0$ by Galilean invariance)
and $k_y = -2 p_y , q_y=p_y$.
Any triad involving 3D modes in the vicinity of these resonant poles (i.e such that either $q_y \simeq - p_y, k_y \ll p_y$ or that $q_y \simeq p_y, k_y \simeq -2 p_y$) is nearly-resonant.
When the 2D manifold has condensed due to the inverse energy cascade, low 2D modes $k_y \ll p$ are favored, and the interactions are naturally nearly-resonant, as triads lie in the vicinity of the first resonant pole ($q_y \simeq - p_y, k_y \ll p_y$).
Furthermore, we can neglect the quasi-resonances around the second pole ($q_y \simeq p_y, k_y \simeq -2 p_y$), as they involve a small-scale 2D mode which is assumed to be of negligible amplitude compared to the largest scale mode.
By neglecting these quasi-resonant triads, we also neglect any catalytic effect, by which a 2D mode energized by quasi-resonant waves acts as a catalyzer for energy transfer between other 3D waves via exact resonances.

Note that the kinetic equation \eqref{eq:E_full} describes the wave energy dynamics in the presence of a strong steady 2D flow with a typical shear rate $\Up$, and not how the rotating 3DNSE dynamically evolved to reach this steady state.
In particular, when the 2D modes are unsteady and weak, e.g in the transient during which 2D modes are first energized, they could also be excited through resonant quartets, which arise at next order in the wave kinetics \citep{bretherton1964resonant, newell1969rossby, smith1999transfer, brunet2020shortcut}.
Here, however, due to the presence of a condensate of strong amplitude, such a quartet mechanism is of sub-leading order: it indeed requires a coupling between a 2D-3D triad and a 3D-3D triad, hence occurs an order $\tau_{nl}^{\rm wave}/U'$ slower than near-resonant condensate-wave interactions. Within our QL approximation, such quartet resonances are therefore neglected.
However, they may occur when only triadic off-resonances $\omega_{\bp\bq}^{s\tilde{s}} > U'$ are present, depending on the weight given to the latter.
Note that when quartet interactions are involved, the helicity of the waves would not be conserved by sign, so there is no a-priori reason for a directional transfer from 3D to 2D.

\subsection{Continuum limit}
\label{app:continuum}
In some of the following we will take the infinite box limit, corresponding to the continuum limit for the wavenumbers.
While such a limit can be taken freely for the 3D waves $\bp$, it is more subtle for the 2D modes $\bk$,
as in the presence of a condensate the 2D energy is concentrated on a few lowest modes, see e.g.\ Fig.~\ref{fig:mf_spec}(a) (for $Ro=0.011, Re=46$). 
Correspondingly, in the continuum limit $L k_f\to \infty$, the condensate can be captured by $U_{\bk} \propto \delta( \bk- \bk_0)$ with a given wavenumber $\bk_0$ (typically the lowest one), where all the mean-flow energy is concentrated.
Then, similarly to the discrete equation \eqref{eq:E_full}, one obtains a secular contribution of the individual condensate mode $\bk_0$
at $O(\Up/\Omega)$ even in the continuum limit,
because
\begin{align}
\Re\left( \frac{1 - e^{-i \omega_{\bp\bq}^{s \tilde{s}} t}}{i\omega_{\bp\bq}^{s \tilde{s}} t}\right)
\underset{\Omega t\to \infty}{\sim} \delta_{\omega_{\bp\bq}^{s \tilde{s}}},
    \label{eq:sin}
\end{align}
where $\delta_{\omega_{\bp\bq}^{s \tilde{s}}}$ is the discrete delta function. Here, the single mode $\bk_0$ interacts with a continuum of 3D waves $\bp$ via near-resonances, which turn into exact resonances in the continuum limit as $k_0\to0$. The sums over the 2D modes $\bk$ in Eq.~(\ref{eq:e_1}) then turn into integrals, yielding a finite result because $U_{\bk} \propto \delta( \bk- \bk_0)$.

This must be contrasted with a configuration without a condensate. There, the 2D spectrum can be approximated as continuous, insensitive to the discretization. 
Here, when the sums over the 2D mode $\bk$ in Eq.~(\ref{eq:e_1}) turn into integrals, the main contribution comes from infinitely many discrete quasi-resonant triads $\omega_{\bp\bq}^{s \tilde{s}} \gg \Delta \omega_{\bp}$ (the frequency step due to the discreteness of the spectral grid).
We can then use the asymptotic result (Riemann-Lebesgue lemma):
\begin{align}
\Re\left( \frac{1 - e^{-i \omega_{\bp\bq}^{s \tilde{s}} t}}{i\omega_{\bp\bq}^{s \tilde{s}}}\right)
\underset{\Omega t\to \infty}{\sim} 2 \pi \delta (\omega_{\bp\bq}^{s \tilde{s}}),
    \label{eq:sin_delta}
\end{align}
that is, the contribution from near resonances around each exact resonance is weighted correctly in order to conserve the integral over them (since $\int_{-\infty}^\infty dx \frac{1 - e^{-i x t}}{i x}  = 2 \pi$). However, this contribution is not secular.
Alternatively, one can use Eq.~\eqref{eq:sin}, where the discrete delta function gives zero contribution upon integration in $\bk$.
Thus, terms at order $O(\Up/\Omega)$, considered when the sum in $\bk$ is discrete (represented through a Dirac delta function in the continuum), are not secular in the absence of a condensate, and in the traditional wave-turbulence approach, one needs to go to the next order in the small parameter.
The classical kinetic equation
\citep{zakharov2012kolmogorov, newell2011wave,galtier2003weak, galtier2023multiple} describing dominant homochiral 2D-3D interactions would be of the form
\begin{align}
    &\partial_{t} e_{\bp}^s
    =
    \Up^2 \int 
   \mathcal{T}_{\bk\bp\bq}
    \delta_{\bk\bp\bq} \delta(\omega_{\bp\bq}),
    \label{eq:KE_continuous}
\end{align}
where $\mathcal{T}_{\bk\bp\bq} = |U_{\bk}|^2/(4U'^2)  \Big(
\Re( V_{\bp\bk\bq}^{ss} V_{\bq \bk \bp}^{ss*} )
 e_{\bp}^s  +  V_{\bp\bk\bq}^{ss} V_{\bp \bk \bq}^{ss*}  e_{\bq}^s
 \Big)$
 in the homogeneous case $\Phi_{\bp\bk}^{ss}=0$
(see \citep{galtier2003weak}).
Note that for scale-separated triads, this kinetic equation is expected to turn into a spectral diffusion equation for waveaction, see \citep{nazarenko2011wave}.

\subsection{Near resonances and 2D-3D decoupling in wave turbulence}
\label{app:decoupling_WT}

In the previous section, we explained why the order of our expansion is one order less in the non-linearity compared to usual wave turbulence theory. However, since 2D modes decouple from 3D modes for exactly resonant interactions, the classical weak-wave turbulence framework cannot be straightforwardly applied to describe 2D-3D interactions, with or without a condensate. Instead, one must consider near-resonant interactions more explicitly, taking into account the finite width around the exact resonance. 
In such near-resonant interactions, the coupling coefficient between the 2D flow and the waves is non-zero, being of the order of the deviation of the frequency sum from exact resonance \cite{shavit2024sign}.

In a system of finite size $L$, 
the minimal frequency broadening that allows for 2D-3D coupling depends on the discreteness of the system, as $\omega^{ss}_{\bp\bq} \propto 1/L$.
Therefore, when $\Omega  t \to \infty$ is taken first, only exact resonances contribute, and 
decoupling necessarily occurs.
This is the case in our example with a 2D condensate, where dealing with near-resonances and decoupling is particularly simple, as the 2D flow is dominated by a few discrete low modes, associated to a well-defined interaction time scale $1/\Up$.
However, decoupling is also expected for any 2D spectrum $U_{\bk}$ (for example, without a condensate).
Taking the limit $\Omega \to \infty$, before the infinite box limit produces a transition from near-resonances contributing to only exact-resonances 
contributing. This corresponds to a transition from continuous to discrete wave turbulence, in the parlance of wave turbulence theory \citep{nazarenko2011wave, lvov2010discrete}.

When the infinite box limit $L\to\infty$ is taken first, however, the occurrence of decoupling is more intricate, as there is a large number of near-resonances $\omega^{ss}_{\bp\bq} t \ll 1$, which in principle enable 2D-3D coupling.
However, in its current formulation, (continuous) weak wave turbulence predicts a kinetic equation \eqref{eq:KE_continuous} restricted to the resonant manifold $\omega_{\bp\bq}^{ss} = 0$, on which the energy transfer to 2D exactly vanishes \citep{galtier2003weak}.
Our procedure gives a hint as to how 
this paradox can be resolved: one needs to broaden the delta function enforcing the resonant condition in the kinetic equation to include all near resonances which are allowed by the time scale separation between the non-linearity and the wave frequency (i.e.\ the frequency broadening), similarly to what is done in~\cite{lvov1997statistical}. This width will in turn determine the magnitude of the interaction coefficient between the 2D and 3D modes. Note that this discussion focuses on capturing interactions between \textit{exactly} zero frequency modes and waves, through \textit{near} resonances, rather than interactions between \textit{approximate} zero modes (slow modes $\omega_{\bp} \to 0$) and waves through \textit{exact} resonances, as done in~\cite{shavit2025turbulent}.

An important question is if such a regularized kinetic theory could provide relevant predictions for what 2D scale is energized and how much energy is transferred to the 2D manifold in the absence of a condensate.
The fraction of energy reaching the 2D manifold is key in dictating how much energy feeds a wave-turbulent cascade due to 3D-3D interactions, 
for which kinetic theories provide relevant estimates of energy spectra and fluxes \citep{shaltiel2024direct}.

\section{Mean-wave kinetic equation in the presence of  scale separation}
\label{app:simplifying}

\begin{figure}[t]
\includegraphics[width=\columnwidth]{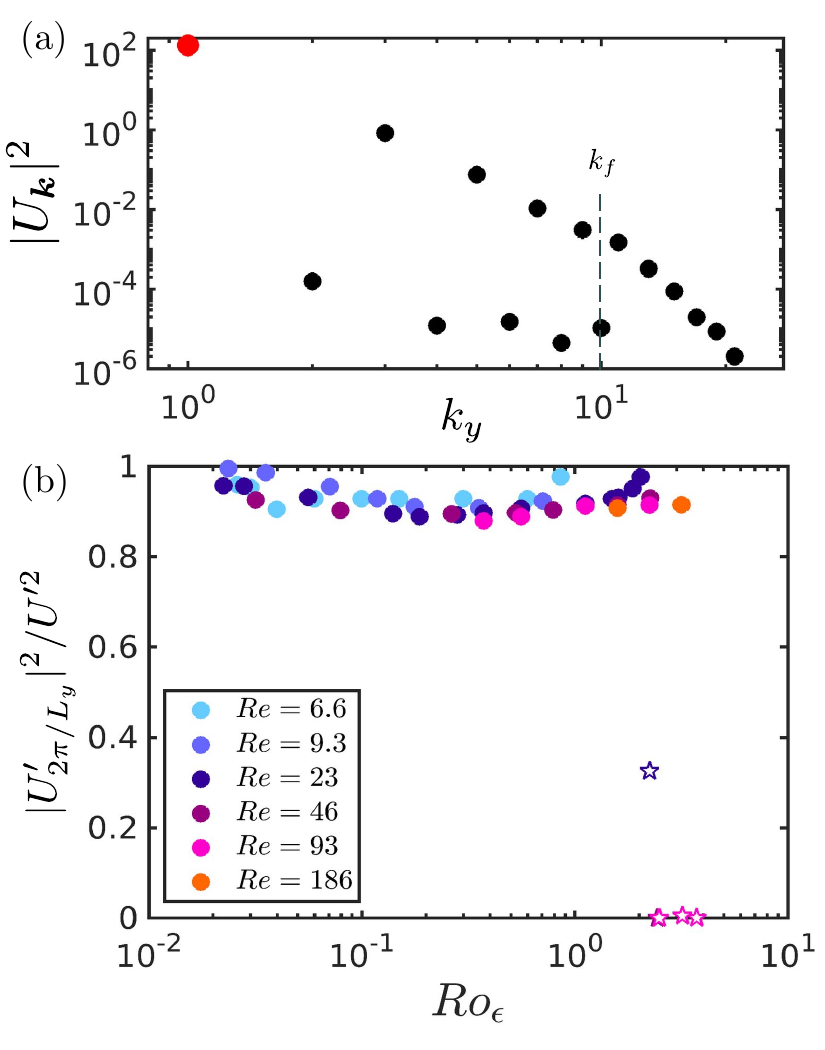} 
\caption{\SG{(a) Condensate energy spectrum at $Ro=0.011$, $Re=46$.
Throughout this paper, the condensate is approximated as the single trigonometric mode $k_y=2\pi/L_y$ of amplitude $U'$ (red point in (a)), which dominates the mean-flow energy spectrum.
(b) Relative amplitude of the first mean-shear mode $|U'_{2\pi/L_y}|^2$ compared to the total mean-shear rate $\Up^2$, as a function of $\Roe$ for all our data points. Pentagrams correspond to a regime of crystallized vortices obtained at low rotation, and not described in our theory.
}
}
\label{fig:mf_spec}
\end{figure}

We now consider homochiral wave interactions with a single 2D mode in the sum in Eq.~\eqref{eq:E_full}, taken at the box scale $k_y =\pm\frac{2 \pi}{L_y}$, giving:
\begin{align}
&\partial_te^s_{\bp} 
 = \epsilon \chi_{\bp}^s   \nonumber\\
& ~ + \frac{1}2 U_{\bk}^* \Vp^{ss} \langle a_{\bp}^s a_{\bq}^s \rangle^*  \delta_{\bk\bp\bq} 
\mathbb{1}_{|\omega_{\bp\bq}^{ss} |< \Up}
+ (\bk \to -\bk) + c.c..
\label{eq:KE_onemode}
\end{align}
Approximating the condensate with a trigonometric function can be justified from the measurement of the mean-flow energy spectrum in the DNS. As seen in Fig.~\ref{fig:mf_spec}(a) for $Ro=0.011, Re=46$, the energy in mode $k_y=2\pi/L_y$ is at least two order of magnitudes larger than any other condensate mode. 
The dominance of the gravest condensate mode is valid for most rotation regime, as shown in Fig.~\ref{fig:mf_spec}(b): the energetic content of the gravest mode represents more than $80 \%$ of the total mean-flow energy.
Note that at very low rotation, the condensate takes the form of localized vortices, occupying a scale different than the domain scale (stars in Fig.~\ref{fig:mf_spec}(b)). This state is outside our current theory.
Note also that this single-mode approximation is not designed to capture the exact spatial dependence of the local mean shear $\partial_y U$.

We now further expand all coupling coefficients $\Vp^{ss}$ in $k_y/k_f$ for each triad $\bk+\bp+\bq=0$), considering only homochiral-wave interactions which are in near resonance with the condensate at the forcing scale $k_f$, i.e such that
$|\omega_{\bp\bq}^{ss}| \delta_{p- k_f} < \Up$.

\subsection{Leading order equation in $k_y/p\sim l_f/L_y $}
\label{app:simplifying_homo}
We expand the interaction coefficient for a triad involving waves $(\bp,s), (\bq,s)$, where $\bq=(-p_x, -p_y-k_y,-p_z)$, in $k_y/p$ as 
\begin{align}
    \Vp^{ss} &=  - i p_x (\bh_{\bp}^{-s}\cdot \bh_{\bq}^{-s}) + i k_y  h_{\bp,x}^{-s} ~ h_{\bq,y}^{-s}, \nonumber\\
    &\approx - i p_x \left(1-is k_y\frac{p_xp_z}{p_\perp^2 p}\right)+ik_y h_{\bp,x}^{-s} ~ h_{-\bp,y}^{-s} \nonumber\\
    &= -i p_x - s  k_y\left(\frac{p_x^2p_z}{p_\perp^2 p} +\frac{p_z}{2p}\right) -i  k_y\frac{p_x p_y}{2p^2}  + O\left(\frac{l_f^2}{L_y^2 }\right),
     \label{eq:Vp_expand}
\end{align}
where we used that
\begin{align}
    h_{\bp,x}^{-s} ~ h_{-\bp,y}^{-s} = -\frac{p_x p_y}{2 p^2}  + i s \frac{p_z}{2 p}
    \label{eq:h_x_h_y}  \\
    (\bh_{\bp}^{-s}\cdot \bh_{\bq}^{-s}) \approx 1 -  i s  k_y \frac{p_x p_z}{p_\perp^2 p}
     \label{eq:h_p_hq}
\end{align}
Note that we have expanded up to order $O(k_y/p)$, including terms of order $O(1)$ and $O(k_y/p)$. Both terms will give contributions of the same order in the end: the former will enter only into correlator differences of order $\sim k_y/p$, while we will keep the latter only for correlator sums $\sim O(1)$. In particular, we will group terms of the form 
\begin{align}
   \langle a_{\bp+\bk}^s a_{-\bp}^s \rangle-\langle a_{\bp}^s a_{-\bp+\bk}^s\rangle 
\approx k_y \partial_{p_y}\langle a_{\bp}^s a_{-\bp+\bk}^s\rangle \\
 \langle a_{\bp}^s a_{-\bp+\bk}^s\rangle 
+
\langle a_{\bp+\bk}^s a_{-\bp}^s \rangle\approx 2\langle a_{\bp}^s a_{-\bp+\bk}^s\rangle .
\end{align}
The two correlators appearing in these sums and differences are related by the transformation $k_y\to -k_y$ together with complex conjugation $(\cdot){\rightarrow} (\cdot)^*$. Correspondingly, terms entering the coefficients $V^{ss}_{\bp\bk\bq}$ in \eqref{eq:Vp_expand} that are odd with respect to this transformation   will give rise to a difference between the correlators in the kinetic equation, while even terms will produce the sum. In particular, the second term in the last line in equation \eqref{eq:Vp_expand} is odd under this symmetry and is proportional to $k_y$, thus will not contribute at leading order.

Thus, using equation \eqref{eq:E_full} and that $U_{-\bk}=\frac{i U'}{\sqrt{2}k_y}$, and $(V^{ss}_{\bp(-\bk)\bq})^*=ip_x-ik_y (p_xp_y)/2p^2$ to leading order, we obtain the result
\begin{align}
    &\partial_t e_p^s =\epsilon \chi^s_p+\frac{i U'}{2\sqrt{2}k_y}ip_x \left(\langle a_{\bp}^s a_{-\bp+\bk}^s\rangle 
-
\langle a_{\bp+\bk}^s a_{-\bp}^s \rangle\right) \nonumber\\
&+\frac{i U'}{2\sqrt{2}k_y}\frac{-ik_y p_x p_y}{2p^2}\left(\langle a_{\bp}^s a_{-\bp+\bk}^s\rangle 
+
\langle a_{\bp+\bk}^s a_{-\bp}^s \rangle\right)+c.c \\
&= \frac{U'p_x}{2\sqrt{2}}\partial_{p_y} \langle a_{\bp}^s a_{-\bp+\bk}^s\rangle +\frac{U' p_x p_y}{2\sqrt{2}p^2}\langle a_{\bp}^s a_{-\bp+\bk}^s\rangle \nonumber \\& 
~~~~~~~~~~~~~~~~~~~ ~~~~~~~ ~~~ ~~~~+c.c+\epsilon\chi^s_p+O\left(\frac{k_y}p\right).
\end{align}

Defining 
\begin{equation}
\begin{split}
    \Phi_{\bp\bk}^s\equiv \Re\left[\langle a_{\bp}^s a_{-\bp+\bk}^s\rangle\right]=\frac{ \langle a_{\bp}^s a_{-\bp+\bk}^s\rangle+c.c}2\\
    = \frac{\langle a_{\bp}^s a_{-\bp+\bk}^s\rangle+\langle a_{\bp+\bk}^s a_{-\bp}^s\rangle}{4}+c.c +O\left(\frac{k_y}p\right) \\
    =\frac{\langle a_{\bp}^s a_{-\bp+\bk}^s\rangle+\langle a_{\bp}^s a_{-\bp-\bk}^s\rangle}{4}+c.c +O\left(\frac{k_y}p\right),
    \end{split}
\end{equation}
we obtain the equation
\begin{align}
    \partial_t e_{\bp}^s=\frac{U'}{\sqrt{2}}\left[p_x\partial_{p_y} \Phi_{\bp\bk}^s +\frac{ p_x p_y}{p^2}\Phi_{\bp\bk}^s\right]+\epsilon \chi^s_{\bp}
    \label{eq:KE_app}
\end{align}
Note that $ \Phi_{\bp\bk}^s$ and $ \Phi_{\bp(-\bk)}^s$ solve the same equation, consistent with there being no distinction between the two at leading order. Indeed, $ 2(\Phi_{\bp\bk}^s+ \Phi_{\bp(-\bk)}^s)$ is the coefficient of the term with wave number $k_y$ in the cosine expansion of the correlator. The symmetry of this correlator is inherited from the assumed anti-symmetry of the mean flow. 

In steady state, the correlator therefore solves the ODE
\begin{align} 
\Big[ p_x \partial_{p_y}  + \frac{p_x p_y}{p^2} \Big] \Phi_{\bp\bk}^s
&= -\frac{\sqrt{2} \epsilon \chi_{\bp}^s}{\Up}
\label{eq:ODE_phi_homo}
\end{align}
for near-resonant modes, i.e.\ modes which satisfy $p_yp_zk_y/p^3<U'/\Omega$. In the following, we will assume that if this condition is not satisfied at the forcing scale for a given $p_x,p_z$, then the corresponding waves do not interact with the condensate (for whichever $p_y$).

\subsection{Reynolds stress}

The wave energy equation \eqref{eq:Econs} is coupled with the mean-flow equation \eqref{eq:global_bal}, obtained by Reynolds averaging over time $T \gg 1/\Up$ and over $z$, in a time window where the ensemble-averaged correlator $\langle a_{\bp}^s a_{\bq}^{s'} \rangle$ is stationary. 
The total energy transfer due to homochiral waves is 
\begin{align}
    \Ttwo 
  = &\sum_{\bp,s}  U'_{-\bk} \langle u_{\bp}^s v_{-\bp + \bk}^s \rangle +  U'_{\bk} \langle u_{\bp}^s v_{-\bp-\bk}^s \rangle    \nonumber \\  &=\frac{U'}{2\sqrt{2}} \sum_{\bp,s}  \left( \langle u_{\bp}^s v_{-\bp + \bk}^s \rangle +\langle u_{-\bp}^s v_{\bp+\bk}^s \rangle+     \right.\nonumber \\ &\left.+\langle u_{\bp}^s v_{-\bp-\bk}^s\rangle+\langle u_{-\bp}^s v_{\bp - \bk}^s \rangle   \right) \nonumber \\
    &= 
 \frac{\Vol U'}{2\sqrt{2}}
 \int 
 {\rm d}\bp \sum_{s=\pm1}\left( \langle u_{\bp}^s v_{-\bp + \bk}^s \rangle +\langle u_{-\bp}^s v_{\bp - \bk}^s \rangle \right. \nonumber\\ &+\left.  \langle u_{\bp}^s v_{-\bp-\bk}^s \rangle + \langle u_{-\bp}^s v_{\bp+\bk}^s \rangle\right) \nonumber \\ &=\frac{\Vol U'}{\sqrt{2}}
 \int 
 {\rm d}\bp \sum_{s=\pm1} \Re\left[\langle u_{\bp}^s v_{-\bp + \bk}^s \rangle\right] +  \Re\left[\langle u_{\bp}^s v_{-\bp-\bk}^s \rangle\right] \nonumber\\
 &\equiv \frac{\Vol \Up}{\sqrt{2}} \int {\rm d}\bp\sum_{s=\pm1} \uv^{ss}_{\bp\bk}  \nonumber
\end{align}
where in the second line we included a factor of $1/2$ since we are double counting modes by including $\bp$ and $-\bp$ for each term in the sum. Here, 
\begin{align}
    \uv_{\bp\bk}^{ss} &\equiv \frac{1}{2} \Big( \langle u_{\bp}^s v^s_{-\bp-\bk} \rangle +  \langle u_{\bp}^s v_{-\bp+\bk}^s \rangle \Big) + c.c\nonumber \\
    &= \Re\left[\langle u_{\bp}^s v^s_{-\bp-\bk} \rangle\right] +\Re\left[\langle u_{\bp}^s v_{-\bp+\bk}^s \rangle\right] \nonumber \\
    &=\Re\left[ h_{\bp,x}^{-s} h_{-\bp +\bk,y}^{-s}  \langle a_{\bp}^{s} a_{-\bp +\bk}^{s} \rangle +h_{\bp,x}^{-s} h_{-\bp -\bk,y}^{-s}  \langle a_{\bp}^{s} a_{-\bp -\bk}^{s} \rangle\right] 
    \nonumber \\
    & =- \frac{p_x p_y}{p^2} ~ \Phi_{\bp \bk}^s ~ + O\left(\frac{1}{L_y k_f}\right),
\end{align}
where in the last line we used that $h_{\bp,x}^{-s} h_{-\bp -\bk,y}^{-s}=h_{\bp,x}^{-s} h_{-\bp ,y}^{-s}+O(k_y/p)$ together with equation \ref{eq:h_x_h_y}, that $\Phi_{\bp \bk}^s=\Phi_{\bp (-\bk)}^s$ and that
the imaginary part of the correlator is sub-leading:
\begin{align}
 & 2\Im\left[\langle a_{\bp}^s a_{-\bp+\bk}^s\rangle 
+
\langle a_{\bp}^s a_{-\bp- \bk}^s \rangle\right]\nonumber\\&=\langle a_{\bp}^s a_{-\bp+\bk}^s\rangle 
+
\langle a_{\bp}^s a_{-\bp- \bk}^s \rangle - c.c  \nonumber\\
&=
\langle a_{\bp}^s a_{-\bp+\bk}^s\rangle
- \langle a_{\bp+\bk}^s a_{-\bp}^s\rangle
+\langle a_{\bp}^s a_{-\bp- \bk}^s \rangle 
- \langle a_{\bp-\bk}^s a_{-\bp}^s\rangle \nonumber \\
& =-k_y\partial_{p_y}\langle a_{\bp-\bk}^s a_{-\bp}^s\rangle+k_y \partial_{p_y}\langle a_{\bp-\bk}^s a_{-\bp}^s\rangle \nonumber \\
& = o\left( \frac{k_y}{p}\right),
\label{eq:imag_O1}
\end{align}
using a Taylor expansion in each correlator. 

The real space Reynolds stress reads
\begin{equation}
    \langle uv\rangle =\left(\sum_{\bp,s} \langle uv\rangle_{\bp\bk}^{ss} \right)\cos(k_y y),
\end{equation}
which inherits the symmetry from the form of the mean shear $\partial_yU= \sqrt{2}U' \cos(k_y y)$, giving that indeed 
\begin{equation}
\begin{split}
  \Ttwo =\langle \langle uv\rangle\partial_y U\rangle_y = \frac{\Vol \Up}{\sqrt{2}} \int {\rm d}\bp\sum_{s=\pm1} \uv^{ss}_{\bp\bk} \\ = -\frac{\Vol \Up}{\sqrt{2}} \int {\rm d}\bp\sum_{s=\pm1} \frac{p_x p_y}{p^2} ~ \Phi_{\bp \bk}^s,
  \end{split}
\end{equation}
which is Eq.~\eqref{eq:uv_homo} in the main text.

\section{2D-2D interactions}
\label{app:2D-2D}
2D-2D interactions are insensitive to rotation (as $\omega_{\bp}^s = 0$ when $p_z=0$), hence are not restricted to resonances and involve both homochiral and heterochiral interactions.
They are therefore characterized by correlations between modes of opposite chiralities $\langle a_{\bp}^s a_{-\bp + \bk}^{-s} \rangle$.
From the triadic system \eqref{eq:triad3D}, the equations for energy $\langle |a_{\bp}^s|^2 \rangle/2$ and correlation $\langle a_{\bp}^s a_{-\bp}^{-s} \rangle$ in the limit of large scale separation $k \ll p$ are
\begin{align}
&\partial_t e_{\bp}^s =
\frac{\Up}{\sqrt{2}} \Bigg( \Big[ p_x \partial_{p_y}   +  \frac{p_x p_y}{p^2} \Big] \Phi_{\bp\bk}^s
- 2 \Re ( H_{\bp}^{s,-s} \Psi_{\bp\bk}^{s,-s*})
\Bigg) \nonumber \label{eq:e_hetero_2D}\\
& ~~~~~~~~~~~~~~~~~~~~~~~~~~~~~~~~~~~~~~~~~~~ 
+  \epsilon \chi_{\bp}^s \\
&\frac{1}2\partial_t \langle a_{\bp}^{s*} a_{\bp}^{-s} \rangle =
\frac{\Up}{\sqrt{2}} \Bigg(  \Big[ p_x \partial_{p_y}   +  \frac{p_x p_y}{p^2} 
\Big] 
(\Psi_{\bp\bk}^{s,-s*}
+\Psi_{\bp\bk}^{-s,s})
)
\nonumber\\
& ~~~~~~~~~~~~~~~~~~~~~~~~~~~~~~~~
-  2 H_{\bp}^{s,-s*} (\Phi_{\bp\bk}^{s*} + 
\Phi_{\bp\bk}^{-s} )
\Bigg),
\label{eq:C_hetero_2D}
\end{align}
with heterochiral correlations
$\Psi_{\bp\bk}^{s,-s} \equiv  \frac{1}2\left(\langle a_{\bp}^s a_{-\bp+\bk}^{-s} \rangle + \langle a_{\bp}^s a_{-\bp-\bk}^{-s} \rangle\right) \in \mathbb{C}$, and coefficients
$H^{s,-s}_{\bp} = h_{\bp,x}^{-s} h_{-\bp,y}^{s} = 
\frac{p_x p_y}{2p^2} $,
written here for 2D fluctuation modes $p_z=0$.

In steady state, and assuming zero helicity injection ($\chi_{\bp}^s = \chi_{\bp}^{-s} \Rightarrow \Psi_{\bp\bk}^{+-*}= \Psi_{\bp\bk}^{-+})$, Eqs.~\eqref{eq:e_hetero_2D}-\eqref{eq:C_hetero_2D} are written as
\begin{align}
    \Up \Big[ p_x \partial_{p_y}   +  \frac{p_x p_y}{p^2} \Big] \Phi_{\bp\bk}^s
-  \frac{p_x p_y}{p^2}\Psi_{\bp\bk}^{s,-s*} &= -  \sqrt{2} \epsilon \chi_{\bp}^s \\
\Up   \Big[ p_x \partial_{p_y}   +  \frac{p_x p_y}{p^2} 
\Big] 
\Psi_{\bp\bk}^{s,-s*}
-  \frac{p_x p_y}{p^2} \Phi_{\bp\bk}^{s*} &=0,
\end{align}
or, equivalently,
\begin{align}
    \Up \Big[ p_x \partial_{p_y}   +  \frac{2p_x p_y}{p^2} \Big] \Big( \Phi_{\bp\bk}^s - 
    \Psi_{\bp\bk}^{s,-s*} \Big)
    &= -  \sqrt{2} \epsilon \chi_{\bp}^s \label{eq:2D_2a} \\
\Up  p_x \partial_{p_y}    
\Big(  \Phi_{\bp\bk}^s  +
\Psi_{\bp\bk}^{s,-s*} \Big)
&=-  \sqrt{2} \epsilon \chi_{\bp}^s. \label{eq:2D_2b}
\end{align}
Eqs.~\eqref{eq:2D_2a}-\eqref{eq:2D_2b} are solved for $p_y > \pyf\equiv (k_f^2 - p_x^2 - p_z^2)^{\frac{1}{2}}$ by:
\begin{align}
    \Phi_{\bp\bk}^s - 
    \Psi_{\bp\bk}^{s,-s*} 
    &= - \frac{ \sqrt{2} \epsilon}{ \Up p_x}
    \int_{-\infty}^{p_y}
    \frac{q^2}{p^2} \chi_{\bp}^s dq_y
    = -\frac{\sqrt{2}\epsilon_{p_x,p_z=0}^s}{\Up p_x} \frac{
    k_f^2}{p^2} \\
    \Phi_{\bp\bk}^s +
    \Psi_{\bp\bk}^{s,-s*} 
    &= - \frac{ \sqrt{2} \epsilon}{ \Up p_x}
    \int_{-\infty}^{p_y}
     \chi_{\bp}^s dq_y = - \frac{\sqrt{2} \epsilon_{p_x,p_z=0}^s }{\Up p_x},
\end{align}
where 
\begin{equation}
\begin{split}
   \epsilon_{p_x,p_z}^s &\equiv  \int_{-\infty}^{\infty} \chi^s_{\bq}~dq_y = \frac{1}{8\pi k_f^2 \Vol}\int_{-\infty}^{\infty} \delta\left(q-k_f\right)~dq_y\\
   & =\frac{  \epsilon }{ 4 \pi k_f\Vol    \sqrt{k_f^2 - p_x^2 - p_z^2 } } 
   \end{split}
\end{equation}
denotes the total energy injection rate in $p_y$  for each chiral sector, and for given $p_x,p_z$.

The small-scale flux due to the advection of the fluctuations by the mean shear is therefore
\begin{align}
    \Pi_{\rm adv}^s & \equiv - \frac{p_x \Up}{\sqrt{2}} \Phi_{\bp\bk}^s \nonumber \\
    &=\frac{\epsilon_{p_x,p_z=0}^s}{2} \left(  \frac{k_f^2}{p^2} + 1  \right)  \underset{|p_y|\to\infty}{\to} \frac{\epsilon_{p_x,p_z=0}^s}{2}, 
    \label{eq:flux2D}
\end{align}
and half the energy injected in the 2D manifold is transferred to small scales. 
The other half energizes the condensate.
This splitting is due to the 2D-2D interactions behaving exactly like in a 2D-3C flow,
which exhibits a forward cascade of the energy of the vertical component, $|w|^2$.
This forward cascade is accompanied with an helicity exchange between the two sectors of opposite chiralities.

Note that the factor $1/2$ in Eq.~\eqref{eq:flux2D} is due to energy being distributed equally between the two independent variables of the 2D3C hydrodynamic equations, vertical velocity $w$ and horizontal enstrophy $\omega_{\parallel}$. These are related to the helical component via 
$w_{\bp} = (a_{\bp}^+ + a_{\bp}^-) /\sqrt{2}$ and 
$\omega_{\parallel\bp} =  p^2 (a_{\bp}^+ - a_{\bp}^-) /\sqrt{2}$.
Equations~\eqref{eq:2D_2a}-\eqref{eq:2D_2b} therefore correspond to the conservation
of horizontal enstrophy 
$|\omega_{\parallel{\bp}}|^2 = \frac{1}{2}(e_{\bp}^+ + e_{\bp}^-  - \langle a_{\bp}^+ a_{-\bp}^- \rangle + \langle a_{\bp}^- a_{-\bp}^+ \rangle)  $
and of vertical energy 
$|w_{\bp}|^2 = \frac{1}{2}(e_{\bp}^+ + e_{\bp}^-  + \langle a_{\bp}^+ a_{-\bp}^- \rangle + \langle a_{\bp}^- a_{-\bp}^+ \rangle)  $, respectively.

The Reynolds stress, which here includes heterochiral correlators, is written as
\begin{align}
   & \uv_{\bp\bk} =  
   -
 \frac{p_x p_y}{  p^2} ~ \Phi_{\bp\bk}^s
 +
  ~ 2\Re( H_{\bp}^{s,-s} \Psi_{\bp\bk}^{s,-s})  \\
   &= 
\left\{
\begin{aligned}
&\frac{\sqrt{2}\epsilon_{p_x,p_z=0}^s k_f^2}{ 2\sigma \Up  } 
\frac{p_y}{p^4}
\text{ if }  -\pyf < \sigma p_y  < \pyf  \label{eq:uv_hetero_sol_2D}
\\
&\frac{\sqrt{2}\epsilon_{p_x,p_z=0}^s k_f^2}{ \sigma \Up  } 
\frac{p_y}{p^4} \text{ if }  \sigma p_y  > \pyf 
\end{aligned}
  \right.
\end{align}
with $\pyf= \sqrt{k_f^2 - p_x^2 - p_z^2}$
and $\sigma= \sgn{ (U' p_x)}$.

Over integration with our isotropic forcing,
\begin{align}
\frac{\Ttwotwo}{\epsilon} = \Vol \int_{p_z=0} \frac{\Up}{\sqrt{2}} \uv_{\bp\bk} d\bp = \frac{1}{2} \epsilon_{\rm 2D} =\frac{l_f}{4L_z}   \label{eq:T2D2D}
\end{align}

\section{Stationary solutions for homochiral waves}
\label{app:stationary}

\subsection{Solution for correlator $\Phi_{\bp\bk}^s$}
\label{app:stationary_homo}
The stationary equation \eqref{eq:ODE_phi_homo} can be written in the form
\begin{equation}
    \partial_{p_y}\left(-sp  \frac{U'p_x}{\sqrt2}\Phi^s_{\bp\bk}\right)\equiv\partial_{p_y}\left(\Pi_{H^s}\right)=sp \epsilon \chi_{\bp}^s,
\end{equation}
which is solved by:
\begin{align}
    \Phi_{\bp\bk}^s &= - \frac{  \sqrt{2} \epsilon}{  \Up p_x} 
    \int_{-\infty}^{p_y} \frac{q}{p}  ~\chi^s_{\bq}  ~dq_y  ~~ \text{ for } \Up p_x < 0   ~(dq_y >0) ; \nonumber \\
     \Phi_{\bp\bk}^s &= - \frac{ \sqrt{2} \epsilon}{\Up p_x} 
    \int_{\infty}^{p_y} \frac{q}{p}  ~\chi^s_{\bq}  ~dq_y ~~
    \text{ for } \Up p_x > 0   ~(dq_y <0),
    \label{eq:sol_stat}
\end{align}
where we have used different boundary conditions depending on the sign of $U' p_x$:
\begin{align}
  \Pi_{H^s}(p_y\to -\infty)&=0 ~~ \text{ for } ~~ \Up p_x < 0 \\
  \Pi_{H^s}(p_y\to \infty)&=0 ~~ \text{ for } ~~ \Up p_x > 0, 
\end{align}
implying that for $\Up p_x < 0$ ($\Up p_x > 0$) the flux of helicity is from the forcing scale to $p_y\to \infty$ ($p_y\to -\infty$), i.e.\ to positive (negative) $p_y$. That these are the right boundary conditions (and not the opposite, as the flux is always from the forcing scale to $p_y\to\pm \infty$) can be seen from the condition that $\Phi_{\bp\bk}^s>0$ by continuity since in the limit $k_y\to0$ the correlator $\Phi_{\bp\bk}^s$ turns into the energy in the mode with wavenumber $\bp$ and helicity sign $s$. 

Put another way, Eq.~\eqref{eq:ODE_phi_homo} is forced differently depending on the sign of $U' p_x$: if $U' p_x<0$, the RHS is positive and $\Phi_{\bp}^s$ increases with $p_y$, and vice-versa if $U' p_x>0$.
We therefore need to select the solution depending on the sign of $\Up p_x$, and we are assuming that the information comes from either $q_y < p_y$ if $\Up p_x <0$, or from $q_y > p_y$ if $\Up p_x >0$.
This choice corresponds to the irreversible dynamics along the characteristic line $p_y(t) = p_y - p_x \Up t$, where the direction of the energy (and helicity) flux in $p_y$ depends on the sign of $\Up p_x$.
The quadrants $\Up p_x p_y <0 $ are selected by the dynamics.
See \citep{kolokolov2020structure} where a similar result is obtained using a dynamical solution of equation \eqref{eq:triad3D}, assuming the interacting waves are complex conjugates (i.e.\ taking the $k_y\to 0$ limit first).

Solution \eqref{eq:sol_stat} is simplified when $\chi_{\bp}^s = \chi_{\bp}^{-s} = \delta(p-k_f) /(8 \pi k_f^2 \Vol)$ into:
\begin{align}
 \Phi_{\bp\bk}^s = 
 \left\{
    \begin{aligned}
     &~ 0 ~~ \text{ if }  \sigma p_y  < -\pyf \\
&   \frac{ \sqrt{2}\epsilon_{p_x,p_z}^s k_f}{   2 |\Up p_x| } \frac{1}{p }   
 ~~ \text{ if }  - \pyf <  \sigma p_y < \pyf \\
& \frac{ \sqrt{2}\epsilon_{p_x,p_z}^s k_f}{    |\Up p_x|  } \frac{1}{p }  
~~ \text{ if } \sigma  p_y  > \pyf
\end{aligned}
  \right.
  \label{eq:sol_homo}
\end{align}
with $\sigma \equiv \sgn (-U' p_x)$ the relative sign of the shear and
\begin{equation}
\begin{split}
   \epsilon_{p_x,p_z}^s &\equiv  \int_{-\infty}^{\infty} \chi^s_{\bq}~dq_y = \frac{1}{8\pi k_f^2 \Vol}\int_{-\infty}^{\infty} \delta\left(q-k_f\right)~dq_y\\&=\frac{1}{8\pi k_f^2 \Vol}\frac{2k_f}{p_y}=\frac{  \epsilon }{ 4 \pi k_f\Vol    \sqrt{k_f^2 - p_x^2 - p_z^2 } } 
   \end{split}
\end{equation}
the energy injection rate in each $(p_x,p_z)$ line.
%
Solution \eqref{eq:sol_homo} reflects the conservation of helicity within each polar sector: $p \Pi_{\rm adv}^s = k_f \epsilon_{p_x,p_z}^s$ with the energy flux $ \Pi_{\rm adv}^s = - U' p_x \Phi_{\bp\bk}^s /\sqrt{2}$.

Solution \eqref{eq:sol_homo} is only valid when the excitation at $k_f$ is in near resonance with the condensate, i.e if $|\omega_{\bp\bq}^{ss}| \delta_{p- k_f} < \Up$.
Otherwise, $\Phi_{\bp}^s=0$ as such excited modes do not interact with the condensate, and the corresponding Reynolds stress is zero.

By integrating solution \eqref{eq:sol_homo} over all waves, one can compute the 
total transfer from 3D to 2D, 
\begin{align}
    &\Tthree = - \sqrt{2} \Up \int_{-K_x,-K_z, -K_y}^{K_x,K_z,K_y} d\bp ~ \frac{p_x p_y}{2 p^2} (\Phi_{\bp}^+ + \Phi_{\bp}^-) \\
    &= 4\epsilon \int_{-K_x}^{0} \int_0^{K_z} dp_x dp_z  \frac{ \mathbb{1}_{|\omega_{\bp\bq}^{ss}| \delta_{p- k_f} < \Up
}}{\sqrt{k_f^2 - p_x^2 - p_z^2}} \nonumber \\
    &\times \Big(
	 \int_{-\pyf}^{\pyf} \frac{p_y}{4\pi p^3} ~dp_y  
     + \int_{ \pyf}^{K_y} \frac{p_y}{2\pi p^3} ~dp_y \Big) \\
	&=\frac{2\epsilon}{\pi \Up} \int_{\pi/2}^{\pi}  {\rm d}\alpha \int_0^{k_f}  {\rm d} p_0   \frac{ p_0 ~ \mathbb{1}_{|\omega_{\bp\bq}^{ss}| \delta_{p- k_f} < \Up}
    }{\sqrt{k_f^2 - p_0^2}} 
    \Big(\frac{1}{k_f} - \frac{1}{\sqrt{K_y^2 + p_0^2}} \Big)
\end{align}
in polar coordinates $p_x=p_0 \cos \alpha, p_z= p_0 \sin \alpha$, $p_0 = \sqrt{p_x^2 + p_z^2}$. Hence, when $\omega_{\bp\bq}^{ss} < \Up$ for all modes in the forcing shell, and in the limit $K_y \to \infty$,
\begin{align}
\Tthree & = \epsilon_{\rm 3D}
\end{align}
and
all the energy injected in 3D is transferred to the condensate in this case.

\subsection{Near-resonant fraction of the injected energy}
\label{app:QR}

\begin{figure}[t]
\subfloat{\includegraphics[width=0.9\columnwidth]{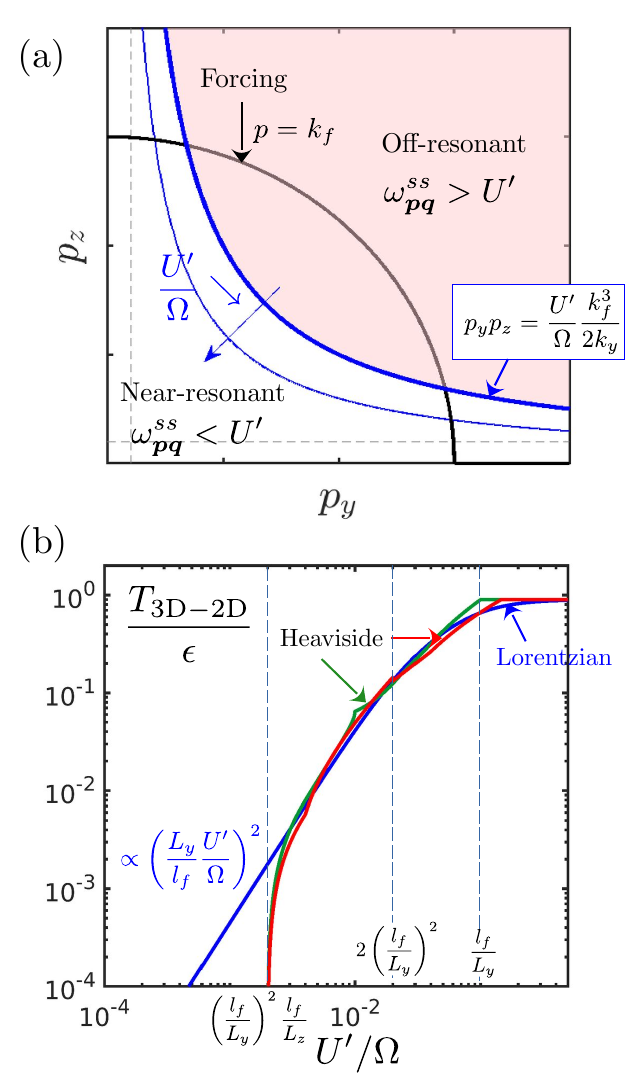}}
\caption{
(a) Within the QL approximation,  only the excited waves in near-resonance with the condensate energize it.
With the stepwise treatment of near-resonances in \eqref{eq:Heaviside}, only the waves within the forcing shell that lie below the hyperbole $p_y p_z = \frac{\Up}{\Omega} \frac{k_f^3}{2 k_y}$ (blue line) contribute.
When $\Omega\to \infty$ ($\Up/\Omega \to 0$), fewer modes energize the 2D condensate as resonances become more and more stringent.
(b) Energy transfer from 3D waves to the 2D condensate as a function of $\Up/\Omega$, with a Heaviside (green, \eqref{eq:eps23_full}) and a Lorentzian filter (blue, \eqref{eq:e32_Lor}).
Red line: asymptotic scaling \eqref{eq:dec_scalings2} approximating \eqref{eq:eps23_full}.
Parameters: $k_f=10, L_y=2L_z=2\pi$.
}
   \label{fig:decoupling}
\end{figure}

In this section, we compute explicitly the energy transfer from the 3D waves to the 2D condensaten $\Tthree$, as a function of the rescaled condensate amplitude $\Up/\Omega$ and the domain geometry, starting from Eq.~\eqref{eq:eps2D_closed}.

Note that Eq.~\eqref{eq:eps2D_closed} holds even if the set of waves is discrete, in the form:
\begin{align}
    \frac{\Tthree}{\epsilon}
    &= \sum_{\bp} \chi_{\bp} ~ \mathbb{1}_{|\omega_{\bp\bq}^{ss}| < \Up},
    \label{eq:T3_discrete}
\end{align}
with the stepwise treatment of the near-resonances \eqref{eq:Heaviside}.
The validity of Eq.~\eqref{eq:T3_discrete} only comes from the conservation of energy and single-sign helicity in the near-resonant mean-wave system at high rotation, the latter being due to scale separation $k\ll p$.
(In the discrete case, the $p_y$ derivative in \eqref{eq:Econs} can be replaced by a discrete difference, which, over summation, yields the bounds $\Pi_{\rm adv} (p_y \to \pm \infty)$, which vanish due to the conservation of single-sign helicity.)

We consider here that $L_y k_f \gg 1$, $L_z k_f \gg 1$, $L_x k_f\gg1$.
When wavenumbers $(p_x,p_y,p_z)$ are large enough (such that $p_i \sim k_f \gg 2\pi/L_i$), the spacing between the Fourier modes is infinitesimal, hence one can consider a continuous set of wavenumbers for the waves $\bp$ (while the interaction occurs with the single discrete 2D mode $k_y$).
However, the continuous approximation fails close to planes $p_z=0, p_y= 0$ and $p_x=0$.
Here, the set of modes is necessarily discrete, unless the dimensions are set to $L_i = \infty$ ($i=x,y,z$).
We therefore truncate the modes at the first nonzero Fourier modes $p_i = 2\pi/L_i$ to account for this discreneness, and
handle modes $p_i > 2\pi/L_i$ continously.
This is a \emph{semi-continuous} approximation, corresponding to a regularization of the $L_i k_f \to \infty$ limit.
We will consider the different limits $L_i k_f \to \infty$ in a later stage.

This stepwise treatment of near-resonances via Heaviside filter $\mathbb{1}_{|\omega_{\bp\bq}^{ss}| < \Up}$ corresponds to a condition on the values of $p_y$ and $p_z$ within the forcing shell,
\begin{numcases}
{\frac{\omega_{\bp\bq}^{ss}}{\Up} < 1 \Leftrightarrow}
 \frac{p_y p_z}{k_f^2} < \frac{\Up}{\Omega} \frac{ k_f}{2 k_y} \text{   for   } |p_y |> k_y  \label{eq:reso1}\\
 \frac{p_z}{k_f} < \frac{\Up}{\Omega} \frac{k_f^2}{k_y^2}  \text{   for  } p_y = 0,  \label{eq:reso2}
\end{numcases}
where we have used the approximated value of $\omega_{\bp\bq}^{ss}$ when $p_y \gg k_y$ in \eqref{eq:reso1}, and 
when $p_y \ll k_y$ in  \eqref{eq:reso2}.
See Fig.~\ref{fig:decoupling}(a) for an illustration of the near-resonant sector.
In the following we consider the case $l_f/L_y, l_f/L_z \ll1$, and do not treat the case of a thin layer compared to the forcing scale $l_f/L_z \geq 1$.

Because the resonant condition in \eqref{eq:reso2} differs in plane $p_y = 0$ and in modes $p_y\neq 0$, we need to consider the integration of the forcing ring $p_y=0$ and of the bulk $p_y\neq 0$ separately.

\subsubsection{Contribution from modes $p_y \neq 0$, $p_z\neq0 $}

We introduce non-dimensional polar coordinates
\begin{align}
&\tilde{p}_z=p_z/k_f, \qquad \tilde{p}_y=\sqrt{1-\tilde{p}_z^2} \sin(\phi), \nonumber \\ &\tilde{p}_x=\sqrt{1-\tilde{p}_z^2} \cos(\phi),
\end{align}
where we consider only $p_y,p_z,p_x>0$, the other quadrants giving the same result by symmetry. We let $p_x\to 0$ continuously but cutoff a strip of width $k_y/k_f$ ($l_f/L_z$) around $p_y=0$ ($p_z=0$) from the integration,  giving a limit on the smallest angle to use in the $\phi$ integration and on the smallest possible $\tilde{p}_z$:
\begin{align}
    &\frac{l_f}{2L_z}<\tilde{p}_z<\sqrt{1-\left(\frac{l_f}{L_y}\right)^2}\approx 1 ,\nonumber \\ &  \phi^-<\phi<\frac{\pi}2 \qquad \phi^-=\arcsin\left[\frac{k_y}{k_f}\frac{1}{\sqrt{1-\tilde{p}_z^2}}\right].
\end{align}

When all 3D modes are resonant (i.e.\ obey condition \eqref{eq:reso1}), their contribution is
\begin{align} \frac{\Tthree^{\rm sphere}}{\epsilon}\approx \frac{2}{\pi}\int_{\frac{l_f}{L_z}}^{1} \int_{\phi^-}^{\pi/2}d\tilde{p}_z d\phi \approx 1-\frac{l_f}{L_z}-\frac{l_f}{L_y},  \qquad (p_y \neq 0)
\end{align}
where we only keep terms up to $O(l_f/L_z), O(l_f/L_y)$ (and in the following contributions of higher order from the integral should be discarded).
We are also assuming that all modes in the sphere are forced with the same weight. 

Nearly-resonant modes $p_y\neq0$ satisfy
\begin{equation}
    |\sin(\phi)| \tilde{p}_z \sqrt{1-\tilde{p}_z^2}<\beta,
\end{equation}
with
\begin{align}
    \beta \equiv  \frac{\Up}{2\Omega} \frac{L_y}{l_f},
\end{align}
which results in a requirement for the angular integration,
\begin{equation}
   \frac{k_y}{k_f}\frac{1}{ \sqrt{1-\tilde{p}_z^2}}< |\sin(\phi)|<\frac{\beta}{\tilde{p}_z \sqrt{1-\tilde{p}_z^2}},
\end{equation}
taking into consideration the $p_y-$cutoff in the lower bound.
This leads to a range of angles $[\phi^- ~ \phi^+]$ defined for each $\tilde{p}_z$ as
\begin{align}
\phi^+=\begin{cases}
        \arcsin\left[\beta\frac{1}{\tilde{p}_z\sqrt{1-\tilde{p}_z^2}}\right], \qquad \frac{\beta}{\tilde{p}_z\sqrt{1-\tilde{p}_z^2}}<1 \\
        \pi/2 \qquad\qquad\qquad\qquad\quad {\rm otherwise}
    \end{cases} 
    \label{eq:phi+}
\end{align}
and
\begin{align}
    \phi^-=\arcsin\left[\frac{k_y}{ k_f}\frac{1}{\sqrt{1-\tilde{p}_z^2}}\right]  
    \label{eq:phi-}
 \end{align}
(we consider $0<\phi<\pi/2$ and multiply the result by 4 to get the contribution from the other quadrants).

Resonant modes exist, i.e.\ $\phi^+>\phi^-$, only for 
\begin{equation}
    \tilde{p}_z< \beta \frac{ k_f}{k_y}
\end{equation}
which restricts the $\tilde{p}_z$ we can consider in this class. Outside this class, modes still need to satisfy 
\begin{equation}
    \tilde{p}_z<\sqrt{1-\left(\frac{k_y}{k_f}\right)^2} 
\end{equation}
for $\tilde{p}_y$ to lie below the cutoff $l_f/L_y$. 

The resonant condition is restrictive, i.e.\ $\phi^+<\pi/2$, for $\tilde{p}_z$ in the range
\begin{equation}
    \sqrt{\frac{1-\sqrt{1-4\beta^2}}2}<\tilde{p}_z<\sqrt{\frac{1+\sqrt{1-4\beta^2}}2}
    \label{eq:pz_limits}
\end{equation}
For $\tilde{p}_z$ outside this range, and satisfying $\tilde{p}_z< \beta k_f/k_y$, all modes up to $\phi=\pi/2$ contribute.

In particular, we see that the waves start to decouple from the condensate, $\Tthree<\epsilon$, when $\beta=1/2$, for $\tilde{p}_z=1/\sqrt{2}$ (the point where $\tilde{p}_z\tilde{p}_y$ is maximized), where the full sphere is resonant. 

Now assuming $\beta \leq 1/2$, we define
\begin{align}
    &{p}_z^- = \max\left(\sqrt{\frac{1-\sqrt{1-4\beta^2}}2}, \frac{l_f}{L_z}\right) \nonumber \\
    &{p}_z^+ = \min\left(\sqrt{\frac{1+\sqrt{1-4\beta^2}}2},\beta \frac{k_f}{k_y}, \right),
    \label{eq:pz_pm}
\end{align}

The contribution from modes $p_y\neq0, p_z\neq 0$ is finally
\begin{align}
        \frac{\Tthree^{\rm sphere}}{\epsilon}&=
        \frac{2}{\pi}\int_{\frac{l_f}{L_z}}^{p_z^-} d\tilde{p}_z 
    \int_{\phi^-}^{\frac{\pi}{2}}d\phi    +\frac{2}{\pi}\int_{p_z^-}^{p_z^+}d\tilde{p}_z \int_{\phi^-}^{\phi^+}d\phi
 \nonumber\\&   +\frac{2}{\pi}\int_{p_z^+}^{\min\left(\beta \frac{k_f}{k_y},\sqrt{1-\left(\frac{k_y}{k_f}\right)^2}\right)} d\tilde{p}_z  \int_{\phi^-}^{\frac{\pi}{2}}d\phi,
 \label{eq:sphere}
\end{align}
where we consider that each integral has increasing bounds, and is otherwise null.

\subsubsection{Contribution from modes $p_y=0$}
In addition, we need to consider separately the contribution from the ring at $\tilde{p}_y=0$. 
The contribution from the full ring at $p_y=0$ is a fraction of energy equal to $\frac{2\pi k_f \Delta k}{(2\pi/L_x) (2\pi/L_z)}\frac{(2\pi/L_x) (2\pi/L_z) (2\pi/L_y)}{4\pi k_f^2 \Delta k} = \frac{k_y}{2k_f}$, where $\Delta k$ is the width of the shell.
Introducing polar coordinates $\tilde{p}_z=\sin\theta$, $\tilde{p}_x =\cos\theta$,
the contribution from the ring is written as
\begin{equation}
\frac{\Tthree^{ring}}{\epsilon}=4\frac{l_f}{2L_y} \int \frac{d\theta}{2\pi}
\end{equation}
(we count 4 times the sector $\theta \in [0 ~\frac{\pi}{2}]$),
where the bounds of the $\theta-$integration 
vary depending on the resonant condition \eqref{eq:reso2}, which reads
 \begin{equation}
\tilde{p}_z=\sin\theta<\frac{U'}{\Omega}\frac{L_y^2}{l_f^2} \equiv2\beta \frac{L_y}{l_f}.
 \end{equation}
The integration on the ring should therefore be performed over
\begin{align}
   \frac{l_f}{L_z} < \sin(\theta)<2\beta \frac{k_f}{k_y}
\end{align}
and we can define
\begin{align}
    \theta^-\approx  \frac{l_f}{L_z} 
    \qquad  \theta^+= \begin{cases}
        \pi/2 \qquad \qquad \qquad\quad  2\beta \frac{L_y}{l_f} >1 \\
        \arcsin \left(2\beta \frac{L_y}{l_f}\right) \qquad  2\beta \frac{L_y}{l_f}<1
\end{cases}
\end{align}
so that 
\begin{equation}
\frac{\Tthree^{\rm ring}}{\epsilon}=\frac{l_f}{L_y} \int_{\theta^-}^{\theta^+} \frac{d\theta}{\pi} =\frac{l_f}{L_y}\frac{(\theta^+-\theta^-)}{\pi}       \label{eq:ring}
\end{equation}

The ring does not contribute  when $\theta^+<\theta^-$, implying the restriction
\begin{equation}
  \beta \geq \frac{1}2\frac{l_f}{L_z}\frac{l_f}{L_y}
\end{equation}

For $\beta < 1/2 (l_f/L_z)(l_f/L_y)$, using the Heaviside filter, the 3D modes inside the forcing ring in the plane $p_y=0$ completely decouple from the condensate.

\subsubsection{Regimes of decoupling}
The total energy contribution from near-resonant 3D modes is finally given by
\begin{equation}
\begin{split}
    \frac{\Tthree}{\epsilon}=\tilde{\epsilon} \left(\frac{\Tthree^{\rm sphere}}{\epsilon}  +\frac{\Tthree^{\rm ring}}{\epsilon} \right)
    \label{eq:energy_tot}
    \end{split}
\end{equation}
with
 \begin{equation}
 \tilde{\epsilon}\equiv\frac{1-\frac{l_f}{2L_z}}{1-\frac{l_f}{2L_y}-\frac{l_f}{L_z}},
\end{equation}
a factor correcting the weight of each point in the truncated sphere and in the ring.
Such a rescaling is necessary to make the continuous integration compatible with the energy injection rate in the discrete setup.
If all 3D modes contribute, 
\begin{align}
    \frac{\Tthree}{\epsilon} =1- \frac{l_f}{2L_z}  \equiv \frac{\epsilon_{\rm 3D}}{\epsilon},
\end{align}
the fraction of the energy injected in 3D modes.
Note that when writing \eqref{eq:energy_tot}, we use the same weight for modes in the ring and in the truncated sphere, which is only an approximation to what happens in the discrete setup with anisotropic dimensions $L_y=2L_x$. 
This can result in errors in $O(l_f/L_y, l_f/L_z)$, which will only slighly alter the leading-order behavior of $\frac{\Tthree}{\epsilon}$.

From \eqref{eq:energy_tot}, \eqref{eq:sphere} and \eqref{eq:ring}, we obtain explicitly:
\begin{align}
    \frac{\Tthree}{\tilde{\epsilon}}&=       \frac{l_f}{\pi L_y}(\theta^+-\theta^-)
    +  \left(p_z^--\frac{l_f}{L_z}\right) \nonumber\\
&    +\frac{2}{\pi}\int_{p_z^-}^{p_z^+}d\tilde{p}_z \phi^+ -\frac{2}{\pi}\int_{p_z^-}^{\min\left(\beta \frac{k_f}{k_y},\sqrt{1-\left(\frac{k_y}{k_f}\right)^2}\right)}\phi^- d\tilde{p}_z
 \nonumber\\&   -\left(p_z^+-\min\left(\beta \frac{k_f}{k_y},\sqrt{1-\left(\frac{k_y}{k_f}\right)^2}\right)\right),
 \label{eq:eps23_full}
\end{align}
with $\beta=  \frac{L_y}{l_f}\frac{\Up}{2 \Omega}$, $p_z^{\pm}$ defined in \eqref{eq:pz_pm}, $\phi^+$ in \eqref{eq:phi+} and $\phi^-$ in \eqref{eq:phi-}.
The energy transfer $\Tthree/\epsilon $ from Eq.~\eqref{eq:eps23_full} is shown as a function of $\Up/\Omega$ in Fig.~\ref{fig:decoupling}(b) (green line).
The energy transfer decreases with decreased $\frac{\Up}{\Omega} = \beta \frac{l_f}{2L_y}$, due to a combined effect of the bounds in the integral being more and more restricted, and the integrand $\phi^+$ decreasing with decreasing $\beta$.

In the following, we consider some asymptotic limits in order to simplify Eq.~\eqref{eq:eps23_full}.

\begin{enumerate}
\item \emph{\underline{Beginning of decoupling: $l_f/L_y, l_f/L_z\ll\beta \ll 1/2$.}} 
Here we can use the asymptotic form $\tilde{p}_z^-\approx \beta$ and $ \tilde{p}_z^+\approx 1-\frac{1}2\beta^2$, and to leading order 
\begin{align}
    &\frac{\Tthree}{\tilde{\epsilon}} =
    \beta - \frac{l_f}{L_z}
    \\&  +\frac{2}{\pi}\int_{\beta}^{1-\frac{1}2\beta^2}\arcsin\left[\frac{\beta}{\tilde{p}_z\sqrt{1-\tilde{p}_z^2}}\right] d\tilde{p}_z\nonumber \\ & +\frac{1}2\left(\beta^2 -\frac{l_f^2}{L_y^2}\right)
    \nonumber \\ &-\frac{2}{\pi}\int_{\beta}^{\sqrt{1-\left(\frac{l_f}{L_y}\right)^2}}\arcsin\left[\frac{k_y}{k_f\sqrt{1-\tilde{p}_z^2}}\right]d\tilde{p}_z.
\end{align}

Outside the regions where $\tilde{p}_z\sim 1$ and $\tilde{p}_z\sim \beta$, the argument of the arcsin is small and one can take a linear approximation:
\begin{equation}
    \int \phi^+ d\tilde{p}_z \approx \int \beta\frac{1}{\tilde{p}_z\sqrt{1-\tilde{p}_z^2}}d\tilde{p}_z=-\beta \tanh^{-1}\left(\sqrt{1-\tilde{p}_z^2}\right)
\end{equation}
and 
\begin{equation}
   \int \phi^- d\tilde{p}_z \approx \int\frac{k_y}{k_f}\frac{1}{\sqrt{1-\tilde{p}_z^2}}d\tilde{p}_z= \frac{k_y}{k_f} \sin^{-1}(\tilde{p}_z).
\end{equation}

We then obtain:
\begin{align}
    \frac{\Tthree}{\epsilon} &=
    \beta - \frac{l_f}{L_z}  - \beta\,\frac{2}{\pi}\left[\tanh^{-1}(\beta)-\tanh^{-1}\!\big(\sqrt{1-\beta^2}\big)\right] \nonumber \\
    &\approx 
    \beta - \frac{l_f}{L_z} + \frac{2}{\pi}
    \beta \log \left( \frac{2}{\beta}\right),
\end{align}
where we have used that 
$$ \left[\tanh^{-1} (\sqrt{1-{\tilde{p}_z}^2})  \right]_{a}^b = \frac{1}{2}\Big[ \log \Big( \frac{1- \sqrt{1 - {\tilde{p}_z}^2}}{1+ \sqrt{1 - {\tilde{p}_z}^2}} \Big)\Big]_{a}^{b} \approx \log (\frac{b}{a}) 
$$
when $b, a \ll 1$ and
$\approx \log (2 /a)$ when $b\sim 1$ and $a \ll 1$.

\item \emph{\underline{Decoupling of bulk modes ($p_y, p_z \gg l_f/L_z, l_f/L_y$)}}
For $l_f/L_y\ll\beta\ll l_f/L_z$ (if $L_z<L_y$)
we obtain that
\begin{align}
    \frac{\Tthree}{\tilde{\epsilon}}& \approx \frac{2}{\pi}\int_{l_f/L_z}^{1-\beta^2}\arcsin\left[\frac{\beta}{\tilde{p}_z\sqrt{1-\tilde{p}_z^2}}\right] d\tilde{p}_z \nonumber \\ & \approx \frac{2}\pi \beta \ln (2L_z/l_f)
\end{align}

Here, the lower boundaries of integration have changed (i.e.\ the smallest possible $p_z$ are now restricted by the resonance condition, hence $p_z^- = l_f/L_z$), and we have neglected terms of the order $O(l_f/L_y)$.  
Note that when $L_z>L_y$, for $l_f/L_z\ll \beta \ll l_f/L_y$ we get
\begin{align}
   \frac{\Tthree}{\tilde{\epsilon}} &\approx
    \frac{2}{\pi}\int_{\beta}^{\beta L_y/l_f}d\tilde{p}_z \phi^+
 \nonumber\\
   &\approx  \frac{2}{\pi} \beta \ln  ( L_y/l_f)
\end{align}

\item \underline{\emph{Energy transfer due to small $p_z$ and $p_y$. }}
In the range $(l_f/L_y)(l_f/L_z)\ll \beta\ll l_f/L_y, l_f/L_z$ we now have $\tilde{p}_z^+=\beta L_y/l_f$, and $\tilde{p}_z^- = l_f/L_z$: 

\begin{equation}
\begin{split}
\frac{\Tthree}{\tilde{\epsilon}}&=
\frac{l_f}{L_y}\int_{l_f/L_z}^{2\beta L_y/l_f}\frac{d\theta}{\pi}
+\frac{2}{\pi}\int_{l_f/L_z}^{\beta \frac{L_y}{l_f}}d\tilde{p}_z \phi^+  
\nonumber\\
&  \approx \frac{2}{\pi}\left(\beta -\frac{1}{2}\frac{l_f}{L_z}\frac{l_f}{L_y}\right)+
      \frac{2}\pi \beta \ln \left(\beta \frac{L_y}{l_f}\frac{ L_z}{l_f}\right) 
    \end{split}
\end{equation}

\item \underline{\emph{{Energy transfer from $p_y=0$ only.}}}
Finally, for $\frac{1}{2}(l_f/L_y)(l_f/L_z)<\beta<(l_f/L_y)(l_f/L_z)$, the tiny region where most of the contribution comes from are modes with $p_y=0$ and they do not decouple fully,
\begin{equation}
\begin{split}
    \frac{\Tthree}{\tilde{\epsilon}}\approx\frac{2}{\pi}\beta -\frac{1}{\pi}\frac{l_f}{L_z}\frac{l_f}{L_y}.
    \end{split}
\end{equation}
For even lower $\beta<\frac{1}{2}(l_f/L_y)(l_f/L_z)$, there is no energy transfer from the 3D waves to the 2D condensate using the Heaviside approximation of the oscillating factor. 
\end{enumerate}

All in all, we obtain the following asymptotic results:
\begin{equation} \frac{\Tthree}{\tilde{\epsilon}}=
  \begin{cases}  
    \frac{\epsilon_{\rm 3D}}{\tilde{\epsilon}}, \qquad \beta \geq \frac{1}{2}, \\
      \frac{2}{\pi}\beta \ln\left(\frac{2}{\beta}\right) + \beta - \frac{l_f}{L_z}, \quad \max\left\{\frac{l_f}{L_z}, \frac{l_f}{L_y} \right\} < \beta <\frac{1}{2}  \\\\
      \frac{2}\pi \beta \ln\left(\min\left(\frac{L_y}{l_f},\frac{ 2L_z}{l_f}\right)\right), \\ \quad \quad\min\left\{\frac{l_f}{L_z},\frac{l_f}{L_y}\right\} < \beta < \max\left\{\frac{l_f}{L_z},\frac{l_f}{L_y}\right\} \\ \\ 
      \frac{2}{\pi}\left(\beta -\frac{1}{2}\frac{l_f}{L_z}\frac{l_f}{L_y}\right)+
      \frac{2}\pi \beta \ln \left(\beta \frac{L_y}{l_f}\frac{ L_z}{l_f}\right), \\
   \quad\quad\quad  \frac{l_f}{L_y}\frac{l_f}{L_z} < \beta < \min\left\{\frac{l_f}{L_z},\frac{l_f}{L_y}\right\} \\\\
   \frac{2}{\pi}\left(\beta -\frac{1}{2}\frac{l_f}{L_z}\frac{l_f}{L_y}\right), \quad\frac{1}{2}\frac{l_f}{L_y}\frac{l_f}{L_z} < \beta <\frac{l_f}{L_y}\frac{l_f}{L_z} \\\\
      0, \qquad \qquad   \beta <\frac{1}{2}\frac{l_f}{L_y}\frac{l_f}{L_z}
  \end{cases}  
  \label{eq:dec_scalings2}
\end{equation}
with $\beta = \frac{U'}{2\Omega}\frac{L_y}{l_f}$.
We have included linear terms in $O(\beta)$ in Eq.~\eqref{eq:dec_scalings2}, so as to obtain better predictions for our DNS case where $L k_f = 10$. 
Note, however, that the asymptotic scalings are valid within each range when $\beta$ is sufficiently far from the bounds of the range. 
Thus, when using inequalities in \eqref{eq:dec_scalings2}, we 
have
derived a good leading-order approximation of the exact function, but where small discontinuities appear at the regime bounds.

The scalings in \eqref{eq:dec_scalings2} are shown in Fig.~\ref{fig:decoupling}(b) as a dashed red line, for the parameters corresponding to our DNS ($L_y/l_f= 10, L_z/l_f=5$), 
and are a very good approximation of the integration of the exact formula \eqref{eq:eps23_full}, shown as a green line.

\subsubsection{Including 2D-2D interactions}

We now need to include the fraction of energy injected into the 2D manifold, which is not affected by the resonant condition, and never decouples if forced.
As established in \S\ref{app:2D-2D}, $\Ttwotwo$ is half the energy injected in the 2D manifold.
With our isotropic forcing,
\begin{align}
\frac{\Ttwotwo}{\epsilon} = \frac{1}{2} \frac{l_f}{2L_z}    
\end{align}

The final asymptotic expression for the total energy transfer is therefore
\begin{equation} \frac{\Ttwo}{\epsilon}=
  \begin{cases}
  \frac{\Ttwotwo}{\epsilon} +
  \frac{\epsilon_{\rm 3D}}{\epsilon}, \qquad \beta \geq \frac{1}{2} \\\\
  \frac{\Ttwotwo}{\epsilon} +
  \tilde{\epsilon} \left(
  \beta - \frac{l_f}{L_z} +
      \frac{2}{\pi}\beta \ln\left(\frac{2}{\beta}\right) \right), \\
      \qquad\qquad 
      \max \{ \frac{l_f}{L_z}, \frac{l_f}{L_y} \} < \beta < \frac{1}{2}  \\\\
      \frac{\Ttwotwo}{\epsilon} + \tilde{\epsilon} \frac{2}\pi \beta \ln\left(2\frac{L_z}{l_f}\right), \\ \quad \quad \min \{ \frac{l_f}{L_z}, \frac{l_f}{L_y} \} <\beta < \max \{ \frac{l_f}{L_z}, \frac{l_f}{L_y} \} \\ \\ 
      \frac{\Ttwotwo}{\epsilon} +
      \tilde{\epsilon} \left(
      \frac{2}{\pi}\left(\beta -\frac{1}{2}\frac{l_f}{L_z}\frac{l_f}{L_y}\right) \right)+
      \frac{2}\pi \beta \ln \left(\beta \frac{L_y}{l_f}\frac{ L_z}{l_f}\right), \\
   \quad\quad\quad  \frac{l_f}{L_y}\frac{l_f}{L_z} < \beta <\min \{ \frac{l_f}{L_z}, \frac{l_f}{L_y} \} \\\\
    \frac{\Ttwotwo}{\epsilon} +
    \tilde{\epsilon} \frac{2}{\pi}\left(\beta -\frac{1}{2}\frac{l_f}{L_z}\frac{l_f}{L_y}\right), \quad\frac{1}{2}\frac{l_f}{L_y}\frac{l_f}{L_z} < \beta <\frac{l_f}{L_y}\frac{l_f}{L_z} \\\\
      \frac{\Ttwo}{\epsilon}, \qquad \qquad   \beta <\frac{1}{2}\frac{l_f}{L_y}\frac{l_f}{L_z}
  \end{cases}  
  \label{eq:eps2D_scalings}
\end{equation}

\subsection{Closure of the energy balance for the mean flow}

Here we derive the expression of both the condensate 
amplitude $\Up/\Omega$ and the energy transfer $\Tthree$ as a function of the control parameters $\Roe$ and $l_f/L_i$ ($i=x,y,z$).
We determine $\Up/\Omega$ from \eqref{eq:eps2D_scalings} by closing with 
the mean-flow energy balance
\begin{equation}
    \left(\frac{U'}{\Omega}\right)^2=4\Roe^2 \frac{\Ttwo}{\epsilon}
    \label{eq:MF2}
\end{equation}
which in terms of variable $ \beta =  \frac{\Up}{2\Omega} \frac{L_y}{l_f},$ reads
\begin{equation}
    \beta^2=\Roe^2 \left(\frac{L_y}{l_f}\right)^2\frac{\Ttwo}{\epsilon}\left(\beta\right).
    \label{eq:MF_beta}
\end{equation}
With the piecewise function $\Ttwo(\beta)$ in \eqref{eq:eps2D_scalings}, we can solve numerically the algebraic equation \eqref{eq:MF_beta}
via a root-finding procedure.
The resulting solution $\Up/\Omega$ is shown as a function of $\Roe$ as a red line in the inset of Fig.~\ref{fig:cond_rescaled}.
In the following, we instead proceed analytically by extracting the leading-order terms in \eqref{eq:eps2D_scalings} and closing 
with \eqref{eq:MF_beta} separately in each range in $\beta$.

\subsubsection{Closure in absence of energy injection into the 2D manifold}
\label{app:closure_wo2D}

It is simpler to first consider the case where $\epsilon_{\rm 2D}=0$,
for which we need to close the mean-flow equation \eqref{eq:MF_beta} with \eqref{eq:dec_scalings2}.
We also assume $L_z<L_y$ for shortness.

\begin{enumerate}
    \item \emph{\underline{For $\beta>\frac{1}2$}}, $\Ttwo=\epsilon_{\rm 3D}$ and the solution is $\beta =\Roe L_y/l_f$, which is valid for $\Roe>l_f/(2 L_y)$. So, in this regime we obtain the result
\begin{equation}
    \frac{U'}{\Omega}=2\frac{\epsilon_{\rm 3D}}{\epsilon} \Roe,
\end{equation}
which is a rotation-independent similar to the scaling of condensates in 2DNSE.

\item \emph{\underline{For $l_f/L_z\ll \beta\ll\frac{1}{2}$}}, consistent with $\Roe\ll l_f/L_y$, we have the balance
\begin{equation}
    \frac{\beta}{\ln(1/\beta)}=\frac{2}{\pi}\Roe^2 \left(\frac{L_y}{l_f}\right)^2 \equiv \alpha
\end{equation}
using Eq.~\eqref{eq:eps2D_scalings} (second row) at the leading order in $\beta$.
Since $\alpha \ll1$, solving this equation perturbatively gives
\begin{equation}
  \beta =\alpha \ln(1/\alpha)= - \frac{2}{\pi}\Roe^2 \left(\frac{L_y}{l_f}\right)^2 \ln\left(\frac{2}{\pi}\Roe^2 \left(\frac{L_y}{l_f}\right)^2\right),
  \end{equation}
with corrections of order $\alpha \ln(-\ln \alpha)$. From the definition of $\beta$ and the condition that $1/2\gg\beta\gg l_f/L_y,l_f/L_z$ we then obtain the solution
\begin{align}
    \frac{U'}{\Omega}=-\frac{8}{\pi}\Roe^2\frac{L_y}{l_f}\ln\left(\sqrt{\frac{2}{\pi}}\Roe\frac{L_y}{l_f}\right) \nonumber \\
   \frac{l_f}{L_y} \max\left(\sqrt{\frac{l_f}{L_y}},\sqrt{\frac{l_f}{L_z}}\right)\ll \Roe \ll \frac{l_f}{L_y}
\end{align}

\item \emph{\underline{For $ \frac{l_f}{L_y} <\beta < \frac{l_f}{L_z}$}}, the closure equation is trivial and we obtain at the leading order:
\begin{align}
    &~~~~~~~~~~~~~~~~ \frac{U'}{\Omega} = \frac{4}{\pi} \Roe^2 \frac{L_y}{l_f}\ln\left(\frac{2 L_z}{l_f}\right),
    \label{eq:UO_logLz}\\
    & \left(\frac{l_f}{L_y}\right)^{3/2}\frac{1}{\sqrt{\ln(L_z/l_f)}}\ll\Roe \ll  \frac{l_f}{L_y}\sqrt{\frac{l_f}{L_z}}\frac{1}{\sqrt{\ln(L_z/l_f)}}. \nonumber
\end{align}

\item  \emph{\underline{For $\frac{l_f}{L_y}\frac{l_f}{L_z}\ll\beta \ll \frac{l_f}{L_y}$}}, we have the balance:
\begin{equation}
    \beta^2=\Roe^2 \left(\frac{L_y}{l_f}\right)^2\frac{2}{\pi} \left[ \beta \ln \left(\beta \frac{L_y}{l_f}\frac{L_z}{l_f}\right)   + \beta - \frac{1}{2} \frac{l_f^2}{L_z L_y} \right].
    \label{eq:blogb+linear}
\end{equation}
We will simplify Eq.~\eqref{eq:blogb+linear} by considering two different regimes, separated by a crossover value $\beta^*$, where either the linear term or the term  $\beta \ln \left(\beta \frac{L_y}{l_f}\frac{L_z}{l_f}\right)$ in \eqref{eq:blogb+linear} is dominant.
Assuming that $\beta \gg \frac{1}{2} \frac{l_f}{L_y}\frac{l_f}{L_z}$, the crossover occurs when
\begin{equation}
    \beta^* \simeq e \frac{l_f}{L_y}\frac{l_f}{L_z}.
    \label{eq:betas}
\end{equation}
For $\beta \gg \beta^*$, the balance at leading order is
\begin{equation}
    \beta^2=\Roe^2 \left(\frac{L_y}{l_f}\right)^2\frac{2}{\pi} \beta \ln \left(\beta \frac{L_y}{l_f}\frac{L_z}{l_f}\right),
\end{equation}
which is of the form
\begin{align}
    &~~~~~~~~~~~~~~~~ \tilde{\beta} \ln \left(1/\tilde{\beta} \right) =  \tilde{\alpha} \ll1, \text{ with} \nonumber \\ &~~~~~ \frac{1}{\tilde{\alpha}}=\Roe^2 \left(\frac{L_y}{l_f}\right)^3\frac{L_z}{l_f}\frac{2}{\pi} \gg 1, \quad \frac{1}{\tilde{\beta}}= \beta \frac{L_y}{l_f}\frac{L_z}{l_f} \gg 1.
    \label{eq:blogb}
\end{align}
To leading order, the solution to Eq.~\eqref{eq:blogb} is
\begin{equation}
\tilde{\beta}= \frac{\tilde{\alpha}}{\log (1/\tilde{\alpha})}.
\end{equation}
(Formally, the solution is  $\tilde{\beta} =  - \tilde{\alpha}/W(-\tilde{\alpha})$ with $W(z)$ the Lambert-$W$ (ProductLog) function solving $W e^W = z$, and using the expansion of the Lambert function when $z\to0$ for its second branch -- the first branch being solved by $\tilde{\beta} =1$.)
We therefore obtain
\begin{equation}
    \beta = \frac{2}{\pi}  \Roe^2 \frac{L_y^2}{l_f^2}
    \ln \left( \frac{2}{\pi}  \Roe^2 \left(  \frac{L_y}{l_f} \right)^3  \frac{L_z}{l_f}\right)
\end{equation}
for $\Roe^2 > \frac{\pi}{2} \frac{l_f^3}{L_y^2 L_z}$, 
which gives
\begin{align}
    &\frac{U'}{\Omega}=\frac{4}{\pi}\Roe^2 \left(\frac{L_y}{l_f}\right) \ln\left(\Roe^2 \left(\frac{L_y}{l_f}\right)^3\frac{L_z}{l_f}\frac{2}{\pi}\right) \\
  &\text{for } \Roe^* < \Roe < \frac{l_f}{L_y}\min\left(\sqrt{\frac{l_f}{L_y}}, \sqrt{\frac{l_f}{L_z}}\right),
  \label{eq:range3}
\end{align}
The lower bound in \eqref{eq:range3} is 
\begin{align}
    \Roe^* =  \sqrt{\frac{\pi e}{2}}  \left(\frac{l_f}{L_y} \right)^{3/2} \left(\frac{l_f}{L_z} \right)^{1/2},
\end{align}
which corresponds to the value where $\beta$ equals the crossover $\beta^*$ \eqref{eq:betas}, hence where $\beta \ln \left(\beta L_y L_z/l_f^2 \right) = \beta $.

\item \emph{\underline{For $\frac{1}{2} \frac{l_f}{L_z} \frac{l_f}{L_y}  \ll \beta < \beta^*$}} (which includes the range $ \frac{1}{2}\frac{l_f}{L_z} \frac{l_f}{L_y}  \beta < \frac{l_f}{L_z} \frac{l_f}{L_y} $, the energy balance reads 
\begin{equation}
    \beta^2 =\Roe^2 \left(\frac{L_y}{l_f}\right)^2 \frac{2}{\pi}\left(\beta -\frac{1}2 \frac{l_f}{L_z}\frac{l_f}{L_y}\right),
\end{equation}
which is solved by
\begin{align}
    &\beta = \Roe^2 \left(\frac{L_y}{l_f}\right)^2 \frac{1}{\pi}\left(1+\sqrt{1-\pi\left(\frac{l_f}{L_y}\right)^3\frac{l_f}{L_z}\Roe^{-2}}\right),\\
    &\text{when }
    \Roe > \sqrt{\pi} \left(\frac{l_f}{L_y}\right)^{3/2}\sqrt{\frac{l_f}{L_z}},
\end{align}
where we chose the larger root for continuity at the upper boundary of the domain. 
Below 
\begin{align}
\Roe^c \equiv \sqrt{\pi} \left(\frac{l_f}{L_y}\right)^{3/2}\sqrt{\frac{l_f}{L_z}},
\end{align}
the mean-flow energy balance has no solution.

\end{enumerate}


Therefore, using the Heaviside as an approximation for the oscillating factor gives that the decoupling of the 2D condensate from 3D modes occurs at a finite $\Roe=\sqrt{\pi} \left(\frac{l_f}{L_y}\right)^{3/2}\sqrt{\frac{l_f}{L_z}}$ and as a first-order phase transition, i.e.\ in the absence of 2D forcing the condensate amplitude jumps from $\frac{U'}{\Omega}=\frac{2}{\pi}\Roe^2 \left(\frac{L_y}{l_f}\right)=2\pi (l_f/L_y)(l_f/L_z)$ to zero. S
Note that we have neglected the effect of finite viscosity on the waves, which should probably introduce a cutoff on the smallest amount of energy they could transfer to the condensate.  

To summarize the cases when $\frac{\Ttwotwo}{\epsilon}=0$,
\begin{widetext}
\begin{align}
    \frac{U'}{\Omega} = \begin{cases}
        2\Roe, \qquad\qquad 1\gg\Roe \gtrsim \frac{l_f}{ 2L_y} \\ \\
        -\frac{8}{\pi}\Roe^2\frac{L_y}{l_f}\ln\left(\sqrt{\frac{2}{\pi}}\Roe\frac{L_y}{l_f}\right),  \qquad\qquad
   \frac{l_f}{2L_y} \max\left(\sqrt{\frac{l_f}{L_y}},\sqrt{\frac{l_f}{L_z}}\right)\ll \Roe \ll \frac{l_f}{2 L_y} \\\\
\frac{4}{\pi} \Roe^2 \frac{L_y}{l_f}\ln\left( \min \left( \frac{L_y}{l_f} , \frac{2 L_z}{l_f}  \right) \right), \qquad\qquad
    \frac{l_f}{L_y}\sqrt{\frac{l_f}{L_z}} \frac{1}{\sqrt{\ln\left( \min \left(  \frac{L_y}{l_f}, \frac{2 L_z}{l_f} \right) \right) )}}\ll\Roe \ll \left(\frac{l_f}{L_y}\right)^{3/2} \frac{1}{\sqrt{\ln\left( \min \left(  \frac{L_y}{l_f}, \frac{2 L_z}{l_f} \right) \right) }} \\\\
     \frac{4}{\pi}\Roe^2 \frac{L_y}{l_f} \ln\left( \frac{2}{\pi} \Roe^2 \left(\frac{L_y}{l_f}\right)^3\frac{L_z}{l_f}\right), \qquad\qquad
     \sqrt{\frac{\pi e}{2}}
  \left(\frac{l_f}{L_y}\right)^{3/2} \sqrt{\frac{l_f}{L_z}} \ll \Roe \ll \frac{l_f}{L_y}\min\left(\sqrt{\frac{l_f}{L_y}}, \sqrt{\frac{2 l_f}{L_z}}\right) \\\\
  \frac{2}{\pi}\Roe^2 \frac{L_y}{l_f} \left(1+\sqrt{1-\pi\left(\frac{l_f}{L_y}\right)^3\frac{l_f}{L_z}\Roe^{-2}}\right), \qquad\qquad
    \sqrt{\pi} \left(\frac{l_f}{L_y}\right)^{3/2}\sqrt{\frac{l_f}{L_z}} <\Roe <  \sqrt{\frac{\pi e}{2}} \left(\frac{l_f}{L_y}\right)^{3/2}\sqrt{\frac{l_f}{L_z}}
    \end{cases}
    \label{eq:closure_wo2D}
\end{align}
\end{widetext}
where the third row generalizes solution \eqref{eq:UO_logLz} to $L_y< L_z$.
The corresponding expression for $\Tthree(\Roe)$ follows from 
$\frac{\Tthree}{\epsilon_{\rm 3D}} 
=
\left(\frac{\Up}{\Omega} \right)^2 \frac{1}{4 \Roe^2}$.

\subsubsection{Including an energy injection in the 2D manifold}
\label{app:closure_w2D}

We now consider the addition of forcing in the 2D manifold, like in our DNS where forcing is isotropic.
In the continuous limit, isotropic forcing leads to $\Ttwotwo = \frac{l_f}{4 L_z}$ (see \S \ref{app:2D-2D}).
This additional energy input is negligible for sufficiently large $\beta$, but dominates when enough waves have decoupled.
With our parameter choice, $\Ttwotwo = \frac{l_f}{4L_z}$ starts to dominate when $ \frac{l_f^2}{L_y L_z}<\beta < l_f/L_y$, that is in the third range in \eqref{eq:eps2D_scalings}.
For $\beta < \max \{ \frac{l_f}{L_z}, \frac{l_f}{L_y} \}$,
We can therefore approximate the decoupling trend as 
$\Tthree \simeq \tilde{\epsilon} \frac{2}{\pi} \ln (2 L_z/l_f)$ (third row in \eqref{eq:eps2D_scalings}), because the less steep part at lower $\beta$, due to $\tilde{\epsilon} \frac{2}{\pi} \log (\beta L_y L_z/l_f^2)$ (fourth row), will be dominated by $\Ttwotwo$.

By closing separately when either $\Ttwotwo$ or $\Tthree$ dominates, we obtain
\begin{align}
    \frac{\Up}{\Omega} &= 2 \Roe \frac{\epsilon_{\rm 3D} + \frac{1}{2} \epsilon_{\rm 2D}}{\epsilon}, 
    \qquad
    \frac{l_f}{2L_y} < \Roe < \frac{l_f}{2L_z}, \nonumber \\
    \frac{\Up}{\Omega} &= \frac{4}{\pi} \Roe^2 \frac{L_y}{l_f} \ln \left( \frac{2 L_z}{l_f}\right), \qquad  \Roe^* < \Roe \lesssim \frac{l_f}{L_y} \sqrt{\frac{l_f}{L_z}}, \nonumber\\
    \frac{\Up}{\Omega} &= 2\Roe \sqrt{ \frac{\Ttwotwo}{\epsilon}},  \qquad \Roe < \Roe^*  
    \label{eq:UO_scaling_w2d}
\end{align}
with the two last scalings intersecting at
\begin{align}
    \Roe^* &= \frac{\pi}{2} \frac{l_f}{L_y}
    \sqrt{\frac{\Ttwotwo}{\epsilon}}
    \frac{1}{\ln (2 L_z/l_f)} \\
    \beta^* &= \frac{\pi}{2} \frac{\Ttwotwo}{ \ln ( 2L_z/l_f)},
\end{align}
above which the input due to 3D-2D interactions dominates over $\Ttwotwo$.
With our DNS parameters,
$\Roe^*\simeq 0.0218$ and $\beta^* = 0.04$. 
Note that we do not derive an analytical solution in the range $\frac{l_f}{L_y} \sqrt{\frac{l_f}{L_z}} <\Roe < \frac{l_f}{2L_y}$, as the asymptotic analysis does not necesarily hold there.

Then we use this approximate solution to estimate the 3D-2D transfer, using the leading-order result for $\beta < l_f/L_z$ in \eqref{eq:dec_scalings2}.
Injecting \eqref{eq:UO_scaling_w2d} into \eqref{eq:dec_scalings2}, we obtain the closed formula
\begin{widetext}
\begin{align}
    \frac{\Tthree}{\epsilon} \left(\Roe \right)=
  \begin{cases}
       \frac{4}{\pi^2} \left(\frac{L_y}{l_f}\right)^2 \Roe^2 (\ln \frac{ \blue{2} L_z}{l_f} )^2, \qquad\qquad 
    \sqrt{\frac{\pi}{2 \ln ( \blue{2} L_z/l_f)}}(\frac{l_f}{L_y})^{3/2} < \Roe \lesssim  \frac{l_f}{L_y} \sqrt{\frac{l_f}{L_z}}\\ \\
    \frac{4}{\pi^2} \Roe^2 \left(\frac{L_y}{l_f}\right)^2 \ln \left(\frac{\blue{2} L_z}{l_f}\right)
    \ln \left[ \frac{2}{\pi} \Roe^2 \left(\frac{L_y}{l_f}\right)^3 \frac{\blue{2} L_z}{l_f} \ln \left( \frac{\blue{2} L_z}{l_f}\right)\right], \qquad\qquad  \Roe^* < \Roe  < 
     \sqrt{\frac{\pi}{2 \ln (\blue{2} L_z/l_f)}}(\frac{l_f}{L_y})^{3/2}
    \\ \\
      \frac{2}\pi \frac{L_y}{l_f} \Roe \sqrt{\frac{\Ttwotwo}{\epsilon}}  \ln \left( \Roe \sqrt{\frac{\Ttwotwo}{\epsilon}} (\frac{L_y}{l_f})^2 \frac{ L_z}{l_f}\right) \blue{+\frac{2}{\pi}\left(\frac{L_y}{l_f} \Roe \sqrt{\frac{\Ttwotwo}{\epsilon}} -\frac{1}{2}\frac{l_f}{L_z}\frac{l_f}{L_y}\right)} , \\
   \quad\quad\quad\qquad 
   \left(\frac{l_f}{L_y} \right)^2 \frac{l_f}{L_z} \sqrt{\frac{\epsilon}{\Ttwotwo}} 
     < \Roe < \Roe^*\\\\
    \frac{2}{\pi}\left(\frac{L_y}{l_f} \Roe \sqrt{\frac{\Ttwotwo}{\epsilon}} -\frac{1}{2}\frac{l_f}{L_z}\frac{l_f}{L_y}\right),  \qquad\qquad  \frac{1}{2} \left(\frac{l_f}{L_y} \right)^2 \frac{l_f}{L_z} \sqrt{\frac{\epsilon}{\Ttwotwo}} < \Roe  <
    \left(\frac{l_f}{L_y} \right)^2 \frac{l_f}{L_z} \sqrt{\frac{\epsilon}{\Ttwotwo}} \\\\
      0, \qquad \qquad   \Roe < \frac{1}{2} \left(\frac{l_f}{L_y} \right)^2 \frac{l_f}{L_z} \sqrt{\frac{\epsilon}{\Ttwotwo}} 
  \end{cases} \label{eq:eps32_leading} \\
  \text{ with }
    \Roe^* = \frac{\pi}{2} \frac{l_f}{L_y}
    \sqrt{\frac{\Ttwotwo}{\epsilon}}
    \frac{1}{\ln ( \blue{2} L_z/l_f)} \nonumber
\end{align}    
\end{widetext}
where the blue terms are $O\left(\beta = \frac{U'}{2\Omega} \frac{L_y}{l_f}\right)$ terms. We considered $L_z < L_y$ here for simplicity.
Solution \eqref{eq:eps32_leading} is visualized as a black solid line in Fig.~\ref{fig:eps32}(a), and fits very well our DNS data points for $\Roe < \frac{l_f}{L_y} \sqrt{\frac{l_f}{L_z}}$.

In presence of 2D forcing, the decoupling is therefore predicted to occur below
\begin{align}
\Roe^c =  \frac{1}{2} \left(\frac{l_f}{L_y} \right)^2 \frac{l_f}{L_z} \sqrt{\frac{\epsilon}{\Ttwotwo}} = \left(\frac{l_f}{L_y} \right)^2 \left(\frac{l_f}{L_z} \right)^{1/2}
\end{align}
and the transition is continuous.

\section{Discussions on the Lorentzian filter}
\label{app:lorentzian}
We can compare our usage of a Heaviside form for the oscillating factor \eqref{eq:Heaviside} with a Lorentzian form \eqref{eq:Lorentzian}, which is continuous in the time-scale ratio $\Up/\Omega$ and better weights the contribution of the oscillating factor over the time window $t \in [0 ~ 1/\Up]$.
We consider here a discrete set of waves and compute
\begin{align}
    \frac{\Tthree}{\epsilon} \left[ \frac{\Up}{\Omega} \right]
    &= \sum_{\bp} \chi_{\bp} ~ F_{k_f} \Big( \frac{\Up}{\Omega}\Big),
    \label{eq:e32_Lor}
\end{align}
with the Lorentzian filter
\begin{align}
     F_{k_f}
     = \frac{1}{ 1+ \left( \frac{\Omega}{\Up}  \frac{4\pi p_z p_y}{L_y k_f^3} \right)^2 }.
    \label{eq:Fkf_Lorentzian}
\end{align}
When writing \eqref{eq:e32_Lor}, we are applying the Lorentzian filter on the forcing rather than on the interaction itself, considering that the rest of the injected energy is transferred to other waves via wave-wave interactions.

The numerical value of \eqref{eq:e32_Lor} is shown as a blue line in Fig.~\ref{fig:decoupling}(b).
At the leading order, it follows a \SG{quadratic} scaling $\sim \left(\frac{\Up}{\Omega} \frac{L_y}{l_f} \right)^2$ (up to prefactors depending on the domain geometry), hence vanishes exactly at $\frac{\Up}{\Omega}=0$, contrary to the computation with the Heaviside filter.
The mean flow balance
\begin{align}
    \left(\frac{\Up}{\Omega}\right)^2 = 4 \Roe^2 \left( \frac{\Tthree}{\epsilon} + \frac{\epsilon_{\rm 2D}}{2 \epsilon} \right)
\end{align}
is then solved numerically with varying parameter $\Roe$, yielding the numerical estimate for $\Tthree(\Roe)$ shown in Fig.~\ref{fig:eps32}(a) (blue line).

With our isotropic forcing, which excites surviving 2D-2D interactions when $\Roe \to 0$, the solution when $\Roe \to 0$ therefore scales as
\begin{align}
    \frac{\Up}{\Omega} \sim 2 \Roe \sqrt{\frac{\epsilon_{\rm 2D}}{2 \epsilon}}, ~~~~
    \frac{\Tthree}{\epsilon} \sim 
    \Roe^2 \left(\frac{L_y}{l_f} \right)^2  \frac{\epsilon_{\rm 2D}}{  \epsilon}.
\end{align}
Therefore, with the inclusion of 
off-resonances
via the tails of the Lorentzian filter, there is always a small 3D-2D coupling until $\Roe=0$ exactly, at which 2D and 3D modes completely decouple.

Meanwhile, the Lorentzian filter better approximates the beginning of the decoupling regime ($\Roe \sim l_f/L_y$), and, in particular, the fact that $\Tthree/\epsilon$ is never 1, even when rotation is low (but $\Roe \lesssim l_f/L_z$ for waves to be  still homochiral).
This is due to the fact that the Lorentzian filter better weights the time window $[0 ~ 1/\Up]$ where the oscillating factor $\Delta(t)$ is not exactly one, but decays with time $t$.
When rotation is low, time-scale separation is limited and it matters to weight this time window appropriately, as achieved with the Lorentzian filter.

Note that with a purely-3D forcing, as $\Tthree(U'/\Omega) \propto (\Up/\Omega)^2$ with the Lorentzian filter, there is no closed solution for the condensate below some finite $\Roe$, similarly to the Heaviside predictions in \S \ref{app:closure_wo2D}. Therefore, full decoupling should also occur below a finite value of $\Roe$ in this case.

\section{Off-resonant terms and validity range of the QL predictions}
\label{app:off-reso}

\begin{figure}
    \centering
    \includegraphics[width=0.9\linewidth]{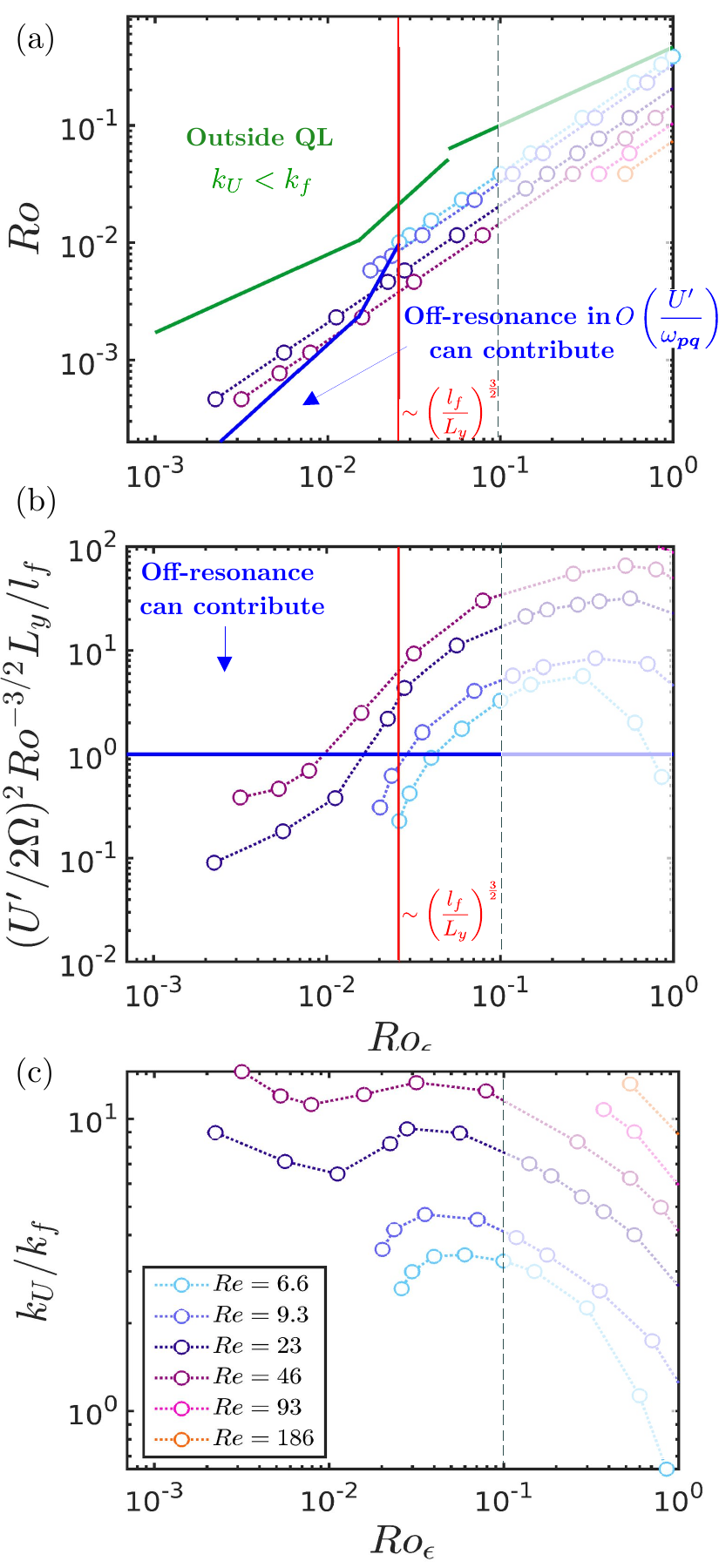}
    \caption{
    Validity range of our QL predictions (a) in the $(\Roe,Ro)$ plane and (b) in the $(U'/\Omega, \Roe)$ plane. 
    \resub{(c) Cutoff wavenumber $k_U/k_f$ \eqref{eq:kU} measured from the DNS.}
    Above the green line in (a), $k_U/k_f <1 $ and the QL approximation does not hold.
    In the sector above the blue line in (a), and below the dashed line in (b), wave-wave interactions dominate over off-resonant condensate-wave interactions, and our predictions are valid.
    The red line shows $\Roe=\sqrt{\frac{\pi}{2 \ln (2L_z/l_f)}} \left(\frac{l_f}{L_y}\right)^{3/2}$, below which $\Tthree$ starts to be $o(U'/\omega_{\bp\bq})$ and the order of off-resonant terms needs to be discussed. Points $\Roe>1$ are outside the current framework and are discussed in \citep{gome2025helicity}.
    }
    \label{fig:validity}
\end{figure}

When enforcing the resonance condition with either a Heaviside \eqref{eq:filter} or a Lorentzian function \eqref{eq:Lorentzian_main} of the single variable $U'$, the contribution from off-resonant terms are set to either zero (with the Heaviside) or to an order $O((U'/\omega_{\bp\bq}^{ss})^2)$ for each mode $\bp$ (with the Lorentzian).
This is a good assumption as long as the number of near-resonant terms is of order one, but it requires justification in the decoupling regime, where the contribution from near-resonant modes is of order $\Tthree\propto U'/\omega_{\bp\bq}^{ss} \sim \beta=\frac{U'}{2\Omega} \frac{L_y}{l_f} $.
There, while resonant terms have an $O(1)$ magnitude, they are few in number, of order $O\left(\frac{U'}{2\Omega} \frac{L_y}{l_f} \right)$, while most of the modes are off-resonant.
Neglecting off-resonances must therefore be taken with caution:
concretely, if off-resonances were to contribute to the energy transfer at order $ O\left(\frac{U'}{2\Omega} \frac{L_y}{l_f} \right)$, they would lead to the same net contribution as near-resonances.

Assuming this to be the case due to an off-resonant 3D-2D interaction rate of order $O\left(\frac{U'^2}{2\Omega} \frac{L_y}{l_f} \right)$, 
our predictions could give the correct leading order if the latter rate was smaller than $1/\tau_{nl}^{\rm wave}$, yielding the condition $(U'/2 \Omega)^2 L_y/l_f < Ro^{3/2}$. 
Here we use an isotropic wave-wave interaction time scale $\tau_{nl}^ {\rm wave} \sim \Omega^{\frac{1}{2}}\epsilon^{-\frac{1}2} k_f^{-1}$, expected for off-resonant modes $p_z\sim p_y \sim k_f$. This corresponds to the assumption that off-resonant 2D-wave contributions are depleted at leading order by wave-wave interactions, which are faster, and instead produce a sub-leading $o\left(\frac{U'}{2\Omega} \frac{L_y}{l_f} \right)$ contribution to the 2D condensate.

In the absence of 2D forcing, we have $U'/\Omega \sim \Roe^2 L_y/l_f$ in the decoupling regime (see Eq.~\eqref{eq:closure_wo2D}), and we obtain the restriction $Re< Ro^{-5/2}(l_f/L_y)^3$ or, equivalently, $\Roe < Ro^{3/8}(l_f/L_y)^{3/4}$. 
In the presence of 2D forcing, we have seen that for $\Roe < \Roe^* \sim l_f/L_y \sqrt{\Ttwotwo/\epsilon}$, $U'/\Omega \sim \Roe \sqrt{\Ttwotwo/\epsilon} $ (see Eq.~\eqref{eq:U_O_w2d}b), giving the validity condition $\Roe < 2 Ro^{3/4}\sqrt{\epsilon/\Ttwotwo}\sqrt{l_f/L_y}$ for wave-wave interactions to occur faster than off-resonant 2D-3D interactions.
If this condition is not respected
below the condition $\Roe < \Roe^*$, purely 2D-2D interactions will still dominate and govern the leading order of $U'/\Omega$, whether off-resonant 3D-2D interactions contribute or not. 
However, 3D-2D off-resonances would then contribute at $O\left(\frac{U'}{\Omega} \frac{L_y}{l_f} \right)$ and alter the scaling of $\Tthree$, which should be at most of $O\left(\frac{U'}{\Omega} \frac{L_y}{l_f} \right)$ for $\Roe< \Roe^*$.
Note that for $\Roe > \Roe^*$, $\Up/\Omega$ scales as $\propto \Roe^2 L_y/l_f$ (Eq.~\eqref{eq:U_O_w2d}a), which modifies the validity condition into $\Roe <  Ro^{3/8} \left(\frac{l_f}{L_y}\right)^{3/4}$.

We show this validity range across the decoupling regime as a blue line in Figure \ref{fig:validity}(a). Most of our DNS data points lie within this validity range, with, in the worse cases, $1/\tau_{nl}^{\rm wave}$ being of the same order as $U'^2 /\Omega (L_y/l_f)$.
In addition, the condition $(U'/2 \Omega)^2 L_y/l_f < Ro^{3/2}$ can be verified directly from the measurement of the condensate amplitude, see Fig.~\ref{fig:validity}(b).
Note that $\Tthree$ lies below $O\left( \frac{U'}{\Omega} \frac{L_y}{l_f}  \right)$ for $\Roe < \sqrt{\frac{\pi}{2 \ln (2 L_z/l_f)}}(\frac{l_f}{L_y})^{3/2}$, marked as a vertical red line in Figs.~\ref{fig:validity}(a) and (b).
Around this line, neglecting off-resonances could lead to errors of the same order as the contribution from near resonances in $\frac{U'}{\Omega} \frac{L_y}{l_f}$, but not modify the leading-order scaling by an order of magnitude. Some points in our DNS lie outside the validity range in this region, for which our predictions should be treated with caution. For $\Tthree >\frac{U'}{\Omega} \frac{L_y}{l_f} $, for $\Roe$ above the vertical red line in Figs.~\ref{fig:validity}(a) and (b), most interactions are near-resonant so the additional condition \eqref{eq:off-reso_condition} needs not be respected for our theory to be valid.

Finally, the QL validity condition, $1/\Up \ll \tau_{nl}^{\rm wave}$, can be verified \resub{in the worse case scenario ($p_z=2\pi/L_z$)} by computing
\begin{equation}
    \frac{k_U}{k_f} = 
     \begin{cases}
  \frac{1}{Ro} \left(\frac{l_f}{L_z} \right)^{\frac{1}{3}} 
  (2 \Roe)^{\frac{2}{3}}, ~~~
  \frac{l_f}{2L_y} < \Roe < \frac{l_f}{2L_z}\\
    \frac{1}{Ro} \left(\frac{l_f}{L_z} \right)^{\frac{1}{3}} 
    \left[
    \frac{4}{\pi} \frac{L_y}{l_f} 
    \ln \left( \frac{2L_z}{l_f}\right) \right]^{\frac{2}{3}}
    \Roe^{\frac{4}{3}}, \\
    ~~~~~~~~~~~~~~~~~~~~~~~~~~~~~~~~~ \Roe^* < \Roe < \frac{l_f}{2L_y} \\    
    \frac{1}{ Ro} \left(\frac{l_f}{L_z} \right)^{\frac{1}{3}} \left(\frac{\epsilon_{\rm 2D}}{2\epsilon}\right)^{\frac{1}{3}}
    (2 \Roe)^{\frac{2}{3}},
    ~~~ \Roe < \Roe^* 
    \end{cases}
\end{equation}
with the predictions obtained with the Heaviside approximation and isotropic forcing.
The line bounding the validity range of the QL theory, $k_U/k_f =1 $, is visualized as a green line in Figure \ref{fig:validity}(a).
We also measure $k_U/k_f$ directly from the DNS in Fig.~\ref{fig:validity}(c).
Our DNS data points \resub{are reasonably well within} the validity range of the QL approximation \resub{at high $Re$}, with $k_U/k_f = O(10)$ for $Re=23$ and 46. However, taking the low-$Re$ limit generally breaks this condition.

With purely-3D forcing, as $\Up/\Omega \sim \Roe^2 L_y/l_f$ in most regimes, the QL approximation holds when 
$\frac{k_U}{k_f} \sim \frac{1}{ Ro} \left( \frac{l_f}{L_z} \right)^{1/3} \left( \frac{L_y}{l_f}\right)^{2/3} \Roe^{4/3} \gg 1$, i.e.\ when $Re \gg Ro^{-1/2} \frac{l_f}{L_y} \left(\frac{L_z}{l_f}\right)^{1/2}$.

\bibliography{bib}
\end{document}